\documentclass[%
reprint,
amsmath,amssymb,
aps,
prl,
]{revtex4-2}

\newif\ifarxiv
\arxivtrue

\usepackage[utf8]{inputenc}
\usepackage{graphicx}
\usepackage{dcolumn}
\usepackage{bm}
\usepackage[hidelinks]{hyperref}
\ifarxiv\else
  \usepackage{xr}
\fi

\newcommand{\runin}[1]{\textit{#1}—}

\begin{document}

\title{Committors and Reaction Rates from Trial Functions That Violate the Boundary Conditions}

\author{Magnus Petersen$^{1, 2}$}
\email{mapetersen@fias.uni-frankfurt.de}
\author{Simon Lichtinger$^{1, 2}$}
\email{slichtinger@fias.uni-frankfurt.de}
\author{Roberto Covino$^{1, 2, 3}$}
\email[Contact author: ]{covino@fias.uni-frankfurt.de}

\affiliation{$^{1}$Institute of Computer Science, Goethe University Frankfurt, Frankfurt am Main, Germany.}
\affiliation{$^{2}$Frankfurt Institute for Advanced Studies, Frankfurt am Main, Germany.}
\affiliation{$^{3}$Cluster of Excellence SubCellular Architecture of Life (SCALE), Goethe University Frankfurt, Frankfurt am Main, Germany.}

\begin{abstract}
The committor is the optimal reaction coordinate for a rare transition: it
pinpoints the transition state and fixes the rate, and it governs events from
protein folding to crystal nucleation. It minimises a Dirichlet energy, whose
value at the minimum is the reactive flux, over functions that vanish on the
reactant state and equal one on the product state. In high dimensions such a
trial space is very hard to build. Here we rewrite the variational principle so that the
boundary conditions are replaced by a normalisation of one boundary observable,
the fidelity. Any trial function is then admissible, including
functions that cannot satisfy the boundary values at all. On this basis we
estimate the high-dimensional committor and rates from one-dimensional profiles along
projected coordinates, taking as input only pre-existing equilibrium or
reweighted configurations, a diffusion constant estimate, and the two state definitions. The optimum has a
closed form that is cheap to evaluate. We obtain committors for AIB9 and villin HP-35 in full
torsion space, and folding and unfolding rates for chignolin from umbrella
sampling alone.
\end{abstract}

\maketitle

The committor $q(\mathbf{x})$ is the probability that a trajectory from
configuration $\mathbf{x}$ reaches a product state $B$ before a reactant state
$A$. It is the optimal reaction coordinate for any rare transition between two
long-lived states~\cite{du_transition_1998,berezhkovskii_one-dimensional_2005,
e_transition-path_2010,bolhuis_transition_2002,hummer_transition_2004,
best_reaction_2005,krivov_protein_2018,berezhkovskii_diffusion_2013}. 
Committors appear well beyond chemical physics: as the fixation probability
of an allele between loss and fixation~\cite{kimura_probability_1962}, and
in forecasts of El Ni\~no~\cite{lucente_committor_2022} and sudden
stratospheric warming~\cite{finkel_learning_2021}. 
In molecular simulations, the committor
serves several purposes. Transition-path theory (TPT) converts it into a reaction
rate~\cite{e_transition-path_2010,vanden-eijnden_transition_2006,roux_transition_2022,
berezhkovskii_committors_2019}.
Compared with candidate coordinates it identifies the degrees of freedom that
govern a transition~\cite{best_reaction_2005,ma_automatic_2005,peters_obtaining_2006,
lechner_nonlinear_2010,jung_machine-guided_2023,strand_high-dimensional_2024}.
As the ideal collective variable it directs enhanced sampling onto the
barrier~\cite{kang_computing_2024,trizio_everything_2025,rotskoff_active_2022}.
It governs the generation of reactive trajectories, both as the optimal
shooting distribution in path sampling~\cite{jung_machine-guided_2023,
lazzeri_molecular_2023,lazzeri_optimal_2025} and, through Doob's
$h$-transform, as the optimal control that produces them
directly~\cite{hartmann_efficient_2012,yuan_optimal_2024,
singh_splitting_2024,holdijk_stochastic_2023}. 
It can also help to analyse single-molecule
experiments~\cite{manuel_reconstructing_2015,covino_molecular_2019}.

Computing the committor in high dimensions is difficult, and not only because of
cost. The committor minimises a
Dirichlet energy over functions that vanish on $A$ and equal one on $B$, and
building a trial space inside that class is the obstacle.
A smooth interpolant between the states must also select a route between them,
which is the reaction coordinate one is trying to find. Building the space cheaply
from low-dimensional pieces fails as well: along almost any single projection the
images of $A$ and $B$ overlap, so no piece can be pinned to zero on one state and
one on the other. Constructing a trial space that is both admissible and accurate therefore
requires anticipating the committor itself.

Existing methods pay for this differently. Discretisations are limited by
dimension: grids~\cite{metzner_illustration_2006},
clusterings~\cite{noe_constructing_2009,metzner_transition_2009} and diffusion
maps~\cite{coifman_diffusion_2006,banisch_diffusion_2020,evans_computing_2023}
confine the committor to a handful of collective variables, while point-cloud
and tensor-network schemes~\cite{lai_point_2018,chen_committor_2023} scale
better but have not reached molecular data. Every other method must supply the
boundary values, and only two routes exist. They can be imposed exactly, on a
trial space that then has to be built: Galerkin expansions of the backward
operator state the requirement directly~\cite{thiede_galerkin_2019}, and it
carries through the dynamical Galerkin
approximation~\cite{strahan_long-time-scale_2021}, its
neural~\cite{strahan_inexact_2023} and kernel~\cite{aristoff_fast_2024}
successors, and reactive-flux principles inside a collective-variable
subspace~\cite{roux_string_2021,he_committor-consistent_2022}. Or they can be
penalised~\cite{khoo_solving_2019,li_computing_2019,rotskoff_active_2022,
chen_discovering_2023,megias_iterative_2025,hasyim_supervised_2022,
contreras-arredondo_learning_2026,strahan_predicting_2023}, but a penalised
objective is not a Dirichlet energy: its minimiser depends on the penalty
weight, and no variational bound remains. Both routes read the operator off
time-lagged pairs, swarms of short trajectories, or shooting
points~\cite{bolhuis_transition_2002,ma_automatic_2005,peters_obtaining_2006},
and iterative schemes require fresh trajectories at every
cycle~\cite{kang_computing_2024,trizio_everything_2025,jung_machine-guided_2023,
lazzeri_molecular_2023,mitchell_committor_2024}. 

In this Letter we show that the boundary conditions can be
removed from the variational principle altogether, neither imposed nor
penalised and without forfeiting the variational bound, and replaced by a
single scalar constraint that pre-existing samples supply. The boundary
conditions enter the variational problem in one place only, a cross-term between
the trial function and the unknown committor, and imposing them on a trial space
makes that term collapse. We evaluate it
instead. Because the reactive current is conserved, the cross-term factorises
into the reactive flux, a single global unknown shared by every trial function,
times a boundary observable of the trial function alone. We call the second
factor the \emph{fidelity}; it equals one when the boundary values hold. The
Dirichlet principle then holds with the boundary conditions replaced by a
normalisation of the fidelity. Since the fidelity enters multiplicatively rather
than as a penalty, a trial space in which the boundary values cannot be satisfied
at all remains admissible. Along a projected coordinate the empirical fidelity has
a closed form in terms of histograms of the projected state samples. We therefore
use the cheapest trial space available: one-dimensional profiles along a set of
projection directions, a ridge-function ansatz~\cite{pinkus_ridge_2015,
friedman_projection_1981} in the spirit of sliced optimal
transport~\cite{rabin_wasserstein_2012,bonneel_sliced_2015,
kolouri_generalized_2019} and score matching~\cite{song_sliced_2020}. The
directions are drawn from a distribution adapted to the state labels alone, never
from dynamical information. The resulting estimator, which we call the
\emph{sliced committor}, costs one tridiagonal solve per profile and one
symmetric linear system for the optimum. The inputs are the two state definitions
and configurations reweightable to equilibrium, so biased data --- from umbrella
sampling~\cite{torrie_nonphysical_1977} or
metadynamics~\cite{laio_escaping_2002} --- serve as well as unbiased dynamics.

\begin{figure*}[t]
\centering
\includegraphics[width=0.95\textwidth]{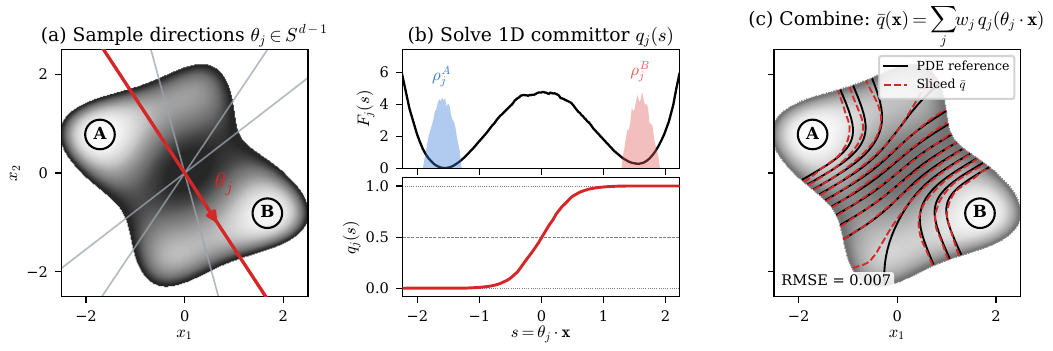}
\caption{\label{fig:schematic}
The sliced committor in three steps.
(a)~Projection directions $\bm\theta_j \in S^{d-1}$ (here isotropic; grey, one highlighted in
red for illustration) are overlaid on equilibrium samples from the rotated
Wolfe--Quapp potential with metastable states $A$ and $B$.
(b)~Along each direction, the projected free energy
$F_j(s)=-\beta^{-1}\log\rho_j(s)$ (in $k_\mathrm{B}T$) and the
state-restricted marginals $\rho_j^A$, $\rho_j^B$ (top) determine the
one-dimensional profile $q_j(s)$
[bottom; the minimiser of Eq.~(\ref{eq:Ekappa})].
(c)~Weighted combination
$\bar q(\mathbf{x}) = c + \sum_j w_j\,q_j(\bm\theta_j\!\cdot\!\mathbf{x})$
with optimal weights and bias from Eq.~(\ref{eq:wstar}); level sets of $\bar q$
(red dashed) match the finite-difference PDE reference (black) with
transition-region root-mean-square error (RMSE) $= 0.007$.
}
\end{figure*}

\runin{The obstruction}Consider reversible overdamped Langevin dynamics on a
domain $\Omega\subseteq\mathbb{R}^d$ at inverse temperature $\beta$, with
equilibrium density $\rho(\mathbf{x})\propto e^{-\beta V(\mathbf{x})}$ for a
potential $V$ and a symmetric positive-definite diffusion tensor $\mathbf{D}$,
taken constant here. The committor is the unique minimiser of the Dirichlet
energy, or reactive-flux functional,
\begin{equation}
\mathcal{E}[u] = \int_\Omega \rho\,\nabla u^\top \mathbf{D}\,\nabla u\,d\mathbf{x}
\label{eq:dirichlet}
\end{equation}
over functions $u$ obeying $u|_A=0$ and $u|_B=1$, and it solves
$\nabla\!\cdot\!(\rho\mathbf{D}\nabla q)=0$ with those
values~\cite{bovier_metastability_2004,bovier_metastability_2015,
lelievre_partial_2016}. Let
$\langle u,v\rangle_{\mathbf{D}}=\int_\Omega\rho\,\nabla u^\top\mathbf{D}\nabla v\,
d\mathbf{x}$ be the bilinear form whose diagonal is $\mathcal{E}$, so that
$\mathcal{E}[u]=\langle u,u\rangle_{\mathbf{D}}$ is the square of a seminorm that annihilates
constants.

We judge a trial function $u$ by the Dirichlet energy of its error, which expands
into three terms,
\begin{equation}
\mathcal{E}[u-q]=
\underbrace{\mathcal{E}[u]}_{\text{computable}}
\;-\;2\,\underbrace{\langle u,q\rangle_{\mathbf{D}}}_{\text{unknown }q}
\;+\;\underbrace{\mathcal{E}[q]}_{\text{constant}} .
\label{eq:error_expansion}
\end{equation}
The first term is a sample average and the third is independent of $u$. The second
couples $u$ to the unknown. If $u$ satisfies the boundary values, integration by
parts reduces it to the constant $\mathcal{E}[q]$ and it drops out. This is the
Galerkin orthogonality that requires trial functions satisfying the boundary conditions. We instead evaluate the cross-term.

\runin{Flux--fidelity identity}Let $\Omega'$ be the transition region between the
two states, on which $q$ satisfies the backward Kolmogorov equation, and let
$\mathbf{J}=\rho\mathbf{D}\nabla q$ be the reactive
current~\cite{e_transition-path_2010,vanden-eijnden_transition_2006}. Since $q$ is
constant on the closed basins, $\mathbf{J}$ vanishes there, and the current leaves
$\partial A$ and enters $\partial B$ with a definite sign. With
$\mathbf{n}_A,\mathbf{n}_B$ the basin normals pointing into $\Omega'$, its total is
the reactive flux,
\begin{equation}
\nu_{AB}=\!\int_{\partial A}\!\mathbf{J}\!\cdot\!\mathbf{n}_A\,dS
=-\!\int_{\partial B}\!\mathbf{J}\!\cdot\!\mathbf{n}_B\,dS=\mathcal{E}[q].
\label{eq:nuab}
\end{equation}
Dividing each boundary integral by $\nu_{AB}$ gives a unit-mass average, written
$\langle\cdot\rangle^{\mathrm{flux}}_{\partial A}$ and
$\langle\cdot\rangle^{\mathrm{flux}}_{\partial B}$. Green's identity
(\hyperref[em:ffi]{End Matter}) then gives, for any $u$ of finite Dirichlet
energy,
\begin{equation}
\langle u,q\rangle_{\mathbf{D}}=\nu_{AB}\,\mathcal{F}[u],
\qquad
\mathcal{F}[u]=\langle u\rangle^{\mathrm{flux}}_{\partial B}
              -\langle u\rangle^{\mathrm{flux}}_{\partial A}.
\label{eq:ffi}
\end{equation}
The interior of $q$ cancels because the current is conserved: its streamlines run
from $\partial A$ to $\partial B$ without closing or ending inside $\Omega'$, so
integrating $\nabla u$ along them leaves only an end-to-end difference. 

The fidelity $\mathcal{F}[u]$ is the value of $u$ where current leaves into $B$
minus its value where current enters from $A$; it depends on $u$ only through
the two state boundaries. It is linear, annihilates constants, and lies in
$[-1,1]$ for $u\in[0,1]$. A trial function satisfying the boundary values has
$\mathcal{F}[u]=1$, and for $u\in[0,1]$ the converse holds as well (End Matter):
unit fidelity \emph{is} the boundary condition. The shortfall $1-\mathcal{F}[u]$ quantifies the violation. Setting $u=q$ gives $\mathcal{F}[q]=1$ and recovers
Eq.~(\ref{eq:nuab}) as a check. In Eq.~(\ref{eq:ffi}) $u$ could be a Galerkin
basis, a neural network or the projection slices below.

\runin{A variational principle without boundary conditions}The identity converts
the Dirichlet principle into one that does not refer to boundary conditions.
Cauchy--Schwarz gives
$\langle u,q\rangle_{\mathbf{D}}^2\le\mathcal{E}[u]\,\mathcal{E}[q]$.
Substituting Eq.~(\ref{eq:ffi}) and $\mathcal{E}[q]=\nu_{AB}$, and dividing by
$\nu_{AB}>0$, gives
$\nu_{AB}\le\mathcal{E}[u]/\mathcal{F}[u]^2$ for $\mathcal{F}[u]\neq0$. The bound
is attained at $u=q$, where $\mathcal{F}[q]=1$ and $\mathcal{E}[q]=\nu_{AB}$.
Hence
\begin{equation}
\nu_{AB}=\min_{u\,:\,\mathcal{F}[u]\neq0}\;\frac{\mathcal{E}[u]}{\mathcal{F}[u]^2}
\label{eq:principle}
\end{equation}
over all trial functions of finite energy, with no boundary conditions imposed.
The Cauchy--Schwarz equality condition identifies the minimisers as
$u=\lambda q+\mathrm{const}$ with $\lambda\neq0$ (\hyperref[em:principle]{End
Matter}). The quotient is scale- and
offset-free.

Restricting to $\{u:\mathcal{F}[u]=1\}$ recovers $\nu_{AB}=\min\mathcal{E}[u]$:
two function-valued boundary conditions have become one scalar equation.
Trial functions violating the boundary values already support a variational upper
bound on the rate in discrete time~\cite{lorpaiboon_exact_2026}. What is new here
is that the functional replacing the boundary conditions is measurable from
labelled configurations alone.

An additive penalty and a multiplicative normalisation differ in what survives the relaxation. A penalty suppresses the boundary violation, driving the trial function
toward the boundary values. The ratio in Eq.~(\ref{eq:principle}) is
scale-invariant and divides the violation out, the family
$\lambda q+\mathrm{const}$ attaining the minimum. Where the boundary values are
merely awkward to impose the two agree. Where they cannot be imposed at all they
do not, and that is the regime of one-dimensional projections: the images of $A$
and $B$ overlap along the directions we use, so no projection takes the boundary values
anywhere.

Restricted to a finite trial space $V_M$ of dimension $M$ the principle still
bounds the flux. Define the attainable fidelity--energy ratio
\begin{equation}
\mathcal{R}_M=\max_{u\in V_M}\frac{\mathcal{F}[u]^2}{\mathcal{E}[u]}
\;\le\;\frac{1}{\nu_{AB}} .
\label{eq:ratio}
\end{equation}
Then $\nu_{AB}\le1/\mathcal{R}_M$, and since adding a direction nests
$V_M\subset V_{M+1}$, enriching the trial space can only tighten the bound, with
$1/\nu_{AB}$ as the ceiling. Monotonicity gives a stopping criterion: what a finer
rung recovers lower-bounds the coarser one's Dirichlet error, so enrich until that
margin is small~\cite{supplemental}.

\runin{One dimensional projections (slices)}Since Eq.~(\ref{eq:principle}) imposes no boundary
conditions, the trial space can be chosen for cost and inductive bias. Molecular committors depend on
few collective coordinates~\cite{best_reaction_2005,bolhuis_reaction_2000,
bittracher_transition_2018,bittracher_dimensionality_2021}, which
suggests a ridge-function ansatz~\cite{pinkus_ridge_2015,friedman_projection_1981}.
Given equilibrium samples $\{\mathbf{x}_n\}_{n=1}^N\sim\rho$, each labelled $A$,
$B$ or neither, and $M$ unit directions $\bm\theta_j\in S^{d-1}$, we write
\begin{equation}
\bar q(\mathbf{x})=c+\sum_{j=1}^{M}w_j\,\tilde q_j(\mathbf{x}),
\qquad
\tilde q_j(\mathbf{x})=q_j(\bm\theta_j\!\cdot\!\mathbf{x}) ,
\label{eq:ansatz}
\end{equation}
a bias $c$ plus one-dimensional profiles $q_j$ of the projected coordinate
$s=\bm\theta_j\!\cdot\!\mathbf{x}$, combined with weights $w_j$
[Fig.~\ref{fig:schematic}(a,c)]. From here $\bar q$ denotes this ansatz, while $u$
remains an arbitrary trial function. The directions are drawn from a
label-adapted law~\cite{supplemental}, never from dynamical information.

Any profile is admissible. Let
$\rho_j(s)=\int\rho\,\delta(s-\bm\theta_j\!\cdot\!\mathbf{x})\,d\mathbf{x}$ be the
marginal density of $s$, and $\rho^A_j,\rho^B_j$ its restrictions to the
configurations labelled $A$ and $B$: three histograms of the $N$ numbers
$\bm\theta_j\!\cdot\!\mathbf{x}_n$. The natural choice is the one-dimensional
committor along $\bm\theta_j$, but the projected states overlap along almost any
direction, and the profile would flatten where the samples
are~\cite{supplemental}. We therefore ask less: not that the profile satisfy the
boundary conditions, but that it pay for violating them. The profile should
minimise
\begin{equation}
\mathcal{E}_\kappa[q_j]=\int\!\rho_j\,(q_j')^2\,ds
+\kappa\!\int\!\rho^A_j\,q_j^{2}\,ds
+\kappa\!\int\!\rho^B_j\,(1-q_j)^2\,ds ,
\label{eq:Ekappa}
\end{equation}
(primes denote $d/ds$), the one-dimensional Dirichlet energy plus a penalty of
strength $\kappa>0$; one tridiagonal solve per direction gives the
profile~\cite{supplemental} [Fig.~\ref{fig:schematic}(b)]. As $\kappa\to\infty$
the penalty dominates where either state has weight and the profile becomes the
Bayesian classifier~\cite{ma_automatic_2005,peters_obtaining_2006,
lechner_nonlinear_2010}
\begin{equation}
q_j(s)\;\longrightarrow\;\frac{\rho^B_j(s)}{\rho^A_j(s)+\rho^B_j(s)} ,
\label{eq:classifier}
\end{equation}
which is insensitive to $\kappa$. This is not the penalty criticised above: it
acts on a basis function rather than the committor, and in a limit where its
weight drops out, so Eq.~(\ref{eq:principle}) is untouched.

\runin{Fidelity from samples}Everything so far is exact. One approximation now
enters, confined to a single input. The fidelity of a slice is a flux-weighted
average over the state boundaries, which equilibrium samples do not resolve. We
replace it by the average over the state samples. With
$\mu_S[f]=N_S^{-1}\sum_{\mathbf{x}_n\in S}f(\mathbf{x}_n)$ the average of $f$ over
the $N_S$ configurations labelled $S$, and the basin moments
$a_j=\mu_A[\tilde q_j]$ and $b_j=\mu_B[\tilde q_j]$,
\begin{equation}
\mathcal{F}_j\equiv\mathcal{F}[\tilde q_j]\;\approx\;\hat{\mathcal{F}}_j=b_j-a_j .
\label{eq:phihat}
\end{equation}
Both averages act on the same $\tilde q_j$, so they differ only through its
variation across a basin. The absorbers hold that variation to $\sim\!10^{-3}$
wherever the projected states separate; where they overlap it is larger, but those
slices carry correspondingly small weight, and the combination retains unit
fidelity to within $0.007$~\cite{supplemental}.

In the $\kappa\to\infty$ limit the empirical fidelity has a closed form. Let
$m^A=\int_A\rho\,d\mathbf{x}$ and $m^B=\int_B\rho\,d\mathbf{x}$ be the state
populations, which no projection alters. Substituting
Eq.~(\ref{eq:classifier}) into the two basin moments leaves the same integral
$\mathcal{I}_j$ in both, supported only where both projected states have weight,
giving $a_j=\mathcal{I}_j/m^A$ and $b_j=1-\mathcal{I}_j/m^B$
(\hyperref[em:basis]{End Matter}), so that
\begin{equation}
\hat{\mathcal{F}}_j\longrightarrow 1-
\Bigl(\frac{1}{m^A}+\frac{1}{m^B}\Bigr)
\!\!\int\limits_{\mathrm{supp}\rho^A_j\cap\,\mathrm{supp}\rho^B_j}\!\!\!\!
\frac{\rho^A_j\rho^B_j}{\rho^A_j+\rho^B_j}\,ds ,
\label{eq:overlap}
\end{equation}
one minus an overlap integral of the projected state densities, half their
harmonic mean over their common support. The empirical fidelity is affine in that
overlap, with a slope shared by every direction: a slice whose projected states
are disjoint has $\hat{\mathcal{F}}_j=1$, and overlap degrades it smoothly. Two
histograms per direction therefore suffice. Equation~(\ref{eq:overlap}) is also
where the boundary values are lost: the projected states overlap along the
directions we use, so no slice attains unit fidelity.

The residual bias in Eq.~(\ref{eq:phihat}) is signed and small. The equilibrium
measure weights the basin interior while the reactive current concentrates on the
saddle-facing boundary, so $\hat{\mathcal{F}}_j$ slightly overestimates
$\mathcal{F}_j$, by $0.015$ both on a two-dimensional benchmark where
$\mathcal{F}_j$ is known exactly and on the deployed AIB9 slices against a
reactive-trajectory estimate. Using
the reactive-flux fidelities in place of the equilibrium ones changes the AIB9
committor by an RMSE of $0.004$ (Ref.~\cite{supplemental}, \emph{Empirical
fidelity on a molecular system}). We collect the per-slice values into
$\mathbf f=(\mathcal{F}_1,\dots,\mathcal{F}_M)$ and
$\hat{\mathbf f}=\mathbf b-\mathbf a$.

\runin{The optimum in closed form}Insert Eq.~(\ref{eq:ansatz}) into
Eq.~(\ref{eq:ratio}). The bias drops from the energy, giving
$\mathcal{E}[\bar q]=\mathbf{w}^\top G\,\mathbf{w}$ with the Gram matrix
\begin{equation}
G_{jk}=(\bm\theta_j^\top\mathbf{D}\,\bm\theta_k)\!\int\!\rho\,q_j'\,q_k'\,
d\mathbf{x} ,
\label{eq:gram}
\end{equation}
projection geometry times an equilibrium average of the slice derivatives. The
fidelity is linear in the weights, $\mathcal{F}[\bar q]=\mathbf f^\top\mathbf{w}$,
so maximising $(\mathbf f^\top\mathbf{w})^2/(\mathbf{w}^\top G\,\mathbf{w})$ is a
rank-one generalised eigenvalue problem with the closed-form solution
(\hyperref[em:solve]{End Matter})
\begin{equation}
\mathbf{w}^\star=\frac{G^{-1}\hat{\mathbf f}}{\hat{\mathcal{R}}_M},
\qquad
c^\star=-\mathbf{a}^\top\mathbf{w}^\star,
\qquad
\hat{\mathcal{R}}_M=\hat{\mathbf f}^\top G^{-1}\hat{\mathbf f},
\label{eq:wstar}
\end{equation}
with attained energy $\mathcal{E}[\bar q^\star]=1/\hat{\mathcal{R}}_M$
[Fig.~\ref{fig:schematic}(c)]. One Cholesky factorisation gives the committor. The
cost is $\mathcal{O}(NdM)$ to project the samples, $\mathcal{O}(NM^2)$ for $G$ and
$\mathcal{O}(M^3)$ to factorise it, linear in sample count and dimension at fixed
$M$.

The scale and the offset in Eq.~(\ref{eq:wstar}) are calibration rather than
optimisation. The quotient fixes only the ray
$\mathbf{w}\propto G^{-1}\hat{\mathbf f}$; along it we take the point with
$\hat{\mathbf f}^\top\mathbf{w}=1$ and then the offset with
$\mu_A[\bar q^\star]=0$. These are the two boundary values imposed weakly, on
sample averages, $\mu_A[\bar q]=0$ and $\mu_B[\bar q]=1$, whose difference is
$\hat{\mathbf f}^\top\mathbf{w}=1$. Unit fidelity is the only scale that returns
$q$ rather than a multiple of it when $q$ lies in the trial space. Minimising the
Dirichlet error instead gives
$\mathbf{w}_{\mathrm{Gal}}=\nu_{AB}G^{-1}\mathbf f$, which needs the unknown flux
and has fidelity $\nu_{AB}\mathcal{R}_M\le1$, so it returns a shrunken committor.
The shrinkage is avoided at the cost of a second-order error penalty
(\hyperref[em:solve]{End Matter}).

The slice space is overcomplete in practice, $M>d$, and $G$ is ill-conditioned, so
$G^{-1}$ denotes a Tikhonov-regularised solve. We fix $\varepsilon$ by
minimising the quotient of Eq.~(\ref{eq:principle}) on held-out samples, so the
method carries no fitted constant~\cite{supplemental}. Since
$(G+\varepsilon I)^{-1}\preceq G^{-1}$ for $\varepsilon>0$, regularisation
decreases $\hat{\mathcal{R}}_M$ and therefore loosens the bound of
Eq.~(\ref{eq:ratio}), which is the safe direction. Using $\hat{\mathbf f}$ in
place of $\mathbf f$ does not. To first order the shift in $\hat{\mathcal{R}}_M$
has the sign of
$(\hat{\mathbf f}-\mathbf f)^\top\mathbf{w}^\star=1-\mathcal{F}[\bar q]$, not that
of the per-slice bias, since $\mathbf{w}^\star$ is sign-indefinite; measured, that
deficit is small and positive ($0.007$ on AIB9, and at most $0.004$ on the
two-dimensional systems where $\mathbf f$ is known
exactly~\cite{supplemental}), so $\hat{\mathcal{R}}_M$ is biased high and
$1/\hat{\mathcal{R}}_M$ biased low. The two biases act in opposite directions. The
bound is exact in $\mathbf f$ and approximate in $\hat{\mathbf f}$.

\runin{An error bound without the answer}For any trial function with
$\mathcal{F}[\bar q]=1$ the inner product in the error expansion is the constant
$\nu_{AB}$, so
\begin{equation}
\mathcal{E}[\bar q-q]=\mathcal{E}[\bar q]-\nu_{AB}\;\ge\;0 .
\label{eq:apost}
\end{equation}
The Dirichlet error is the difference of two numbers, neither of which requires
$q$. The first is the ansatz energy, a sample average, equal to
$1/\hat{\mathcal{R}}_M$ at the optimum. The second is measured from the same samples by
current conservation: the reactive current through every isocommittor surface is
the same constant, so stratifying
$\rho\,\nabla\bar q^\top\mathbf{D}\nabla\bar q$ on $\bar q$ and reading the
plateau across the transition region estimates $\nu_{AB}$. This filters the
basin-side contamination that makes the whole-volume energy
overshoot~\cite{berezhkovskii_diffusion_2013}.

Error estimates for dynamical quantities already exist: a computable lower
bound on the committor's mean-squared error from two lag
times~\cite{lorpaiboon_exact_2026}, and a bound on the error of dynamical spectral
estimation~\cite{webber_error_2021}. Both are framed for trajectory-based
estimators and need dynamical input.
Equation~(\ref{eq:apost}) is the counterpart in the Dirichlet seminorm, the metric
this construction minimises at unit fidelity, and it uses only the samples at
hand. The flux is the estimated term, and Eq.~(\ref{eq:ratio}) tests it: any trial
function caps $\nu_{AB}$, so a plateau above the cap of a finer fit is
inadmissible, as it is on AIB9~\cite{supplemental}. Substituting that cap turns
the gap into a lower bound on the error rather than a certificate; a large gap
calls for more directions.

\begin{figure*}[!t]
\centering
\includegraphics[width=0.95\textwidth]{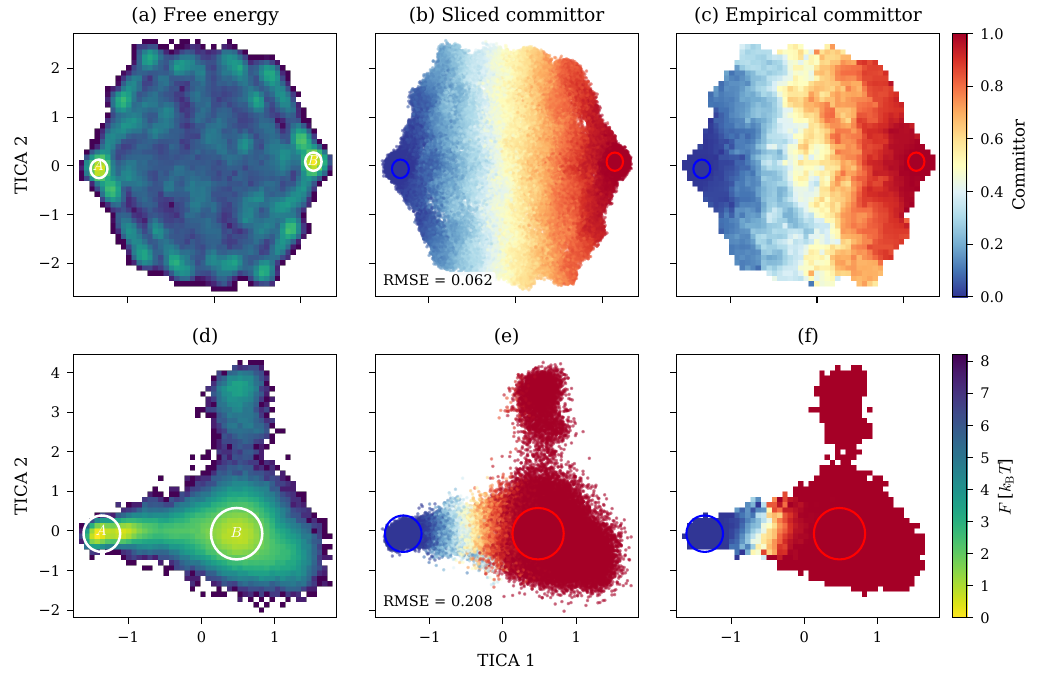}
\caption{\label{fig:molecular}
Sliced committor for molecular systems. In both rows the committor is computed in
the full sine and cosine torsion feature space, $d=52$ for AIB9 and $d=350$ for
villin; the time-lagged independent component analysis (TICA) plane is a
two-dimensional viewing projection and the frame in which $A$ and $B$ are
defined, and is never an input to the fit. Top row, AIB9 peptide:
(a)~free-energy landscape in the leading TICA subspace, with states $A$ and $B$;
(b)~sliced committor with broad Fisher linear-discriminant (LDA) directions,
projected onto the leading TICA components;
(c)~reference empirical committor from forward-tracing.
Bottom row, villin HP-35, with directions from a tightly concentrated Fisher LDA
sampler:
(d)~free-energy landscape in the leading TICA subspace with states $A$ and $B$;
(e)~sliced committor, projected onto the leading TICA components;
(f)~empirical forward-tracing reference.
Parameters are in Tables~\ref{tab:aib9_params} and~\ref{tab:villin_params};
simulation details in Ref.~\cite{supplemental}.
}
\end{figure*}

\runin{AIB9 and villin HP-35}We first validated the method on the
two-dimensional model against a finite-difference reference. From exact Boltzmann
samples the transition-region RMSE was $0.007$
[Fig.~\ref{fig:schematic}(c)], a best case with perfect sampling. We then computed
committors for two molecular systems from full torsion-angle representations.

AIB9 is a 9-residue $\alpha$-aminoisobutyric acid peptide; villin HP-35 is a
35-residue three-helix bundle, here the wild type (PDB 1YRF) at $345$~K. Their
feature spaces had $d=52$ and $d=350$, with $M=512$ and $M=2048$ directions.
For both, $A$ and $B$ were ellipsoidal basins
[Fig.~\ref{fig:molecular}(a,d)] in the leading TICA
subspace~\cite{perez-hernandez_identification_2013,schwantes_improvements_2013}
built from sine- and cosine-encoded torsions, the directions followed a Fisher LDA
sampler, and we took $\mathbf{D}=\mathbf{I}$ in this feature space. The sampler's
two parameters were fixed the same way as $\varepsilon$, by minimising the
held-out quotient of Eq.~(\ref{eq:principle}); no reference committor, and no
error measured against one, enters the choice of any parameter reported
here~\cite{supplemental}. The reference
was the empirical committor, obtained by forward-tracing each equilibrium frame to
its first hit of $A$ or $B$. The sliced committor tracked it closely for AIB9, with
transition-region RMSE $0.062$ [Fig.~\ref{fig:molecular}(b,c)]. For villin HP-35
the RMSE was $0.21$ [Fig.~\ref{fig:molecular}(e,f)], where the reference was the
limiting factor: only about $15$ committed folding events occurred over the
$398~\mu\mathrm{s}$ trajectory, so the $q\!\to\!0$ side is thin and the reactive
region undersampled~\cite{supplemental}.

\begin{figure}[t]
\centering
\includegraphics[width=0.95\columnwidth]{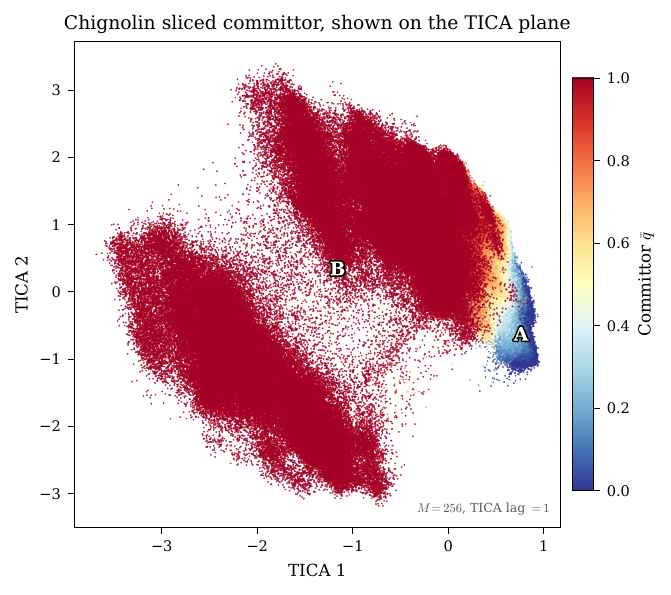}
\caption{\label{fig:chignolin}
Chignolin (CLN025) sliced committor, computed in the torsion-angle
representation from three pooled umbrella-sampling replicates. Shown on a
leading TICA plane and coloured from folded ($A$, $\bar q\!\to\!0$) to unfolded
($B$, $\bar q\!\to\!1$).}
\end{figure}

\runin{Committor and rate for chignolin}Our last example recovered a rate from
biased data, where the bias improves barrier sampling but corrupts the global
first-passage time~\cite{tiwary_metadynamics_2013}. We used $340$~K umbrella-sampling simulations of chignolin
(CLN025)~\cite{honda_crystal_2008}, a fast-folding 10-residue $\beta$-hairpin,
reweighted to equilibrium with the multistate Bennett acceptance ratio
(MBAR)~\cite{shirts_statistically_2008}. The committor was built from the
reweighted ensemble alone, by the same procedure and torsion representation as
before ($d=86$, $M=256$) but with the states defined by the umbrella coordinate
itself, $Q>0.85$ and $Q<0.30$ [Fig.~\ref{fig:chignolin}]. No long unbiased
trajectory was available to trace, so we validated through the rate.

The TPT rate identity turns the flux into a rate, $k_{A\to B}=\nu_{AB}/p_A$ and
$k_{B\to A}=\nu_{AB}/p_B$, with $p_A=\int\!\rho\,q^-\,d\mathbf{x}$ and
$p_B=\int\!\rho\,q\,d\mathbf{x}$ the probabilities of last having been in $A$ and
in $B$, $q^-=1-q$ the backward committor, and the same flux in both directions by
detailed balance. We took $A$ as the folded state, so $q$ increased toward
unfolded and $k_{B\to A}$ was the folding rate. The flux was read from the
isocommittor plateau: the whole-volume energy $\mathcal{E}[\bar q]$ is
contaminated in the basins, where $\bar q$ is least accurate and $\rho$ largest,
so we instead took the current $\Phi(z)$ through each surface $\{\bar q=z\}$, with
$\mathcal{E}[\bar q]=\int_0^1\Phi(z)\,dz$~\cite{berezhkovskii_diffusion_2013}, and
read $\nu_{AB}$ as the band median of $\Phi(z)$ across the transition region.

Computing the sliced committor does not require a diffusion estimate, while the rate does. We
took it along the umbrella coordinate $Q$, the fraction of native
contacts~\cite{best_native_2013}, which already defines the biased ensemble, so no
new information entered. We obtained the diffusivity along $Q$ from the
within-window variance and integrated autocorrelation time,
$D_Q=\mathrm{Var}(Q)/\tau_{\mathrm{int}}(Q)$~\cite{hummer_position-dependent_2005},
and converted it to the configuration-space value by
$D_0=D_Q/\langle|\nabla Q|^2\rangle$~\cite{berezhkovskii_time_2011,
best_coordinate_2010}. Collapsing the position-dependent $D_Q(Q)$ to one constant
is the main approximation for calculating the rate: $k\propto D_0$, so it sets the overall
scale. The folding and unfolding rates were $2.2$ and $0.20~\mu\mathrm{s}^{-1}$,
pooled over three replicates, each carrying a factor $2$--$3$ uncertainty from the
incompletely converged unfolded free energy~\cite{supplemental}. Both agreed with
two independent sets of $340$~K equilibrium simulations to within a factor of 2.5.
Against the reference rates of Lazzeri \emph{et al.}~\cite{lazzeri_molecular_2023},
obtained in the same native-contact coordinate, ours were $0.9\times$
and $0.7\times$; against those of Lindorff-Larsen \emph{et
al.}~\cite{lindorff-larsen_how_2011}, $1.3\times$ and $0.4\times$.

\runin{Summary and outlook}The boundary conditions enter the Dirichlet error of
a trial committor in one place, and what they do there is measurable. Removing
them lets the trial space be chosen for cost: cheap one-dimensional profiles
combined by one symmetric linear system, on configurations reweightable to
equilibrium and two state definitions, with no clustering, basis, training or lag
selection. The committor reproduced a finite-difference reference in two
dimensions, tracked forward-tracing references for AIB9 and villin HP-35, and for
chignolin gave folding and unfolding rates from umbrella-sampling data alone,
within a small factor of two independent estimates. What one more refinement
recovers lower-bounds its Dirichlet error, and says when to stop.

Two things limit the method. The flux-weighted fidelity is replaced by a
basin-sample average, which is where the theorems become estimates; the
substitution is quantified rather than assumed. And a projection-based trial space
is only as good as its directions: isotropic draws on $S^{d-1}$ miss a narrow
reactive subspace when one exists, which a label-adapted
sampler~\cite{decao_power_2020} repairs at the cost of an added component.
Non-equilibrium steady states, with their irreversible currents, and
non-stationary landscapes remain open. The identity of Eq.~(\ref{eq:ffi}) is tied
neither to slices nor to chemistry: it needs only a reversible process with two
absorbing sets, and it gives any variational committor method a boundary
diagnostic in place of a boundary constraint.


\begin{acknowledgments}
We thank the Goethe University Frankfurt, the Frankfurt Institute for Advanced Studies, and the Johanna Quandt foundation for support.
M.P.\ acknowledges funding from the Stiftung Polytechnische Gesellschaft.
We acknowledge support from the Cluster of Excellence SubCellular Architecture
of Life (SCALE) and from hessian.AI.
The authors gratefully acknowledge the computing time provided to them at the
NHR Center NHR@SW at Goethe University Frankfurt and at the the Gauss Centre for Supercomputing e.V.on the supercomputer JUPITER.
We were assisted by Claude (Opus 4.8 and 5).
\end{acknowledgments}

\runin{Data and Code Availability}
The code implementing the sliced committor and the analysis scripts that
reproduce the results are openly available in an archived reproduction package,
\mbox{doi:10.5281/zenodo.21703983}~\cite{ourcode}, which also contains the AIB9
trajectory and the chignolin umbrella-sampling inputs.
The villin HP-35 trajectory is the property of D.~E.~Shaw Research and is
available from them upon request~\cite{piana_protein_2012}.

\bibliography{references}

\begin{thebibliography}{96}%
\makeatletter
\providecommand \@ifxundefined [1]{%
 \@ifx{#1\undefined}
}%
\providecommand \@ifnum [1]{%
 \ifnum #1\expandafter \@firstoftwo
 \else \expandafter \@secondoftwo
 \fi
}%
\providecommand \@ifx [1]{%
 \ifx #1\expandafter \@firstoftwo
 \else \expandafter \@secondoftwo
 \fi
}%
\providecommand \natexlab [1]{#1}%
\providecommand \enquote  [1]{``#1''}%
\providecommand \bibnamefont  [1]{#1}%
\providecommand \bibfnamefont [1]{#1}%
\providecommand \citenamefont [1]{#1}%
\providecommand \href@noop [0]{\@secondoftwo}%
\providecommand \href [0]{\begingroup \@sanitize@url \@href}%
\providecommand \@href[1]{\@@startlink{#1}\@@href}%
\providecommand \@@href[1]{\endgroup#1\@@endlink}%
\providecommand \@sanitize@url [0]{\catcode `\\12\catcode `\$12\catcode `\&12\catcode `\#12\catcode `\^12\catcode `\_12\catcode `\%12\relax}%
\providecommand \@@startlink[1]{}%
\providecommand \@@endlink[0]{}%
\providecommand \url  [0]{\begingroup\@sanitize@url \@url }%
\providecommand \@url [1]{\endgroup\@href {#1}{\urlprefix }}%
\providecommand \urlprefix  [0]{URL }%
\providecommand \Eprint [0]{\href }%
\providecommand \doibase [0]{https://doi.org/}%
\providecommand \selectlanguage [0]{\@gobble}%
\providecommand \bibinfo  [0]{\@secondoftwo}%
\providecommand \bibfield  [0]{\@secondoftwo}%
\providecommand \translation [1]{[#1]}%
\providecommand \BibitemOpen [0]{}%
\providecommand \bibitemStop [0]{}%
\providecommand \bibitemNoStop [0]{.\EOS\space}%
\providecommand \EOS [0]{\spacefactor3000\relax}%
\providecommand \BibitemShut  [1]{\csname bibitem#1\endcsname}%
\let\auto@bib@innerbib\@empty
\bibitem [{\citenamefont {Du}\ \emph {et~al.}(1998)\citenamefont {Du}, \citenamefont {Pande}, \citenamefont {Grosberg}, \citenamefont {Tanaka},\ and\ \citenamefont {Shakhnovich}}]{du_transition_1998}%
  \BibitemOpen
  \bibfield  {author} {\bibinfo {author} {\bibfnamefont {R.}~\bibnamefont {Du}}, \bibinfo {author} {\bibfnamefont {V.~S.}\ \bibnamefont {Pande}}, \bibinfo {author} {\bibfnamefont {A.~Y.}\ \bibnamefont {Grosberg}}, \bibinfo {author} {\bibfnamefont {T.}~\bibnamefont {Tanaka}},\ and\ \bibinfo {author} {\bibfnamefont {E.~S.}\ \bibnamefont {Shakhnovich}},\ }\bibfield  {title} {\bibinfo {title} {On the transition coordinate for protein folding},\ }\href {https://doi.org/10.1063/1.475393} {\bibfield  {journal} {\bibinfo  {journal} {The Journal of Chemical Physics}\ }\textbf {\bibinfo {volume} {108}},\ \bibinfo {pages} {334} (\bibinfo {year} {1998})}\BibitemShut {NoStop}%
\bibitem [{\citenamefont {Berezhkovskii}\ and\ \citenamefont {Szabo}(2005)}]{berezhkovskii_one-dimensional_2005}%
  \BibitemOpen
  \bibfield  {author} {\bibinfo {author} {\bibfnamefont {A.~M.}\ \bibnamefont {Berezhkovskii}}\ and\ \bibinfo {author} {\bibfnamefont {A.}~\bibnamefont {Szabo}},\ }\bibfield  {title} {\bibinfo {title} {One-dimensional reaction coordinates for diffusive activated rate processes in many dimensions},\ }\href {https://doi.org/10.1063/1.1818091} {\bibfield  {journal} {\bibinfo  {journal} {The Journal of Chemical Physics}\ }\textbf {\bibinfo {volume} {122}},\ \bibinfo {pages} {014503} (\bibinfo {year} {2005})}\BibitemShut {NoStop}%
\bibitem [{\citenamefont {E}\ and\ \citenamefont {Vanden-Eijnden}(2010)}]{e_transition-path_2010}%
  \BibitemOpen
  \bibfield  {author} {\bibinfo {author} {\bibfnamefont {W.}~\bibnamefont {E}}\ and\ \bibinfo {author} {\bibfnamefont {E.}~\bibnamefont {Vanden-Eijnden}},\ }\bibfield  {title} {\bibinfo {title} {Transition-{Path} {Theory} and {Path}-{Finding} {Algorithms} for the {Study} of {Rare} {Events}},\ }\href {https://doi.org/10.1146/annurev.physchem.040808.090412} {\bibfield  {journal} {\bibinfo  {journal} {Annual Review of Physical Chemistry}\ }\textbf {\bibinfo {volume} {61}},\ \bibinfo {pages} {391} (\bibinfo {year} {2010})}\BibitemShut {NoStop}%
\bibitem [{\citenamefont {Bolhuis}\ \emph {et~al.}(2002)\citenamefont {Bolhuis}, \citenamefont {Chandler}, \citenamefont {Dellago},\ and\ \citenamefont {Geissler}}]{bolhuis_transition_2002}%
  \BibitemOpen
  \bibfield  {author} {\bibinfo {author} {\bibfnamefont {P.~G.}\ \bibnamefont {Bolhuis}}, \bibinfo {author} {\bibfnamefont {D.}~\bibnamefont {Chandler}}, \bibinfo {author} {\bibfnamefont {C.}~\bibnamefont {Dellago}},\ and\ \bibinfo {author} {\bibfnamefont {P.~L.}\ \bibnamefont {Geissler}},\ }\bibfield  {title} {\bibinfo {title} {Transition {Path} {Sampling}: {Throwing} {Ropes} over {Rough} {Mountain} {Passes}, in the {Dark}},\ }\href {https://doi.org/10.1146/annurev.physchem.53.082301.113146} {\bibfield  {journal} {\bibinfo  {journal} {Annual review of physical chemistry}\ }\textbf {\bibinfo {volume} {53}},\ \bibinfo {pages} {291} (\bibinfo {year} {2002})}\BibitemShut {NoStop}%
\bibitem [{\citenamefont {Hummer}(2004)}]{hummer_transition_2004}%
  \BibitemOpen
  \bibfield  {author} {\bibinfo {author} {\bibfnamefont {G.}~\bibnamefont {Hummer}},\ }\bibfield  {title} {\bibinfo {title} {From transition paths to transition states and rate coefficients},\ }\href {https://doi.org/10.1063/1.1630572} {\bibfield  {journal} {\bibinfo  {journal} {The Journal of Chemical Physics}\ }\textbf {\bibinfo {volume} {120}},\ \bibinfo {pages} {516} (\bibinfo {year} {2004})}\BibitemShut {NoStop}%
\bibitem [{\citenamefont {Best}\ and\ \citenamefont {Hummer}(2005)}]{best_reaction_2005}%
  \BibitemOpen
  \bibfield  {author} {\bibinfo {author} {\bibfnamefont {R.~B.}\ \bibnamefont {Best}}\ and\ \bibinfo {author} {\bibfnamefont {G.}~\bibnamefont {Hummer}},\ }\bibfield  {title} {\bibinfo {title} {Reaction coordinates and rates from transition paths},\ }\href {https://doi.org/10.1073/pnas.0408098102} {\bibfield  {journal} {\bibinfo  {journal} {Proceedings of the National Academy of Sciences}\ }\textbf {\bibinfo {volume} {102}},\ \bibinfo {pages} {6732} (\bibinfo {year} {2005})}\BibitemShut {NoStop}%
\bibitem [{\citenamefont {Krivov}(2018)}]{krivov_protein_2018}%
  \BibitemOpen
  \bibfield  {author} {\bibinfo {author} {\bibfnamefont {S.~V.}\ \bibnamefont {Krivov}},\ }\bibfield  {title} {\bibinfo {title} {Protein {Folding} {Free} {Energy} {Landscape} along the {Committor} - the {Optimal} {Folding} {Coordinate}},\ }\href {https://doi.org/10.1021/acs.jctc.8b00101} {\bibfield  {journal} {\bibinfo  {journal} {Journal of Chemical Theory and Computation}\ }\textbf {\bibinfo {volume} {14}},\ \bibinfo {pages} {3418} (\bibinfo {year} {2018})}\BibitemShut {NoStop}%
\bibitem [{\citenamefont {Berezhkovskii}\ and\ \citenamefont {Szabo}(2013)}]{berezhkovskii_diffusion_2013}%
  \BibitemOpen
  \bibfield  {author} {\bibinfo {author} {\bibfnamefont {A.~M.}\ \bibnamefont {Berezhkovskii}}\ and\ \bibinfo {author} {\bibfnamefont {A.}~\bibnamefont {Szabo}},\ }\bibfield  {title} {\bibinfo {title} {Diffusion along the {Splitting}/{Commitment} {Probability} {Reaction} {Coordinate}},\ }\href {https://doi.org/10.1021/jp403043a} {\bibfield  {journal} {\bibinfo  {journal} {The Journal of Physical Chemistry B}\ }\textbf {\bibinfo {volume} {117}},\ \bibinfo {pages} {13115} (\bibinfo {year} {2013})}\BibitemShut {NoStop}%
\bibitem [{\citenamefont {Kimura}(1962)}]{kimura_probability_1962}%
  \BibitemOpen
  \bibfield  {author} {\bibinfo {author} {\bibfnamefont {M.}~\bibnamefont {Kimura}},\ }\bibfield  {title} {\bibinfo {title} {On the probability of fixation of mutant genes in a population},\ }\href {https://doi.org/10.1093/genetics/47.6.713} {\bibfield  {journal} {\bibinfo  {journal} {Genetics}\ }\textbf {\bibinfo {volume} {47}},\ \bibinfo {pages} {713} (\bibinfo {year} {1962})}\BibitemShut {NoStop}%
\bibitem [{\citenamefont {Lucente}\ \emph {et~al.}(2022)\citenamefont {Lucente}, \citenamefont {Herbert},\ and\ \citenamefont {Bouchet}}]{lucente_committor_2022}%
  \BibitemOpen
  \bibfield  {author} {\bibinfo {author} {\bibfnamefont {D.}~\bibnamefont {Lucente}}, \bibinfo {author} {\bibfnamefont {C.}~\bibnamefont {Herbert}},\ and\ \bibinfo {author} {\bibfnamefont {F.}~\bibnamefont {Bouchet}},\ }\bibfield  {title} {\bibinfo {title} {Committor {Functions} for {Climate} {Phenomena} at the {Predictability} {Margin}: {The} {Example} of {El} {Ni\~no}--{Southern} {Oscillation} in the {Jin} and {Timmermann} {Model}},\ }\href {https://doi.org/10.1175/JAS-D-22-0038.1} {\bibfield  {journal} {\bibinfo  {journal} {Journal of the Atmospheric Sciences}\ }\textbf {\bibinfo {volume} {79}},\ \bibinfo {pages} {2387} (\bibinfo {year} {2022})}\BibitemShut {NoStop}%
\bibitem [{\citenamefont {Finkel}\ \emph {et~al.}(2021)\citenamefont {Finkel}, \citenamefont {Webber}, \citenamefont {Gerber}, \citenamefont {Abbot},\ and\ \citenamefont {Weare}}]{finkel_learning_2021}%
  \BibitemOpen
  \bibfield  {author} {\bibinfo {author} {\bibfnamefont {J.}~\bibnamefont {Finkel}}, \bibinfo {author} {\bibfnamefont {R.~J.}\ \bibnamefont {Webber}}, \bibinfo {author} {\bibfnamefont {E.~P.}\ \bibnamefont {Gerber}}, \bibinfo {author} {\bibfnamefont {D.~S.}\ \bibnamefont {Abbot}},\ and\ \bibinfo {author} {\bibfnamefont {J.}~\bibnamefont {Weare}},\ }\bibfield  {title} {\bibinfo {title} {Learning {Forecasts} of {Rare} {Stratospheric} {Transitions} from {Short} {Simulations}},\ }\href {https://doi.org/10.1175/MWR-D-21-0024.1} {\bibfield  {journal} {\bibinfo  {journal} {Monthly Weather Review}\ }\textbf {\bibinfo {volume} {149}},\ \bibinfo {pages} {3647} (\bibinfo {year} {2021})}\BibitemShut {NoStop}%
\bibitem [{\citenamefont {Vanden-Eijnden}(2006)}]{vanden-eijnden_transition_2006}%
  \BibitemOpen
  \bibfield  {author} {\bibinfo {author} {\bibfnamefont {E.}~\bibnamefont {Vanden-Eijnden}},\ }\bibfield  {title} {\bibinfo {title} {Transition {Path} {Theory}},\ }in\ \href {https://doi.org/10.1007/3-540-35273-2_13} {\emph {\bibinfo {booktitle} {Computer {Simulations} in {Condensed} {Matter} {Systems}: {From} {Materials} to {Chemical} {Biology} {Volume} 1}}},\ \bibinfo {editor} {edited by\ \bibinfo {editor} {\bibfnamefont {M.}~\bibnamefont {Ferrario}}, \bibinfo {editor} {\bibfnamefont {G.}~\bibnamefont {Ciccotti}},\ and\ \bibinfo {editor} {\bibfnamefont {K.}~\bibnamefont {Binder}}}\ (\bibinfo  {publisher} {Springer},\ \bibinfo {address} {Berlin, Heidelberg},\ \bibinfo {year} {2006})\ pp.\ \bibinfo {pages} {453--493}\BibitemShut {NoStop}%
\bibitem [{\citenamefont {Roux}(2022)}]{roux_transition_2022}%
  \BibitemOpen
  \bibfield  {author} {\bibinfo {author} {\bibfnamefont {B.}~\bibnamefont {Roux}},\ }\bibfield  {title} {\bibinfo {title} {Transition rate theory, spectral analysis, and reactive paths},\ }\href {https://doi.org/10.1063/5.0084209} {\bibfield  {journal} {\bibinfo  {journal} {The Journal of Chemical Physics}\ }\textbf {\bibinfo {volume} {156}},\ \bibinfo {pages} {134111} (\bibinfo {year} {2022})}\BibitemShut {NoStop}%
\bibitem [{\citenamefont {Berezhkovskii}\ and\ \citenamefont {Szabo}(2019)}]{berezhkovskii_committors_2019}%
  \BibitemOpen
  \bibfield  {author} {\bibinfo {author} {\bibfnamefont {A.~M.}\ \bibnamefont {Berezhkovskii}}\ and\ \bibinfo {author} {\bibfnamefont {A.}~\bibnamefont {Szabo}},\ }\bibfield  {title} {\bibinfo {title} {Committors, first-passage times, fluxes, {Markov} states, milestones, and all that},\ }\href {https://doi.org/10.1063/1.5079742} {\bibfield  {journal} {\bibinfo  {journal} {The Journal of Chemical Physics}\ }\textbf {\bibinfo {volume} {150}},\ \bibinfo {pages} {054106} (\bibinfo {year} {2019})}\BibitemShut {NoStop}%
\bibitem [{\citenamefont {Ma}\ and\ \citenamefont {Dinner}(2005)}]{ma_automatic_2005}%
  \BibitemOpen
  \bibfield  {author} {\bibinfo {author} {\bibfnamefont {A.}~\bibnamefont {Ma}}\ and\ \bibinfo {author} {\bibfnamefont {A.~R.}\ \bibnamefont {Dinner}},\ }\bibfield  {title} {\bibinfo {title} {Automatic {Method} for {Identifying} {Reaction} {Coordinates} in {Complex} {Systems}},\ }\href {https://doi.org/10.1021/jp045546c} {\bibfield  {journal} {\bibinfo  {journal} {The Journal of Physical Chemistry B}\ }\textbf {\bibinfo {volume} {109}},\ \bibinfo {pages} {6769} (\bibinfo {year} {2005})}\BibitemShut {NoStop}%
\bibitem [{\citenamefont {Peters}\ and\ \citenamefont {Trout}(2006)}]{peters_obtaining_2006}%
  \BibitemOpen
  \bibfield  {author} {\bibinfo {author} {\bibfnamefont {B.}~\bibnamefont {Peters}}\ and\ \bibinfo {author} {\bibfnamefont {B.~L.}\ \bibnamefont {Trout}},\ }\bibfield  {title} {\bibinfo {title} {Obtaining reaction coordinates by likelihood maximization},\ }\href {https://doi.org/10.1063/1.2234477} {\bibfield  {journal} {\bibinfo  {journal} {The Journal of Chemical Physics}\ }\textbf {\bibinfo {volume} {125}},\ \bibinfo {pages} {054108} (\bibinfo {year} {2006})}\BibitemShut {NoStop}%
\bibitem [{\citenamefont {Lechner}\ \emph {et~al.}(2010)\citenamefont {Lechner}, \citenamefont {Rogal}, \citenamefont {Juraszek}, \citenamefont {Ensing},\ and\ \citenamefont {Bolhuis}}]{lechner_nonlinear_2010}%
  \BibitemOpen
  \bibfield  {author} {\bibinfo {author} {\bibfnamefont {W.}~\bibnamefont {Lechner}}, \bibinfo {author} {\bibfnamefont {J.}~\bibnamefont {Rogal}}, \bibinfo {author} {\bibfnamefont {J.}~\bibnamefont {Juraszek}}, \bibinfo {author} {\bibfnamefont {B.}~\bibnamefont {Ensing}},\ and\ \bibinfo {author} {\bibfnamefont {P.~G.}\ \bibnamefont {Bolhuis}},\ }\bibfield  {title} {\bibinfo {title} {Nonlinear reaction coordinate analysis in the reweighted path ensemble},\ }\href {https://doi.org/10.1063/1.3491818} {\bibfield  {journal} {\bibinfo  {journal} {The Journal of Chemical Physics}\ }\textbf {\bibinfo {volume} {133}},\ \bibinfo {pages} {174110} (\bibinfo {year} {2010})}\BibitemShut {NoStop}%
\bibitem [{\citenamefont {Jung}\ \emph {et~al.}(2023)\citenamefont {Jung}, \citenamefont {Covino}, \citenamefont {Arjun}, \citenamefont {Leitold}, \citenamefont {Dellago}, \citenamefont {Bolhuis},\ and\ \citenamefont {Hummer}}]{jung_machine-guided_2023}%
  \BibitemOpen
  \bibfield  {author} {\bibinfo {author} {\bibfnamefont {H.}~\bibnamefont {Jung}}, \bibinfo {author} {\bibfnamefont {R.}~\bibnamefont {Covino}}, \bibinfo {author} {\bibfnamefont {A.}~\bibnamefont {Arjun}}, \bibinfo {author} {\bibfnamefont {C.}~\bibnamefont {Leitold}}, \bibinfo {author} {\bibfnamefont {C.}~\bibnamefont {Dellago}}, \bibinfo {author} {\bibfnamefont {P.~G.}\ \bibnamefont {Bolhuis}},\ and\ \bibinfo {author} {\bibfnamefont {G.}~\bibnamefont {Hummer}},\ }\bibfield  {title} {\bibinfo {title} {Machine-guided path sampling to discover mechanisms of molecular self-organization},\ }\href {https://doi.org/10.1038/s43588-023-00428-z} {\bibfield  {journal} {\bibinfo  {journal} {Nature Computational Science}\ }\textbf {\bibinfo {volume} {3}},\ \bibinfo {pages} {334} (\bibinfo {year} {2023})}\BibitemShut {NoStop}%
\bibitem [{\citenamefont {Strand}\ \emph {et~al.}(2024)\citenamefont {Strand}, \citenamefont {Nicholson}, \citenamefont {Vroylandt},\ and\ \citenamefont {Gingrich}}]{strand_high-dimensional_2024}%
  \BibitemOpen
  \bibfield  {author} {\bibinfo {author} {\bibfnamefont {N.~E.}\ \bibnamefont {Strand}}, \bibinfo {author} {\bibfnamefont {S.~B.}\ \bibnamefont {Nicholson}}, \bibinfo {author} {\bibfnamefont {H.}~\bibnamefont {Vroylandt}},\ and\ \bibinfo {author} {\bibfnamefont {T.~R.}\ \bibnamefont {Gingrich}},\ }\bibfield  {title} {\bibinfo {title} {From high-dimensional committors to reactive insights},\ }\href {https://doi.org/10.1063/5.0232705} {\bibfield  {journal} {\bibinfo  {journal} {The Journal of Chemical Physics}\ }\textbf {\bibinfo {volume} {161}},\ \bibinfo {pages} {224109} (\bibinfo {year} {2024})}\BibitemShut {NoStop}%
\bibitem [{\citenamefont {Kang}\ \emph {et~al.}(2024)\citenamefont {Kang}, \citenamefont {Trizio},\ and\ \citenamefont {Parrinello}}]{kang_computing_2024}%
  \BibitemOpen
  \bibfield  {author} {\bibinfo {author} {\bibfnamefont {P.}~\bibnamefont {Kang}}, \bibinfo {author} {\bibfnamefont {E.}~\bibnamefont {Trizio}},\ and\ \bibinfo {author} {\bibfnamefont {M.}~\bibnamefont {Parrinello}},\ }\bibfield  {title} {\bibinfo {title} {Computing the committor with the committor to study the transition state ensemble},\ }\href {https://doi.org/10.1038/s43588-024-00645-0} {\bibfield  {journal} {\bibinfo  {journal} {Nature Computational Science}\ }\textbf {\bibinfo {volume} {4}},\ \bibinfo {pages} {451} (\bibinfo {year} {2024})}\BibitemShut {NoStop}%
\bibitem [{\citenamefont {Trizio}\ \emph {et~al.}(2025)\citenamefont {Trizio}, \citenamefont {Kang},\ and\ \citenamefont {Parrinello}}]{trizio_everything_2025}%
  \BibitemOpen
  \bibfield  {author} {\bibinfo {author} {\bibfnamefont {E.}~\bibnamefont {Trizio}}, \bibinfo {author} {\bibfnamefont {P.}~\bibnamefont {Kang}},\ and\ \bibinfo {author} {\bibfnamefont {M.}~\bibnamefont {Parrinello}},\ }\bibfield  {title} {\bibinfo {title} {Everything everywhere all at once: a probability-based enhanced sampling approach to rare events},\ }\href {https://doi.org/10.1038/s43588-025-00799-5} {\bibfield  {journal} {\bibinfo  {journal} {Nature Computational Science}\ }\textbf {\bibinfo {volume} {5}},\ \bibinfo {pages} {582} (\bibinfo {year} {2025})}\BibitemShut {NoStop}%
\bibitem [{\citenamefont {Rotskoff}\ \emph {et~al.}(2022)\citenamefont {Rotskoff}, \citenamefont {Mitchell},\ and\ \citenamefont {Vanden-Eijnden}}]{rotskoff_active_2022}%
  \BibitemOpen
  \bibfield  {author} {\bibinfo {author} {\bibfnamefont {G.~M.}\ \bibnamefont {Rotskoff}}, \bibinfo {author} {\bibfnamefont {A.~R.}\ \bibnamefont {Mitchell}},\ and\ \bibinfo {author} {\bibfnamefont {E.}~\bibnamefont {Vanden-Eijnden}},\ }\bibfield  {title} {\bibinfo {title} {Active {Importance} {Sampling} for {Variational} {Objectives} {Dominated} by {Rare} {Events}: {Consequences} for {Optimization} and {Generalization}},\ }in\ \href {https://doi.org/10.48550/arXiv.2008.06334} {\emph {\bibinfo {booktitle} {Proceedings of the 2nd {Mathematical} and {Scientific} {Machine} {Learning} {Conference}}}},\ Vol.\ \bibinfo {volume} {145}\ (\bibinfo  {publisher} {PMLR},\ \bibinfo {year} {2022})\ pp.\ \bibinfo {pages} {757--780}\BibitemShut {NoStop}%
\bibitem [{\citenamefont {Lazzeri}\ \emph {et~al.}(2023)\citenamefont {Lazzeri}, \citenamefont {Jung}, \citenamefont {Bolhuis},\ and\ \citenamefont {Covino}}]{lazzeri_molecular_2023}%
  \BibitemOpen
  \bibfield  {author} {\bibinfo {author} {\bibfnamefont {G.}~\bibnamefont {Lazzeri}}, \bibinfo {author} {\bibfnamefont {H.}~\bibnamefont {Jung}}, \bibinfo {author} {\bibfnamefont {P.~G.}\ \bibnamefont {Bolhuis}},\ and\ \bibinfo {author} {\bibfnamefont {R.}~\bibnamefont {Covino}},\ }\bibfield  {title} {\bibinfo {title} {Molecular {Free} {Energies}, {Rates}, and {Mechanisms} from {Data}-{Efficient} {Path} {Sampling} {Simulations}},\ }\href {https://doi.org/10.1021/acs.jctc.3c00821} {\bibfield  {journal} {\bibinfo  {journal} {Journal of Chemical Theory and Computation}\ }\textbf {\bibinfo {volume} {19}},\ \bibinfo {pages} {9060} (\bibinfo {year} {2023})}\BibitemShut {NoStop}%
\bibitem [{\citenamefont {Lazzeri}\ \emph {et~al.}(2025)\citenamefont {Lazzeri}, \citenamefont {Bolhuis},\ and\ \citenamefont {Covino}}]{lazzeri_optimal_2025}%
  \BibitemOpen
  \bibfield  {author} {\bibinfo {author} {\bibfnamefont {G.}~\bibnamefont {Lazzeri}}, \bibinfo {author} {\bibfnamefont {P.~G.}\ \bibnamefont {Bolhuis}},\ and\ \bibinfo {author} {\bibfnamefont {R.}~\bibnamefont {Covino}},\ }\href {https://doi.org/10.48550/arXiv.2503.21037} {\bibinfo {title} {Optimal {Rejection}-{Free} {Path} {Sampling}}} (\bibinfo {year} {2025}),\ \bibinfo {note} {arXiv:2503.21037 [physics]}\BibitemShut {NoStop}%
\bibitem [{\citenamefont {Hartmann}\ and\ \citenamefont {Sch{\"u}tte}(2012)}]{hartmann_efficient_2012}%
  \BibitemOpen
  \bibfield  {author} {\bibinfo {author} {\bibfnamefont {C.}~\bibnamefont {Hartmann}}\ and\ \bibinfo {author} {\bibfnamefont {C.}~\bibnamefont {Sch{\"u}tte}},\ }\bibfield  {title} {\bibinfo {title} {Efficient rare event simulation by optimal nonequilibrium forcing},\ }\href {https://doi.org/10.1088/1742-5468/2012/11/P11004} {\bibfield  {journal} {\bibinfo  {journal} {Journal of Statistical Mechanics: Theory and Experiment}\ }\textbf {\bibinfo {volume} {2012}},\ \bibinfo {pages} {P11004} (\bibinfo {year} {2012})}\BibitemShut {NoStop}%
\bibitem [{\citenamefont {Yuan}\ \emph {et~al.}(2024)\citenamefont {Yuan}, \citenamefont {Shah}, \citenamefont {Bentz},\ and\ \citenamefont {Cameron}}]{yuan_optimal_2024}%
  \BibitemOpen
  \bibfield  {author} {\bibinfo {author} {\bibfnamefont {J.}~\bibnamefont {Yuan}}, \bibinfo {author} {\bibfnamefont {A.}~\bibnamefont {Shah}}, \bibinfo {author} {\bibfnamefont {C.}~\bibnamefont {Bentz}},\ and\ \bibinfo {author} {\bibfnamefont {M.~K.}\ \bibnamefont {Cameron}},\ }\bibfield  {title} {\bibinfo {title} {Optimal control for sampling the transition path process and estimating rates},\ }\href {https://doi.org/10.1016/j.cnsns.2023.107701} {\bibfield  {journal} {\bibinfo  {journal} {Communications in Nonlinear Science and Numerical Simulation}\ }\textbf {\bibinfo {volume} {129}},\ \bibinfo {pages} {107701} (\bibinfo {year} {2024})}\BibitemShut {NoStop}%
\bibitem [{\citenamefont {Singh}\ and\ \citenamefont {Limmer}(2024)}]{singh_splitting_2024}%
  \BibitemOpen
  \bibfield  {author} {\bibinfo {author} {\bibfnamefont {A.~N.}\ \bibnamefont {Singh}}\ and\ \bibinfo {author} {\bibfnamefont {D.~T.}\ \bibnamefont {Limmer}},\ }\bibfield  {title} {\bibinfo {title} {Splitting probabilities as optimal controllers of rare reactive events},\ }\href {https://doi.org/10.1063/5.0203840} {\bibfield  {journal} {\bibinfo  {journal} {The Journal of Chemical Physics}\ }\textbf {\bibinfo {volume} {161}},\ \bibinfo {pages} {054113} (\bibinfo {year} {2024})}\BibitemShut {NoStop}%
\bibitem [{\citenamefont {Holdijk}\ \emph {et~al.}(2023)\citenamefont {Holdijk}, \citenamefont {Du}, \citenamefont {Hooft}, \citenamefont {Jaini}, \citenamefont {Ensing},\ and\ \citenamefont {Welling}}]{holdijk_stochastic_2023}%
  \BibitemOpen
  \bibfield  {author} {\bibinfo {author} {\bibfnamefont {L.}~\bibnamefont {Holdijk}}, \bibinfo {author} {\bibfnamefont {Y.}~\bibnamefont {Du}}, \bibinfo {author} {\bibfnamefont {F.}~\bibnamefont {Hooft}}, \bibinfo {author} {\bibfnamefont {P.}~\bibnamefont {Jaini}}, \bibinfo {author} {\bibfnamefont {B.}~\bibnamefont {Ensing}},\ and\ \bibinfo {author} {\bibfnamefont {M.}~\bibnamefont {Welling}},\ }\href {https://doi.org/10.48550/arXiv.2207.02149} {\bibinfo {title} {Stochastic {Optimal} {Control} for {Collective} {Variable} {Free} {Sampling} of {Molecular} {Transition} {Paths}}} (\bibinfo {year} {2023}),\ \bibinfo {note} {arXiv:2207.02149 [physics, q-bio]}\BibitemShut {NoStop}%
\bibitem [{\citenamefont {Manuel}\ \emph {et~al.}(2015)\citenamefont {Manuel}, \citenamefont {Lambert},\ and\ \citenamefont {Woodside}}]{manuel_reconstructing_2015}%
  \BibitemOpen
  \bibfield  {author} {\bibinfo {author} {\bibfnamefont {A.~P.}\ \bibnamefont {Manuel}}, \bibinfo {author} {\bibfnamefont {J.}~\bibnamefont {Lambert}},\ and\ \bibinfo {author} {\bibfnamefont {M.~T.}\ \bibnamefont {Woodside}},\ }\bibfield  {title} {\bibinfo {title} {Reconstructing folding energy landscapes from splitting probability analysis of single-molecule trajectories},\ }\href {https://doi.org/10.1073/pnas.1419490112} {\bibfield  {journal} {\bibinfo  {journal} {Proceedings of the National Academy of Sciences}\ }\textbf {\bibinfo {volume} {112}},\ \bibinfo {pages} {7183} (\bibinfo {year} {2015})}\BibitemShut {NoStop}%
\bibitem [{\citenamefont {Covino}\ \emph {et~al.}(2019)\citenamefont {Covino}, \citenamefont {Woodside}, \citenamefont {Hummer}, \citenamefont {Szabo},\ and\ \citenamefont {Cossio}}]{covino_molecular_2019}%
  \BibitemOpen
  \bibfield  {author} {\bibinfo {author} {\bibfnamefont {R.}~\bibnamefont {Covino}}, \bibinfo {author} {\bibfnamefont {M.~T.}\ \bibnamefont {Woodside}}, \bibinfo {author} {\bibfnamefont {G.}~\bibnamefont {Hummer}}, \bibinfo {author} {\bibfnamefont {A.}~\bibnamefont {Szabo}},\ and\ \bibinfo {author} {\bibfnamefont {P.}~\bibnamefont {Cossio}},\ }\bibfield  {title} {\bibinfo {title} {Molecular free energy profiles from force spectroscopy experiments by inversion of observed committors},\ }\href {https://doi.org/10.1063/1.5118362} {\bibfield  {journal} {\bibinfo  {journal} {The Journal of Chemical Physics}\ }\textbf {\bibinfo {volume} {151}},\ \bibinfo {pages} {154115} (\bibinfo {year} {2019})}\BibitemShut {NoStop}%
\bibitem [{\citenamefont {Metzner}\ \emph {et~al.}(2006)\citenamefont {Metzner}, \citenamefont {Schütte},\ and\ \citenamefont {Vanden-Eijnden}}]{metzner_illustration_2006}%
  \BibitemOpen
  \bibfield  {author} {\bibinfo {author} {\bibfnamefont {P.}~\bibnamefont {Metzner}}, \bibinfo {author} {\bibfnamefont {C.}~\bibnamefont {Schütte}},\ and\ \bibinfo {author} {\bibfnamefont {E.}~\bibnamefont {Vanden-Eijnden}},\ }\bibfield  {title} {\bibinfo {title} {Illustration of transition path theory on a collection of simple examples},\ }\href {https://doi.org/10.1063/1.2335447} {\bibfield  {journal} {\bibinfo  {journal} {The Journal of Chemical Physics}\ }\textbf {\bibinfo {volume} {125}},\ \bibinfo {pages} {084110} (\bibinfo {year} {2006})}\BibitemShut {NoStop}%
\bibitem [{\citenamefont {Noé}\ \emph {et~al.}(2009)\citenamefont {Noé}, \citenamefont {Schütte}, \citenamefont {Vanden-Eijnden}, \citenamefont {Reich},\ and\ \citenamefont {Weikl}}]{noe_constructing_2009}%
  \BibitemOpen
  \bibfield  {author} {\bibinfo {author} {\bibfnamefont {F.}~\bibnamefont {Noé}}, \bibinfo {author} {\bibfnamefont {C.}~\bibnamefont {Schütte}}, \bibinfo {author} {\bibfnamefont {E.}~\bibnamefont {Vanden-Eijnden}}, \bibinfo {author} {\bibfnamefont {L.}~\bibnamefont {Reich}},\ and\ \bibinfo {author} {\bibfnamefont {T.~R.}\ \bibnamefont {Weikl}},\ }\bibfield  {title} {\bibinfo {title} {Constructing the equilibrium ensemble of folding pathways from short off-equilibrium simulations},\ }\href {https://doi.org/10.1073/pnas.0905466106} {\bibfield  {journal} {\bibinfo  {journal} {Proceedings of the National Academy of Sciences}\ }\textbf {\bibinfo {volume} {106}},\ \bibinfo {pages} {19011} (\bibinfo {year} {2009})}\BibitemShut {NoStop}%
\bibitem [{\citenamefont {Metzner}\ \emph {et~al.}(2009)\citenamefont {Metzner}, \citenamefont {Schütte},\ and\ \citenamefont {Vanden-Eijnden}}]{metzner_transition_2009}%
  \BibitemOpen
  \bibfield  {author} {\bibinfo {author} {\bibfnamefont {P.}~\bibnamefont {Metzner}}, \bibinfo {author} {\bibfnamefont {C.}~\bibnamefont {Schütte}},\ and\ \bibinfo {author} {\bibfnamefont {E.}~\bibnamefont {Vanden-Eijnden}},\ }\bibfield  {title} {\bibinfo {title} {Transition {Path} {Theory} for {Markov} {Jump} {Processes}},\ }\href {https://doi.org/10.1137/070699500} {\bibfield  {journal} {\bibinfo  {journal} {Multiscale Modeling \& Simulation}\ }\textbf {\bibinfo {volume} {7}},\ \bibinfo {pages} {1192} (\bibinfo {year} {2009})}\BibitemShut {NoStop}%
\bibitem [{\citenamefont {Coifman}\ and\ \citenamefont {Lafon}(2006)}]{coifman_diffusion_2006}%
  \BibitemOpen
  \bibfield  {author} {\bibinfo {author} {\bibfnamefont {R.~R.}\ \bibnamefont {Coifman}}\ and\ \bibinfo {author} {\bibfnamefont {S.}~\bibnamefont {Lafon}},\ }\bibfield  {title} {\bibinfo {title} {Diffusion maps},\ }\href {https://doi.org/10.1016/j.acha.2006.04.006} {\bibfield  {journal} {\bibinfo  {journal} {Applied and Computational Harmonic Analysis}\ }\textbf {\bibinfo {volume} {21}},\ \bibinfo {pages} {5} (\bibinfo {year} {2006})}\BibitemShut {NoStop}%
\bibitem [{\citenamefont {Banisch}\ \emph {et~al.}(2020)\citenamefont {Banisch}, \citenamefont {Trstanova}, \citenamefont {Bittracher}, \citenamefont {Klus},\ and\ \citenamefont {Koltai}}]{banisch_diffusion_2020}%
  \BibitemOpen
  \bibfield  {author} {\bibinfo {author} {\bibfnamefont {R.}~\bibnamefont {Banisch}}, \bibinfo {author} {\bibfnamefont {Z.}~\bibnamefont {Trstanova}}, \bibinfo {author} {\bibfnamefont {A.}~\bibnamefont {Bittracher}}, \bibinfo {author} {\bibfnamefont {S.}~\bibnamefont {Klus}},\ and\ \bibinfo {author} {\bibfnamefont {P.}~\bibnamefont {Koltai}},\ }\bibfield  {title} {\bibinfo {title} {Diffusion maps tailored to arbitrary non-degenerate {Itô} processes},\ }\href {https://doi.org/10.1016/j.acha.2018.05.001} {\bibfield  {journal} {\bibinfo  {journal} {Applied and Computational Harmonic Analysis}\ }\textbf {\bibinfo {volume} {48}},\ \bibinfo {pages} {242} (\bibinfo {year} {2020})}\BibitemShut {NoStop}%
\bibitem [{\citenamefont {Evans}\ \emph {et~al.}(2023)\citenamefont {Evans}, \citenamefont {Cameron},\ and\ \citenamefont {Tiwary}}]{evans_computing_2023}%
  \BibitemOpen
  \bibfield  {author} {\bibinfo {author} {\bibfnamefont {L.}~\bibnamefont {Evans}}, \bibinfo {author} {\bibfnamefont {M.~K.}\ \bibnamefont {Cameron}},\ and\ \bibinfo {author} {\bibfnamefont {P.}~\bibnamefont {Tiwary}},\ }\bibfield  {title} {\bibinfo {title} {Computing committors in collective variables via {Mahalanobis} diffusion maps},\ }\href {https://doi.org/10.1016/j.acha.2023.01.001} {\bibfield  {journal} {\bibinfo  {journal} {Applied and Computational Harmonic Analysis}\ }\textbf {\bibinfo {volume} {64}},\ \bibinfo {pages} {62} (\bibinfo {year} {2023})}\BibitemShut {NoStop}%
\bibitem [{\citenamefont {Lai}\ and\ \citenamefont {Lu}(2018)}]{lai_point_2018}%
  \BibitemOpen
  \bibfield  {author} {\bibinfo {author} {\bibfnamefont {R.}~\bibnamefont {Lai}}\ and\ \bibinfo {author} {\bibfnamefont {J.}~\bibnamefont {Lu}},\ }\bibfield  {title} {\bibinfo {title} {Point {Cloud} {Discretization} of {Fokker}--{Planck} {Operators} for {Committor} {Functions}},\ }\href {https://doi.org/10.1137/17M1123018} {\bibfield  {journal} {\bibinfo  {journal} {Multiscale Modeling \& Simulation}\ }\textbf {\bibinfo {volume} {16}},\ \bibinfo {pages} {710} (\bibinfo {year} {2018})}\BibitemShut {NoStop}%
\bibitem [{\citenamefont {Chen}\ \emph {et~al.}(2023{\natexlab{a}})\citenamefont {Chen}, \citenamefont {Hoskins}, \citenamefont {Khoo},\ and\ \citenamefont {Lindsey}}]{chen_committor_2023}%
  \BibitemOpen
  \bibfield  {author} {\bibinfo {author} {\bibfnamefont {Y.}~\bibnamefont {Chen}}, \bibinfo {author} {\bibfnamefont {J.}~\bibnamefont {Hoskins}}, \bibinfo {author} {\bibfnamefont {Y.}~\bibnamefont {Khoo}},\ and\ \bibinfo {author} {\bibfnamefont {M.}~\bibnamefont {Lindsey}},\ }\bibfield  {title} {\bibinfo {title} {Committor functions via tensor networks},\ }\href {https://doi.org/10.1016/j.jcp.2022.111646} {\bibfield  {journal} {\bibinfo  {journal} {Journal of Computational Physics}\ }\textbf {\bibinfo {volume} {472}},\ \bibinfo {pages} {111646} (\bibinfo {year} {2023}{\natexlab{a}})}\BibitemShut {NoStop}%
\bibitem [{\citenamefont {Thiede}\ \emph {et~al.}(2019)\citenamefont {Thiede}, \citenamefont {Giannakis}, \citenamefont {Dinner},\ and\ \citenamefont {Weare}}]{thiede_galerkin_2019}%
  \BibitemOpen
  \bibfield  {author} {\bibinfo {author} {\bibfnamefont {E.~H.}\ \bibnamefont {Thiede}}, \bibinfo {author} {\bibfnamefont {D.}~\bibnamefont {Giannakis}}, \bibinfo {author} {\bibfnamefont {A.~R.}\ \bibnamefont {Dinner}},\ and\ \bibinfo {author} {\bibfnamefont {J.}~\bibnamefont {Weare}},\ }\bibfield  {title} {\bibinfo {title} {Galerkin approximation of dynamical quantities using trajectory data},\ }\href {https://doi.org/10.1063/1.5063730} {\bibfield  {journal} {\bibinfo  {journal} {The Journal of Chemical Physics}\ }\textbf {\bibinfo {volume} {150}},\ \bibinfo {pages} {244111} (\bibinfo {year} {2019})}\BibitemShut {NoStop}%
\bibitem [{\citenamefont {Strahan}\ \emph {et~al.}(2021)\citenamefont {Strahan}, \citenamefont {Antoszewski}, \citenamefont {Lorpaiboon}, \citenamefont {Vani}, \citenamefont {Weare},\ and\ \citenamefont {Dinner}}]{strahan_long-time-scale_2021}%
  \BibitemOpen
  \bibfield  {author} {\bibinfo {author} {\bibfnamefont {J.}~\bibnamefont {Strahan}}, \bibinfo {author} {\bibfnamefont {A.}~\bibnamefont {Antoszewski}}, \bibinfo {author} {\bibfnamefont {C.}~\bibnamefont {Lorpaiboon}}, \bibinfo {author} {\bibfnamefont {B.~P.}\ \bibnamefont {Vani}}, \bibinfo {author} {\bibfnamefont {J.}~\bibnamefont {Weare}},\ and\ \bibinfo {author} {\bibfnamefont {A.~R.}\ \bibnamefont {Dinner}},\ }\bibfield  {title} {\bibinfo {title} {Long-{Time}-{Scale} {Predictions} from {Short}-{Trajectory} {Data}: {A} {Benchmark} {Analysis} of the {Trp}-{Cage} {Miniprotein}},\ }\href {https://doi.org/10.1021/acs.jctc.0c00933} {\bibfield  {journal} {\bibinfo  {journal} {Journal of Chemical Theory and Computation}\ }\textbf {\bibinfo {volume} {17}},\ \bibinfo {pages} {2948} (\bibinfo {year} {2021})}\BibitemShut {NoStop}%
\bibitem [{\citenamefont {Strahan}\ \emph {et~al.}(2023{\natexlab{a}})\citenamefont {Strahan}, \citenamefont {Guo}, \citenamefont {Lorpaiboon}, \citenamefont {Dinner},\ and\ \citenamefont {Weare}}]{strahan_inexact_2023}%
  \BibitemOpen
  \bibfield  {author} {\bibinfo {author} {\bibfnamefont {J.}~\bibnamefont {Strahan}}, \bibinfo {author} {\bibfnamefont {S.~C.}\ \bibnamefont {Guo}}, \bibinfo {author} {\bibfnamefont {C.}~\bibnamefont {Lorpaiboon}}, \bibinfo {author} {\bibfnamefont {A.~R.}\ \bibnamefont {Dinner}},\ and\ \bibinfo {author} {\bibfnamefont {J.}~\bibnamefont {Weare}},\ }\bibfield  {title} {\bibinfo {title} {Inexact iterative numerical linear algebra for neural network-based spectral estimation and rare-event prediction},\ }\href {https://doi.org/10.1063/5.0151309} {\bibfield  {journal} {\bibinfo  {journal} {The Journal of Chemical Physics}\ }\textbf {\bibinfo {volume} {159}},\ \bibinfo {pages} {014110} (\bibinfo {year} {2023}{\natexlab{a}})}\BibitemShut {NoStop}%
\bibitem [{\citenamefont {Aristoff}\ \emph {et~al.}(2024)\citenamefont {Aristoff}, \citenamefont {Johnson}, \citenamefont {Simpson},\ and\ \citenamefont {Webber}}]{aristoff_fast_2024}%
  \BibitemOpen
  \bibfield  {author} {\bibinfo {author} {\bibfnamefont {D.}~\bibnamefont {Aristoff}}, \bibinfo {author} {\bibfnamefont {M.}~\bibnamefont {Johnson}}, \bibinfo {author} {\bibfnamefont {G.}~\bibnamefont {Simpson}},\ and\ \bibinfo {author} {\bibfnamefont {R.~J.}\ \bibnamefont {Webber}},\ }\bibfield  {title} {\bibinfo {title} {The fast committor machine: {I}nterpretable prediction with kernels},\ }\href {https://doi.org/10.1063/5.0222798} {\bibfield  {journal} {\bibinfo  {journal} {The Journal of Chemical Physics}\ }\textbf {\bibinfo {volume} {161}},\ \bibinfo {pages} {084113} (\bibinfo {year} {2024})},\ \bibinfo {note} {arXiv:2405.10410}\BibitemShut {NoStop}%
\bibitem [{\citenamefont {Roux}(2021)}]{roux_string_2021}%
  \BibitemOpen
  \bibfield  {author} {\bibinfo {author} {\bibfnamefont {B.}~\bibnamefont {Roux}},\ }\bibfield  {title} {\bibinfo {title} {String method with swarms-of-trajectories, mean drifts, lag time, and committor},\ }\href {https://doi.org/10.1021/acs.jpca.1c04110} {\bibfield  {journal} {\bibinfo  {journal} {The Journal of Physical Chemistry A}\ }\textbf {\bibinfo {volume} {125}},\ \bibinfo {pages} {7558} (\bibinfo {year} {2021})}\BibitemShut {NoStop}%
\bibitem [{\citenamefont {He}\ \emph {et~al.}(2022)\citenamefont {He}, \citenamefont {Chipot},\ and\ \citenamefont {Roux}}]{he_committor-consistent_2022}%
  \BibitemOpen
  \bibfield  {author} {\bibinfo {author} {\bibfnamefont {Z.}~\bibnamefont {He}}, \bibinfo {author} {\bibfnamefont {C.}~\bibnamefont {Chipot}},\ and\ \bibinfo {author} {\bibfnamefont {B.}~\bibnamefont {Roux}},\ }\bibfield  {title} {\bibinfo {title} {Committor-{Consistent} {Variational} {String} {Method}},\ }\href {https://doi.org/10.1021/acs.jpclett.2c02529} {\bibfield  {journal} {\bibinfo  {journal} {The Journal of Physical Chemistry Letters}\ }\textbf {\bibinfo {volume} {13}},\ \bibinfo {pages} {9263} (\bibinfo {year} {2022})}\BibitemShut {NoStop}%
\bibitem [{\citenamefont {Khoo}\ \emph {et~al.}(2019)\citenamefont {Khoo}, \citenamefont {Lu},\ and\ \citenamefont {Ying}}]{khoo_solving_2019}%
  \BibitemOpen
  \bibfield  {author} {\bibinfo {author} {\bibfnamefont {Y.}~\bibnamefont {Khoo}}, \bibinfo {author} {\bibfnamefont {J.}~\bibnamefont {Lu}},\ and\ \bibinfo {author} {\bibfnamefont {L.}~\bibnamefont {Ying}},\ }\bibfield  {title} {\bibinfo {title} {Solving for high-dimensional committor functions using artificial neural networks},\ }\href {https://doi.org/10.1007/s40687-018-0160-2} {\bibfield  {journal} {\bibinfo  {journal} {Research in the Mathematical Sciences}\ }\textbf {\bibinfo {volume} {6}},\ \bibinfo {pages} {1} (\bibinfo {year} {2019})}\BibitemShut {NoStop}%
\bibitem [{\citenamefont {Li}\ \emph {et~al.}(2019)\citenamefont {Li}, \citenamefont {Lin},\ and\ \citenamefont {Ren}}]{li_computing_2019}%
  \BibitemOpen
  \bibfield  {author} {\bibinfo {author} {\bibfnamefont {Q.}~\bibnamefont {Li}}, \bibinfo {author} {\bibfnamefont {B.}~\bibnamefont {Lin}},\ and\ \bibinfo {author} {\bibfnamefont {W.}~\bibnamefont {Ren}},\ }\bibfield  {title} {\bibinfo {title} {Computing committor functions for the study of rare events using deep learning},\ }\href {https://doi.org/10.1063/1.5110439} {\bibfield  {journal} {\bibinfo  {journal} {The Journal of Chemical Physics}\ }\textbf {\bibinfo {volume} {151}},\ \bibinfo {pages} {054112} (\bibinfo {year} {2019})},\ \bibinfo {note} {arXiv:1906.06285 [physics]}\BibitemShut {NoStop}%
\bibitem [{\citenamefont {Chen}\ \emph {et~al.}(2023{\natexlab{b}})\citenamefont {Chen}, \citenamefont {Roux},\ and\ \citenamefont {Chipot}}]{chen_discovering_2023}%
  \BibitemOpen
  \bibfield  {author} {\bibinfo {author} {\bibfnamefont {H.}~\bibnamefont {Chen}}, \bibinfo {author} {\bibfnamefont {B.}~\bibnamefont {Roux}},\ and\ \bibinfo {author} {\bibfnamefont {C.}~\bibnamefont {Chipot}},\ }\bibfield  {title} {\bibinfo {title} {Discovering {Reaction} {Pathways}, {Slow} {Variables}, and {Committor} {Probabilities} with {Machine} {Learning}},\ }\href {https://doi.org/10.1021/acs.jctc.3c00028} {\bibfield  {journal} {\bibinfo  {journal} {Journal of Chemical Theory and Computation}\ }\textbf {\bibinfo {volume} {19}},\ \bibinfo {pages} {4414} (\bibinfo {year} {2023}{\natexlab{b}})}\BibitemShut {NoStop}%
\bibitem [{\citenamefont {Megías}\ \emph {et~al.}(2025)\citenamefont {Megías}, \citenamefont {Contreras~Arredondo}, \citenamefont {Chen}, \citenamefont {Tang}, \citenamefont {Roux},\ and\ \citenamefont {Chipot}}]{megias_iterative_2025}%
  \BibitemOpen
  \bibfield  {author} {\bibinfo {author} {\bibfnamefont {A.}~\bibnamefont {Megías}}, \bibinfo {author} {\bibfnamefont {S.}~\bibnamefont {Contreras~Arredondo}}, \bibinfo {author} {\bibfnamefont {C.~G.}\ \bibnamefont {Chen}}, \bibinfo {author} {\bibfnamefont {C.}~\bibnamefont {Tang}}, \bibinfo {author} {\bibfnamefont {B.}~\bibnamefont {Roux}},\ and\ \bibinfo {author} {\bibfnamefont {C.}~\bibnamefont {Chipot}},\ }\bibfield  {title} {\bibinfo {title} {Iterative variational learning of committor-consistent transition pathways using artificial neural networks},\ }\href {https://doi.org/10.1038/s43588-025-00828-3} {\bibfield  {journal} {\bibinfo  {journal} {Nature Computational Science}\ }\textbf {\bibinfo {volume} {5}},\ \bibinfo {pages} {592} (\bibinfo {year} {2025})}\BibitemShut {NoStop}%
\bibitem [{\citenamefont {Hasyim}\ \emph {et~al.}(2022)\citenamefont {Hasyim}, \citenamefont {Batton},\ and\ \citenamefont {Mandadapu}}]{hasyim_supervised_2022}%
  \BibitemOpen
  \bibfield  {author} {\bibinfo {author} {\bibfnamefont {M.~R.}\ \bibnamefont {Hasyim}}, \bibinfo {author} {\bibfnamefont {C.~H.}\ \bibnamefont {Batton}},\ and\ \bibinfo {author} {\bibfnamefont {K.~K.}\ \bibnamefont {Mandadapu}},\ }\bibfield  {title} {\bibinfo {title} {Supervised learning and the finite-temperature string method for computing committor functions and reaction rates},\ }\href {https://doi.org/10.1063/5.0102423} {\bibfield  {journal} {\bibinfo  {journal} {The Journal of Chemical Physics}\ }\textbf {\bibinfo {volume} {157}},\ \bibinfo {pages} {184111} (\bibinfo {year} {2022})}\BibitemShut {NoStop}%
\bibitem [{\citenamefont {{Contreras Arredondo}}\ \emph {et~al.}(2026)\citenamefont {{Contreras Arredondo}}, \citenamefont {Tang}, \citenamefont {Talmazan}, \citenamefont {Meg{\'i}as}, \citenamefont {Chen},\ and\ \citenamefont {Chipot}}]{contreras-arredondo_learning_2026}%
  \BibitemOpen
  \bibfield  {author} {\bibinfo {author} {\bibfnamefont {S.}~\bibnamefont {{Contreras Arredondo}}}, \bibinfo {author} {\bibfnamefont {C.}~\bibnamefont {Tang}}, \bibinfo {author} {\bibfnamefont {R.~A.}\ \bibnamefont {Talmazan}}, \bibinfo {author} {\bibfnamefont {A.}~\bibnamefont {Meg{\'i}as}}, \bibinfo {author} {\bibfnamefont {C.~G.}\ \bibnamefont {Chen}},\ and\ \bibinfo {author} {\bibfnamefont {C.}~\bibnamefont {Chipot}},\ }\bibfield  {title} {\bibinfo {title} {Learning the committor without collective variables},\ }\href {https://doi.org/10.1038/s43588-026-00958-2} {\bibfield  {journal} {\bibinfo  {journal} {Nature Computational Science}\ }\textbf {\bibinfo {volume} {6}},\ \bibinfo {pages} {350} (\bibinfo {year} {2026})}\BibitemShut {NoStop}%
\bibitem [{\citenamefont {Strahan}\ \emph {et~al.}(2023{\natexlab{b}})\citenamefont {Strahan}, \citenamefont {Finkel}, \citenamefont {Dinner},\ and\ \citenamefont {Weare}}]{strahan_predicting_2023}%
  \BibitemOpen
  \bibfield  {author} {\bibinfo {author} {\bibfnamefont {J.}~\bibnamefont {Strahan}}, \bibinfo {author} {\bibfnamefont {J.}~\bibnamefont {Finkel}}, \bibinfo {author} {\bibfnamefont {A.~R.}\ \bibnamefont {Dinner}},\ and\ \bibinfo {author} {\bibfnamefont {J.}~\bibnamefont {Weare}},\ }\bibfield  {title} {\bibinfo {title} {Predicting rare events using neural networks and short-trajectory data},\ }\href {https://doi.org/10.1016/j.jcp.2023.112152} {\bibfield  {journal} {\bibinfo  {journal} {Journal of Computational Physics}\ }\textbf {\bibinfo {volume} {488}},\ \bibinfo {pages} {112152} (\bibinfo {year} {2023}{\natexlab{b}})}\BibitemShut {NoStop}%
\bibitem [{\citenamefont {Mitchell}\ and\ \citenamefont {Rotskoff}(2024)}]{mitchell_committor_2024}%
  \BibitemOpen
  \bibfield  {author} {\bibinfo {author} {\bibfnamefont {A.~R.}\ \bibnamefont {Mitchell}}\ and\ \bibinfo {author} {\bibfnamefont {G.~M.}\ \bibnamefont {Rotskoff}},\ }\bibfield  {title} {\bibinfo {title} {Committor {Guided} {Estimates} of {Molecular} {Transition} {Rates}},\ }\href {https://doi.org/10.1021/acs.jctc.4c00997} {\bibfield  {journal} {\bibinfo  {journal} {Journal of Chemical Theory and Computation}\ }\textbf {\bibinfo {volume} {20}},\ \bibinfo {pages} {9378} (\bibinfo {year} {2024})}\BibitemShut {NoStop}%
\bibitem [{\citenamefont {Pinkus}(2015)}]{pinkus_ridge_2015}%
  \BibitemOpen
  \bibfield  {author} {\bibinfo {author} {\bibfnamefont {A.}~\bibnamefont {Pinkus}},\ }\href {https://doi.org/10.1017/CBO9781316408124} {\emph {\bibinfo {title} {Ridge {Functions}}}},\ Cambridge {Tracts} in {Mathematics}\ (\bibinfo  {publisher} {Cambridge University Press},\ \bibinfo {address} {Cambridge},\ \bibinfo {year} {2015})\BibitemShut {NoStop}%
\bibitem [{\citenamefont {Friedman}\ and\ \citenamefont {Stuetzle}(1981)}]{friedman_projection_1981}%
  \BibitemOpen
  \bibfield  {author} {\bibinfo {author} {\bibfnamefont {J.~H.}\ \bibnamefont {Friedman}}\ and\ \bibinfo {author} {\bibfnamefont {W.}~\bibnamefont {Stuetzle}},\ }\bibfield  {title} {\bibinfo {title} {Projection {Pursuit} {Regression}},\ }\href {https://doi.org/10.1080/01621459.1981.10477729} {\bibfield  {journal} {\bibinfo  {journal} {Journal of the American Statistical Association}\ }\textbf {\bibinfo {volume} {76}},\ \bibinfo {pages} {817} (\bibinfo {year} {1981})}\BibitemShut {NoStop}%
\bibitem [{\citenamefont {Rabin}\ \emph {et~al.}(2012)\citenamefont {Rabin}, \citenamefont {Peyré}, \citenamefont {Delon},\ and\ \citenamefont {Bernot}}]{rabin_wasserstein_2012}%
  \BibitemOpen
  \bibfield  {author} {\bibinfo {author} {\bibfnamefont {J.}~\bibnamefont {Rabin}}, \bibinfo {author} {\bibfnamefont {G.}~\bibnamefont {Peyré}}, \bibinfo {author} {\bibfnamefont {J.}~\bibnamefont {Delon}},\ and\ \bibinfo {author} {\bibfnamefont {M.}~\bibnamefont {Bernot}},\ }\bibfield  {title} {\bibinfo {title} {Wasserstein {Barycenter} and {Its} {Application} to {Texture} {Mixing}},\ }in\ \href {https://doi.org/10.1007/978-3-642-24785-9_37} {\emph {\bibinfo {booktitle} {Scale {Space} and {Variational} {Methods} in {Computer} {Vision}}}},\ \bibinfo {editor} {edited by\ \bibinfo {editor} {\bibfnamefont {A.~M.}\ \bibnamefont {Bruckstein}}, \bibinfo {editor} {\bibfnamefont {B.~M.}\ \bibnamefont {ter Haar~Romeny}}, \bibinfo {editor} {\bibfnamefont {A.~M.}\ \bibnamefont {Bronstein}},\ and\ \bibinfo {editor} {\bibfnamefont {M.~M.}\ \bibnamefont {Bronstein}}}\ (\bibinfo  {publisher} {Springer},\ \bibinfo {address} {Berlin, Heidelberg},\ \bibinfo {year} {2012})\ pp.\ \bibinfo {pages} {435--446}\BibitemShut {NoStop}%
\bibitem [{\citenamefont {Bonneel}\ \emph {et~al.}(2015)\citenamefont {Bonneel}, \citenamefont {Rabin}, \citenamefont {Peyré},\ and\ \citenamefont {Pfister}}]{bonneel_sliced_2015}%
  \BibitemOpen
  \bibfield  {author} {\bibinfo {author} {\bibfnamefont {N.}~\bibnamefont {Bonneel}}, \bibinfo {author} {\bibfnamefont {J.}~\bibnamefont {Rabin}}, \bibinfo {author} {\bibfnamefont {G.}~\bibnamefont {Peyré}},\ and\ \bibinfo {author} {\bibfnamefont {H.}~\bibnamefont {Pfister}},\ }\bibfield  {title} {\bibinfo {title} {Sliced and {Radon} {Wasserstein} {Barycenters} of {Measures}},\ }\href {https://doi.org/10.1007/s10851-014-0506-3} {\bibfield  {journal} {\bibinfo  {journal} {Journal of Mathematical Imaging and Vision}\ }\textbf {\bibinfo {volume} {51}},\ \bibinfo {pages} {22} (\bibinfo {year} {2015})}\BibitemShut {NoStop}%
\bibitem [{\citenamefont {Kolouri}\ \emph {et~al.}(2019)\citenamefont {Kolouri}, \citenamefont {Nadjahi}, \citenamefont {Simsekli}, \citenamefont {Badeau},\ and\ \citenamefont {Rohde}}]{kolouri_generalized_2019}%
  \BibitemOpen
  \bibfield  {author} {\bibinfo {author} {\bibfnamefont {S.}~\bibnamefont {Kolouri}}, \bibinfo {author} {\bibfnamefont {K.}~\bibnamefont {Nadjahi}}, \bibinfo {author} {\bibfnamefont {U.}~\bibnamefont {Simsekli}}, \bibinfo {author} {\bibfnamefont {R.}~\bibnamefont {Badeau}},\ and\ \bibinfo {author} {\bibfnamefont {G.~K.}\ \bibnamefont {Rohde}},\ }\bibfield  {title} {\bibinfo {title} {Generalized sliced {Wasserstein} distances},\ }in\ \href {https://proceedings.neurips.cc/paper/2019/hash/f0935e4cd5920aa6c7c996a5ee53a70f-Abstract.html} {\emph {\bibinfo {booktitle} {Advances in Neural Information Processing Systems}}},\ Vol.~\bibinfo {volume} {32}\ (\bibinfo {year} {2019})\ \bibinfo {note} {arXiv:1902.00434}\BibitemShut {NoStop}%
\bibitem [{\citenamefont {Song}\ \emph {et~al.}(2020)\citenamefont {Song}, \citenamefont {Garg}, \citenamefont {Shi},\ and\ \citenamefont {Ermon}}]{song_sliced_2020}%
  \BibitemOpen
  \bibfield  {author} {\bibinfo {author} {\bibfnamefont {Y.}~\bibnamefont {Song}}, \bibinfo {author} {\bibfnamefont {S.}~\bibnamefont {Garg}}, \bibinfo {author} {\bibfnamefont {J.}~\bibnamefont {Shi}},\ and\ \bibinfo {author} {\bibfnamefont {S.}~\bibnamefont {Ermon}},\ }\bibfield  {title} {\bibinfo {title} {Sliced score matching: A scalable approach to density and score estimation},\ }in\ \href {https://proceedings.mlr.press/v115/song20a.html} {\emph {\bibinfo {booktitle} {Proceedings of the 35th Conference on Uncertainty in Artificial Intelligence (UAI)}}},\ \bibinfo {series} {PMLR}, Vol.\ \bibinfo {volume} {115}\ (\bibinfo {year} {2020})\ pp.\ \bibinfo {pages} {574--584},\ \bibinfo {note} {arXiv:1905.07088}\BibitemShut {NoStop}%
\bibitem [{\citenamefont {Torrie}\ and\ \citenamefont {Valleau}(1977)}]{torrie_nonphysical_1977}%
  \BibitemOpen
  \bibfield  {author} {\bibinfo {author} {\bibfnamefont {G.~M.}\ \bibnamefont {Torrie}}\ and\ \bibinfo {author} {\bibfnamefont {J.~P.}\ \bibnamefont {Valleau}},\ }\bibfield  {title} {\bibinfo {title} {Nonphysical sampling distributions in {Monte} {Carlo} free-energy estimation: {Umbrella} sampling},\ }\href {https://doi.org/10.1016/0021-9991(77)90121-8} {\bibfield  {journal} {\bibinfo  {journal} {Journal of Computational Physics}\ }\textbf {\bibinfo {volume} {23}},\ \bibinfo {pages} {187} (\bibinfo {year} {1977})}\BibitemShut {NoStop}%
\bibitem [{\citenamefont {Laio}\ and\ \citenamefont {Parrinello}(2002)}]{laio_escaping_2002}%
  \BibitemOpen
  \bibfield  {author} {\bibinfo {author} {\bibfnamefont {A.}~\bibnamefont {Laio}}\ and\ \bibinfo {author} {\bibfnamefont {M.}~\bibnamefont {Parrinello}},\ }\bibfield  {title} {\bibinfo {title} {Escaping free-energy minima},\ }\href {https://doi.org/10.1073/pnas.202427399} {\bibfield  {journal} {\bibinfo  {journal} {Proceedings of the National Academy of Sciences}\ }\textbf {\bibinfo {volume} {99}},\ \bibinfo {pages} {12562} (\bibinfo {year} {2002})}\BibitemShut {NoStop}%
\bibitem [{\citenamefont {Bovier}\ \emph {et~al.}(2004)\citenamefont {Bovier}, \citenamefont {Eckhoff}, \citenamefont {Gayrard},\ and\ \citenamefont {Klein}}]{bovier_metastability_2004}%
  \BibitemOpen
  \bibfield  {author} {\bibinfo {author} {\bibfnamefont {A.}~\bibnamefont {Bovier}}, \bibinfo {author} {\bibfnamefont {M.}~\bibnamefont {Eckhoff}}, \bibinfo {author} {\bibfnamefont {V.}~\bibnamefont {Gayrard}},\ and\ \bibinfo {author} {\bibfnamefont {M.}~\bibnamefont {Klein}},\ }\bibfield  {title} {\bibinfo {title} {Metastability in reversible diffusion processes {I}: Sharp asymptotics for capacities and exit times},\ }\href {https://doi.org/10.4171/JEMS/14} {\bibfield  {journal} {\bibinfo  {journal} {Journal of the European Mathematical Society}\ }\textbf {\bibinfo {volume} {6}},\ \bibinfo {pages} {399} (\bibinfo {year} {2004})}\BibitemShut {NoStop}%
\bibitem [{\citenamefont {Bovier}\ and\ \citenamefont {den Hollander}(2015)}]{bovier_metastability_2015}%
  \BibitemOpen
  \bibfield  {author} {\bibinfo {author} {\bibfnamefont {A.}~\bibnamefont {Bovier}}\ and\ \bibinfo {author} {\bibfnamefont {F.}~\bibnamefont {den Hollander}},\ }\href {https://doi.org/10.1007/978-3-319-24777-9} {\emph {\bibinfo {title} {Metastability: A Potential-Theoretic Approach}}},\ \bibinfo {series} {Grundlehren der mathematischen Wissenschaften}, Vol.\ \bibinfo {volume} {351}\ (\bibinfo  {publisher} {Springer},\ \bibinfo {year} {2015})\BibitemShut {NoStop}%
\bibitem [{\citenamefont {Leli{\`e}vre}\ and\ \citenamefont {Stoltz}(2016)}]{lelievre_partial_2016}%
  \BibitemOpen
  \bibfield  {author} {\bibinfo {author} {\bibfnamefont {T.}~\bibnamefont {Leli{\`e}vre}}\ and\ \bibinfo {author} {\bibfnamefont {G.}~\bibnamefont {Stoltz}},\ }\bibfield  {title} {\bibinfo {title} {Partial differential equations and stochastic methods in molecular dynamics},\ }\href {https://doi.org/10.1017/S0962492916000039} {\bibfield  {journal} {\bibinfo  {journal} {Acta Numerica}\ }\textbf {\bibinfo {volume} {25}},\ \bibinfo {pages} {681} (\bibinfo {year} {2016})}\BibitemShut {NoStop}%
\bibitem [{\citenamefont {Lorpaiboon}\ \emph {et~al.}(2026)\citenamefont {Lorpaiboon}, \citenamefont {Weare},\ and\ \citenamefont {Dinner}}]{lorpaiboon_exact_2026}%
  \BibitemOpen
  \bibfield  {author} {\bibinfo {author} {\bibfnamefont {C.}~\bibnamefont {Lorpaiboon}}, \bibinfo {author} {\bibfnamefont {J.}~\bibnamefont {Weare}},\ and\ \bibinfo {author} {\bibfnamefont {A.~R.}\ \bibnamefont {Dinner}},\ }\bibfield  {title} {\bibinfo {title} {An exact multiple-time-step variational formulation for the committor and the transition rate},\ }\href {https://doi.org/10.1021/acs.jpcb.5c06047} {\bibfield  {journal} {\bibinfo  {journal} {The Journal of Physical Chemistry B}\ }\textbf {\bibinfo {volume} {130}},\ \bibinfo {pages} {155} (\bibinfo {year} {2026})}\BibitemShut {NoStop}%
\bibitem [{sup()}]{supplemental}%
  \BibitemOpen
  \href@noop {} {}\bibinfo {note} {See Supplemental Material at [URL will be inserted by publisher] for the reaction-diffusion slice formulation, weight derivation, direction-sampling schemes, and implementation details.}\BibitemShut {Stop}%
\bibitem [{\citenamefont {Bolhuis}\ \emph {et~al.}(2000)\citenamefont {Bolhuis}, \citenamefont {Dellago},\ and\ \citenamefont {Chandler}}]{bolhuis_reaction_2000}%
  \BibitemOpen
  \bibfield  {author} {\bibinfo {author} {\bibfnamefont {P.~G.}\ \bibnamefont {Bolhuis}}, \bibinfo {author} {\bibfnamefont {C.}~\bibnamefont {Dellago}},\ and\ \bibinfo {author} {\bibfnamefont {D.}~\bibnamefont {Chandler}},\ }\bibfield  {title} {\bibinfo {title} {Reaction coordinates of biomolecular isomerization},\ }\href {https://doi.org/10.1073/pnas.100127697} {\bibfield  {journal} {\bibinfo  {journal} {Proceedings of the National Academy of Sciences}\ }\textbf {\bibinfo {volume} {97}},\ \bibinfo {pages} {5877} (\bibinfo {year} {2000})}\BibitemShut {NoStop}%
\bibitem [{\citenamefont {Bittracher}\ \emph {et~al.}(2018)\citenamefont {Bittracher}, \citenamefont {Koltai}, \citenamefont {Klus}, \citenamefont {Banisch}, \citenamefont {Dellnitz},\ and\ \citenamefont {Sch{\"u}tte}}]{bittracher_transition_2018}%
  \BibitemOpen
  \bibfield  {author} {\bibinfo {author} {\bibfnamefont {A.}~\bibnamefont {Bittracher}}, \bibinfo {author} {\bibfnamefont {P.}~\bibnamefont {Koltai}}, \bibinfo {author} {\bibfnamefont {S.}~\bibnamefont {Klus}}, \bibinfo {author} {\bibfnamefont {R.}~\bibnamefont {Banisch}}, \bibinfo {author} {\bibfnamefont {M.}~\bibnamefont {Dellnitz}},\ and\ \bibinfo {author} {\bibfnamefont {C.}~\bibnamefont {Sch{\"u}tte}},\ }\bibfield  {title} {\bibinfo {title} {Transition manifolds of complex metastable systems: Theory and data-driven computation of effective dynamics},\ }\href {https://doi.org/10.1007/s00332-017-9415-0} {\bibfield  {journal} {\bibinfo  {journal} {Journal of Nonlinear Science}\ }\textbf {\bibinfo {volume} {28}},\ \bibinfo {pages} {471} (\bibinfo {year} {2018})}\BibitemShut {NoStop}%
\bibitem [{\citenamefont {Bittracher}\ \emph {et~al.}(2021)\citenamefont {Bittracher}, \citenamefont {Klus}, \citenamefont {Hamzi}, \citenamefont {Koltai},\ and\ \citenamefont {Sch{\"u}tte}}]{bittracher_dimensionality_2021}%
  \BibitemOpen
  \bibfield  {author} {\bibinfo {author} {\bibfnamefont {A.}~\bibnamefont {Bittracher}}, \bibinfo {author} {\bibfnamefont {S.}~\bibnamefont {Klus}}, \bibinfo {author} {\bibfnamefont {B.}~\bibnamefont {Hamzi}}, \bibinfo {author} {\bibfnamefont {P.}~\bibnamefont {Koltai}},\ and\ \bibinfo {author} {\bibfnamefont {C.}~\bibnamefont {Sch{\"u}tte}},\ }\bibfield  {title} {\bibinfo {title} {Dimensionality reduction of complex metastable systems via kernel embeddings of transition manifolds},\ }\href {https://doi.org/10.1007/s00332-020-09668-z} {\bibfield  {journal} {\bibinfo  {journal} {Journal of Nonlinear Science}\ }\textbf {\bibinfo {volume} {31}},\ \bibinfo {pages} {3} (\bibinfo {year} {2021})}\BibitemShut {NoStop}%
\bibitem [{\citenamefont {Webber}\ \emph {et~al.}(2021)\citenamefont {Webber}, \citenamefont {Thiede}, \citenamefont {Dow}, \citenamefont {Dinner},\ and\ \citenamefont {Weare}}]{webber_error_2021}%
  \BibitemOpen
  \bibfield  {author} {\bibinfo {author} {\bibfnamefont {R.~J.}\ \bibnamefont {Webber}}, \bibinfo {author} {\bibfnamefont {E.~H.}\ \bibnamefont {Thiede}}, \bibinfo {author} {\bibfnamefont {D.}~\bibnamefont {Dow}}, \bibinfo {author} {\bibfnamefont {A.~R.}\ \bibnamefont {Dinner}},\ and\ \bibinfo {author} {\bibfnamefont {J.}~\bibnamefont {Weare}},\ }\bibfield  {title} {\bibinfo {title} {Error bounds for dynamical spectral estimation},\ }\href {https://doi.org/10.1137/20M1335984} {\bibfield  {journal} {\bibinfo  {journal} {SIAM Journal on Mathematics of Data Science}\ }\textbf {\bibinfo {volume} {3}},\ \bibinfo {pages} {225} (\bibinfo {year} {2021})}\BibitemShut {NoStop}%
\bibitem [{\citenamefont {Pérez-Hernández}\ \emph {et~al.}(2013)\citenamefont {Pérez-Hernández}, \citenamefont {Paul}, \citenamefont {Giorgino}, \citenamefont {De~Fabritiis},\ and\ \citenamefont {Noé}}]{perez-hernandez_identification_2013}%
  \BibitemOpen
  \bibfield  {author} {\bibinfo {author} {\bibfnamefont {G.}~\bibnamefont {Pérez-Hernández}}, \bibinfo {author} {\bibfnamefont {F.}~\bibnamefont {Paul}}, \bibinfo {author} {\bibfnamefont {T.}~\bibnamefont {Giorgino}}, \bibinfo {author} {\bibfnamefont {G.}~\bibnamefont {De~Fabritiis}},\ and\ \bibinfo {author} {\bibfnamefont {F.}~\bibnamefont {Noé}},\ }\bibfield  {title} {\bibinfo {title} {Identification of slow molecular order parameters for {Markov} model construction},\ }\href {https://doi.org/10.1063/1.4811489} {\bibfield  {journal} {\bibinfo  {journal} {The Journal of Chemical Physics}\ }\textbf {\bibinfo {volume} {139}},\ \bibinfo {pages} {015102} (\bibinfo {year} {2013})}\BibitemShut {NoStop}%
\bibitem [{\citenamefont {Schwantes}\ and\ \citenamefont {Pande}(2013)}]{schwantes_improvements_2013}%
  \BibitemOpen
  \bibfield  {author} {\bibinfo {author} {\bibfnamefont {C.~R.}\ \bibnamefont {Schwantes}}\ and\ \bibinfo {author} {\bibfnamefont {V.~S.}\ \bibnamefont {Pande}},\ }\bibfield  {title} {\bibinfo {title} {Improvements in {Markov} {State} {Model} {Construction} {Reveal} {Many} {Non}-{Native} {Interactions} in the {Folding} of {NTL9}},\ }\href {https://doi.org/10.1021/ct300878a} {\bibfield  {journal} {\bibinfo  {journal} {Journal of Chemical Theory and Computation}\ }\textbf {\bibinfo {volume} {9}},\ \bibinfo {pages} {2000} (\bibinfo {year} {2013})}\BibitemShut {NoStop}%
\bibitem [{\citenamefont {Tiwary}\ and\ \citenamefont {Parrinello}(2013)}]{tiwary_metadynamics_2013}%
  \BibitemOpen
  \bibfield  {author} {\bibinfo {author} {\bibfnamefont {P.}~\bibnamefont {Tiwary}}\ and\ \bibinfo {author} {\bibfnamefont {M.}~\bibnamefont {Parrinello}},\ }\bibfield  {title} {\bibinfo {title} {From metadynamics to dynamics},\ }\href {https://doi.org/10.1103/PhysRevLett.111.230602} {\bibfield  {journal} {\bibinfo  {journal} {Physical Review Letters}\ }\textbf {\bibinfo {volume} {111}},\ \bibinfo {pages} {230602} (\bibinfo {year} {2013})}\BibitemShut {NoStop}%
\bibitem [{\citenamefont {Honda}\ \emph {et~al.}(2008)\citenamefont {Honda}, \citenamefont {Akiba}, \citenamefont {Kato}, \citenamefont {Sawada}, \citenamefont {Sekijima}, \citenamefont {Ishimura}, \citenamefont {Ooishi}, \citenamefont {Watanabe}, \citenamefont {Odahara},\ and\ \citenamefont {Harata}}]{honda_crystal_2008}%
  \BibitemOpen
  \bibfield  {author} {\bibinfo {author} {\bibfnamefont {S.}~\bibnamefont {Honda}}, \bibinfo {author} {\bibfnamefont {T.}~\bibnamefont {Akiba}}, \bibinfo {author} {\bibfnamefont {Y.~S.}\ \bibnamefont {Kato}}, \bibinfo {author} {\bibfnamefont {Y.}~\bibnamefont {Sawada}}, \bibinfo {author} {\bibfnamefont {M.}~\bibnamefont {Sekijima}}, \bibinfo {author} {\bibfnamefont {M.}~\bibnamefont {Ishimura}}, \bibinfo {author} {\bibfnamefont {A.}~\bibnamefont {Ooishi}}, \bibinfo {author} {\bibfnamefont {H.}~\bibnamefont {Watanabe}}, \bibinfo {author} {\bibfnamefont {T.}~\bibnamefont {Odahara}},\ and\ \bibinfo {author} {\bibfnamefont {K.}~\bibnamefont {Harata}},\ }\bibfield  {title} {\bibinfo {title} {Crystal structure of a ten-amino acid protein},\ }\href {https://doi.org/10.1021/ja8030533} {\bibfield  {journal} {\bibinfo  {journal} {Journal of the American Chemical Society}\ }\textbf {\bibinfo {volume} {130}},\ \bibinfo {pages} {15327} (\bibinfo {year} {2008})}\BibitemShut {NoStop}%
\bibitem [{\citenamefont {Shirts}\ and\ \citenamefont {Chodera}(2008)}]{shirts_statistically_2008}%
  \BibitemOpen
  \bibfield  {author} {\bibinfo {author} {\bibfnamefont {M.~R.}\ \bibnamefont {Shirts}}\ and\ \bibinfo {author} {\bibfnamefont {J.~D.}\ \bibnamefont {Chodera}},\ }\bibfield  {title} {\bibinfo {title} {Statistically optimal analysis of samples from multiple equilibrium states},\ }\href {https://doi.org/10.1063/1.2978177} {\bibfield  {journal} {\bibinfo  {journal} {The Journal of Chemical Physics}\ }\textbf {\bibinfo {volume} {129}},\ \bibinfo {pages} {124105} (\bibinfo {year} {2008})}\BibitemShut {NoStop}%
\bibitem [{\citenamefont {Best}\ \emph {et~al.}(2013)\citenamefont {Best}, \citenamefont {Hummer},\ and\ \citenamefont {Eaton}}]{best_native_2013}%
  \BibitemOpen
  \bibfield  {author} {\bibinfo {author} {\bibfnamefont {R.~B.}\ \bibnamefont {Best}}, \bibinfo {author} {\bibfnamefont {G.}~\bibnamefont {Hummer}},\ and\ \bibinfo {author} {\bibfnamefont {W.~A.}\ \bibnamefont {Eaton}},\ }\bibfield  {title} {\bibinfo {title} {Native contacts determine protein folding mechanisms in atomistic simulations},\ }\href {https://doi.org/10.1073/pnas.1311599110} {\bibfield  {journal} {\bibinfo  {journal} {Proceedings of the National Academy of Sciences}\ }\textbf {\bibinfo {volume} {110}},\ \bibinfo {pages} {17874} (\bibinfo {year} {2013})}\BibitemShut {NoStop}%
\bibitem [{\citenamefont {Hummer}(2005)}]{hummer_position-dependent_2005}%
  \BibitemOpen
  \bibfield  {author} {\bibinfo {author} {\bibfnamefont {G.}~\bibnamefont {Hummer}},\ }\bibfield  {title} {\bibinfo {title} {Position-dependent diffusion coefficients and free energies from {B}ayesian analysis of equilibrium and replica molecular dynamics simulations},\ }\href {https://doi.org/10.1088/1367-2630/7/1/034} {\bibfield  {journal} {\bibinfo  {journal} {New Journal of Physics}\ }\textbf {\bibinfo {volume} {7}},\ \bibinfo {pages} {34} (\bibinfo {year} {2005})}\BibitemShut {NoStop}%
\bibitem [{\citenamefont {Berezhkovskii}\ and\ \citenamefont {Szabo}(2011)}]{berezhkovskii_time_2011}%
  \BibitemOpen
  \bibfield  {author} {\bibinfo {author} {\bibfnamefont {A.~M.}\ \bibnamefont {Berezhkovskii}}\ and\ \bibinfo {author} {\bibfnamefont {A.}~\bibnamefont {Szabo}},\ }\bibfield  {title} {\bibinfo {title} {Time scale separation leads to position-dependent diffusion along a slow coordinate},\ }\href {https://doi.org/10.1063/1.3626215} {\bibfield  {journal} {\bibinfo  {journal} {The Journal of Chemical Physics}\ }\textbf {\bibinfo {volume} {135}},\ \bibinfo {pages} {074108} (\bibinfo {year} {2011})}\BibitemShut {NoStop}%
\bibitem [{\citenamefont {Best}\ and\ \citenamefont {Hummer}(2010)}]{best_coordinate_2010}%
  \BibitemOpen
  \bibfield  {author} {\bibinfo {author} {\bibfnamefont {R.~B.}\ \bibnamefont {Best}}\ and\ \bibinfo {author} {\bibfnamefont {G.}~\bibnamefont {Hummer}},\ }\bibfield  {title} {\bibinfo {title} {Coordinate-dependent diffusion in protein folding},\ }\href {https://doi.org/10.1073/pnas.0910390107} {\bibfield  {journal} {\bibinfo  {journal} {Proceedings of the National Academy of Sciences}\ }\textbf {\bibinfo {volume} {107}},\ \bibinfo {pages} {1088} (\bibinfo {year} {2010})}\BibitemShut {NoStop}%
\bibitem [{\citenamefont {Lindorff-Larsen}\ \emph {et~al.}(2011)\citenamefont {Lindorff-Larsen}, \citenamefont {Piana}, \citenamefont {Dror},\ and\ \citenamefont {Shaw}}]{lindorff-larsen_how_2011}%
  \BibitemOpen
  \bibfield  {author} {\bibinfo {author} {\bibfnamefont {K.}~\bibnamefont {Lindorff-Larsen}}, \bibinfo {author} {\bibfnamefont {S.}~\bibnamefont {Piana}}, \bibinfo {author} {\bibfnamefont {R.~O.}\ \bibnamefont {Dror}},\ and\ \bibinfo {author} {\bibfnamefont {D.~E.}\ \bibnamefont {Shaw}},\ }\bibfield  {title} {\bibinfo {title} {How {Fast}-{Folding} {Proteins} {Fold}},\ }\href {https://doi.org/10.1126/science.1208351} {\bibfield  {journal} {\bibinfo  {journal} {Science}\ }\textbf {\bibinfo {volume} {334}},\ \bibinfo {pages} {517} (\bibinfo {year} {2011})}\BibitemShut {NoStop}%
\bibitem [{\citenamefont {De~Cao}\ and\ \citenamefont {Aziz}(2020)}]{decao_power_2020}%
  \BibitemOpen
  \bibfield  {author} {\bibinfo {author} {\bibfnamefont {N.}~\bibnamefont {De~Cao}}\ and\ \bibinfo {author} {\bibfnamefont {W.}~\bibnamefont {Aziz}},\ }\href@noop {} {\bibinfo {title} {The power spherical distribution}} (\bibinfo {year} {2020}),\ \bibinfo {note} {{ICML} 2020 Workshop INNF+},\ \Eprint {https://arxiv.org/abs/2006.04437} {arXiv:2006.04437 [stat.ML]} \BibitemShut {NoStop}%
\bibitem [{\citenamefont {Petersen}\ \emph {et~al.}(2026)\citenamefont {Petersen}, \citenamefont {Lichtinger},\ and\ \citenamefont {Covino}}]{ourcode}%
  \BibitemOpen
  \bibfield  {author} {\bibinfo {author} {\bibfnamefont {M.}~\bibnamefont {Petersen}}, \bibinfo {author} {\bibfnamefont {S.}~\bibnamefont {Lichtinger}},\ and\ \bibinfo {author} {\bibfnamefont {R.}~\bibnamefont {Covino}},\ }\href {https://doi.org/10.5281/zenodo.21703983} {\bibinfo {title} {Reproduction package for ``{Committors} and {Reaction} {Rates} from {Trial} {Functions} {That} {Violate} the {Boundary} {Conditions}''}} (\bibinfo {year} {2026}),\ \bibinfo {note} {bundles the \texttt{sliced\_committor} library, the analysis scripts and the input data. Development repository: \url{https://github.com/MagnusPetersen/slicedcommittor}}\BibitemShut {NoStop}%
\bibitem [{\citenamefont {Piana}\ \emph {et~al.}(2012)\citenamefont {Piana}, \citenamefont {Lindorff-Larsen},\ and\ \citenamefont {Shaw}}]{piana_protein_2012}%
  \BibitemOpen
  \bibfield  {author} {\bibinfo {author} {\bibfnamefont {S.}~\bibnamefont {Piana}}, \bibinfo {author} {\bibfnamefont {K.}~\bibnamefont {Lindorff-Larsen}},\ and\ \bibinfo {author} {\bibfnamefont {D.~E.}\ \bibnamefont {Shaw}},\ }\bibfield  {title} {\bibinfo {title} {Protein folding kinetics and thermodynamics from atomistic simulation},\ }\href {https://doi.org/10.1073/pnas.1201811109} {\bibfield  {journal} {\bibinfo  {journal} {Proceedings of the National Academy of Sciences}\ }\textbf {\bibinfo {volume} {109}},\ \bibinfo {pages} {17845} (\bibinfo {year} {2012})}\BibitemShut {NoStop}%
\bibitem [{\citenamefont {Krivov}\ and\ \citenamefont {Karplus}(2006)}]{krivov_one-dimensional_2006}%
  \BibitemOpen
  \bibfield  {author} {\bibinfo {author} {\bibfnamefont {S.~V.}\ \bibnamefont {Krivov}}\ and\ \bibinfo {author} {\bibfnamefont {M.}~\bibnamefont {Karplus}},\ }\bibfield  {title} {\bibinfo {title} {One-dimensional free-energy profiles of complex systems: Progress variables that preserve the barriers},\ }\href {https://doi.org/10.1021/jp060039b} {\bibfield  {journal} {\bibinfo  {journal} {The Journal of Physical Chemistry B}\ }\textbf {\bibinfo {volume} {110}},\ \bibinfo {pages} {12689} (\bibinfo {year} {2006})}\BibitemShut {NoStop}%
\bibitem [{\citenamefont {Krivov}\ and\ \citenamefont {Karplus}(2008)}]{krivov_diffusive_2008}%
  \BibitemOpen
  \bibfield  {author} {\bibinfo {author} {\bibfnamefont {S.~V.}\ \bibnamefont {Krivov}}\ and\ \bibinfo {author} {\bibfnamefont {M.}~\bibnamefont {Karplus}},\ }\bibfield  {title} {\bibinfo {title} {Diffusive reaction dynamics on invariant free energy profiles},\ }\href {https://doi.org/10.1073/pnas.0800228105} {\bibfield  {journal} {\bibinfo  {journal} {Proceedings of the National Academy of Sciences}\ }\textbf {\bibinfo {volume} {105}},\ \bibinfo {pages} {13841} (\bibinfo {year} {2008})}\BibitemShut {NoStop}%
\bibitem [{\citenamefont {Li}(1991)}]{li_sliced_1991}%
  \BibitemOpen
  \bibfield  {author} {\bibinfo {author} {\bibfnamefont {K.-C.}\ \bibnamefont {Li}},\ }\bibfield  {title} {\bibinfo {title} {Sliced inverse regression for dimension reduction},\ }\href {https://doi.org/10.1080/01621459.1991.10475035} {\bibfield  {journal} {\bibinfo  {journal} {Journal of the American Statistical Association}\ }\textbf {\bibinfo {volume} {86}},\ \bibinfo {pages} {316} (\bibinfo {year} {1991})}\BibitemShut {NoStop}%
\bibitem [{\citenamefont {Bradbury}\ \emph {et~al.}(2018)\citenamefont {Bradbury}, \citenamefont {Frostig}, \citenamefont {Hawkins}, \citenamefont {Johnson}, \citenamefont {Katariya}, \citenamefont {Leary}, \citenamefont {Maclaurin}, \citenamefont {Necula}, \citenamefont {Paszke}, \citenamefont {Vander{P}las}, \citenamefont {Wanderman-{M}ilne},\ and\ \citenamefont {Zhang}}]{jax2018github}%
  \BibitemOpen
  \bibfield  {author} {\bibinfo {author} {\bibfnamefont {J.}~\bibnamefont {Bradbury}}, \bibinfo {author} {\bibfnamefont {R.}~\bibnamefont {Frostig}}, \bibinfo {author} {\bibfnamefont {P.}~\bibnamefont {Hawkins}}, \bibinfo {author} {\bibfnamefont {M.~J.}\ \bibnamefont {Johnson}}, \bibinfo {author} {\bibfnamefont {Y.}~\bibnamefont {Katariya}}, \bibinfo {author} {\bibfnamefont {C.}~\bibnamefont {Leary}}, \bibinfo {author} {\bibfnamefont {D.}~\bibnamefont {Maclaurin}}, \bibinfo {author} {\bibfnamefont {G.}~\bibnamefont {Necula}}, \bibinfo {author} {\bibfnamefont {A.}~\bibnamefont {Paszke}}, \bibinfo {author} {\bibfnamefont {J.}~\bibnamefont {Vander{P}las}}, \bibinfo {author} {\bibfnamefont {S.}~\bibnamefont {Wanderman-{M}ilne}},\ and\ \bibinfo {author} {\bibfnamefont {Q.}~\bibnamefont {Zhang}},\ }\href {http://github.com/jax-ml/jax} {\bibinfo {title} {{JAX}: composable transformations of {P}ython+{N}um{P}y programs}} (\bibinfo {year} {2018})\BibitemShut {NoStop}%
\bibitem [{\citenamefont {Lorpaiboon}\ \emph {et~al.}(2022)\citenamefont {Lorpaiboon}, \citenamefont {Weare},\ and\ \citenamefont {Dinner}}]{lorpaiboon_augmented_2022}%
  \BibitemOpen
  \bibfield  {author} {\bibinfo {author} {\bibfnamefont {C.}~\bibnamefont {Lorpaiboon}}, \bibinfo {author} {\bibfnamefont {J.}~\bibnamefont {Weare}},\ and\ \bibinfo {author} {\bibfnamefont {A.~R.}\ \bibnamefont {Dinner}},\ }\bibfield  {title} {\bibinfo {title} {Augmented transition path theory for sequences of events},\ }\href {https://doi.org/10.1063/5.0098587} {\bibfield  {journal} {\bibinfo  {journal} {The Journal of Chemical Physics}\ }\textbf {\bibinfo {volume} {157}},\ \bibinfo {pages} {094115} (\bibinfo {year} {2022})}\BibitemShut {NoStop}%
\bibitem [{\citenamefont {Faradjian}\ and\ \citenamefont {Elber}(2004)}]{faradjian_computing_2004}%
  \BibitemOpen
  \bibfield  {author} {\bibinfo {author} {\bibfnamefont {A.~K.}\ \bibnamefont {Faradjian}}\ and\ \bibinfo {author} {\bibfnamefont {R.}~\bibnamefont {Elber}},\ }\bibfield  {title} {\bibinfo {title} {Computing time scales from reaction coordinates by milestoning},\ }\href {https://doi.org/10.1063/1.1738640} {\bibfield  {journal} {\bibinfo  {journal} {The Journal of Chemical Physics}\ }\textbf {\bibinfo {volume} {120}},\ \bibinfo {pages} {10880} (\bibinfo {year} {2004})}\BibitemShut {NoStop}%
\bibitem [{\citenamefont {Bello-Rivas}\ and\ \citenamefont {Elber}(2015)}]{bello-rivas_exact_2015}%
  \BibitemOpen
  \bibfield  {author} {\bibinfo {author} {\bibfnamefont {J.~M.}\ \bibnamefont {Bello-Rivas}}\ and\ \bibinfo {author} {\bibfnamefont {R.}~\bibnamefont {Elber}},\ }\bibfield  {title} {\bibinfo {title} {Exact milestoning},\ }\href {https://doi.org/10.1063/1.4913399} {\bibfield  {journal} {\bibinfo  {journal} {The Journal of Chemical Physics}\ }\textbf {\bibinfo {volume} {142}},\ \bibinfo {pages} {094102} (\bibinfo {year} {2015})}\BibitemShut {NoStop}%
\bibitem [{\citenamefont {Zuckerman}\ and\ \citenamefont {Chong}(2017)}]{zuckerman_weighted_2017}%
  \BibitemOpen
  \bibfield  {author} {\bibinfo {author} {\bibfnamefont {D.~M.}\ \bibnamefont {Zuckerman}}\ and\ \bibinfo {author} {\bibfnamefont {L.~T.}\ \bibnamefont {Chong}},\ }\bibfield  {title} {\bibinfo {title} {Weighted ensemble simulation: Review of methodology, applications, and software},\ }\href {https://doi.org/10.1146/annurev-biophys-070816-033834} {\bibfield  {journal} {\bibinfo  {journal} {Annual Review of Biophysics}\ }\textbf {\bibinfo {volume} {46}},\ \bibinfo {pages} {43} (\bibinfo {year} {2017})}\BibitemShut {NoStop}%
\bibitem [{\citenamefont {Strahan}\ \emph {et~al.}(2024)\citenamefont {Strahan}, \citenamefont {Lorpaiboon}, \citenamefont {Weare},\ and\ \citenamefont {Dinner}}]{strahan_bad-neus_2024}%
  \BibitemOpen
  \bibfield  {author} {\bibinfo {author} {\bibfnamefont {J.}~\bibnamefont {Strahan}}, \bibinfo {author} {\bibfnamefont {C.}~\bibnamefont {Lorpaiboon}}, \bibinfo {author} {\bibfnamefont {J.}~\bibnamefont {Weare}},\ and\ \bibinfo {author} {\bibfnamefont {A.~R.}\ \bibnamefont {Dinner}},\ }\bibfield  {title} {\bibinfo {title} {{BAD-NEUS}: Rapidly converging trajectory stratification},\ }\href {https://doi.org/10.1063/5.0215975} {\bibfield  {journal} {\bibinfo  {journal} {The Journal of Chemical Physics}\ }\textbf {\bibinfo {volume} {161}},\ \bibinfo {pages} {084109} (\bibinfo {year} {2024})}\BibitemShut {NoStop}%
\bibitem [{\citenamefont {Geyer}(1992)}]{geyer_practical_1992}%
  \BibitemOpen
  \bibfield  {author} {\bibinfo {author} {\bibfnamefont {C.~J.}\ \bibnamefont {Geyer}},\ }\bibfield  {title} {\bibinfo {title} {Practical {M}arkov chain {M}onte {C}arlo},\ }\href {https://doi.org/10.1214/ss/1177011137} {\bibfield  {journal} {\bibinfo  {journal} {Statistical Science}\ }\textbf {\bibinfo {volume} {7}},\ \bibinfo {pages} {473} (\bibinfo {year} {1992})}\BibitemShut {NoStop}%
\bibitem [{\citenamefont {Bogetti}\ \emph {et~al.}(2020)\citenamefont {Bogetti}, \citenamefont {Piston}, \citenamefont {Leung}, \citenamefont {Cabalteja}, \citenamefont {Yang}, \citenamefont {DeGrave}, \citenamefont {Debiec}, \citenamefont {Cerutti}, \citenamefont {Case}, \citenamefont {Horne},\ and\ \citenamefont {Chong}}]{bogetti_twist_2020}%
  \BibitemOpen
  \bibfield  {author} {\bibinfo {author} {\bibfnamefont {A.~T.}\ \bibnamefont {Bogetti}}, \bibinfo {author} {\bibfnamefont {H.~E.}\ \bibnamefont {Piston}}, \bibinfo {author} {\bibfnamefont {J.~M.~G.}\ \bibnamefont {Leung}}, \bibinfo {author} {\bibfnamefont {C.~C.}\ \bibnamefont {Cabalteja}}, \bibinfo {author} {\bibfnamefont {D.~T.}\ \bibnamefont {Yang}}, \bibinfo {author} {\bibfnamefont {A.~J.}\ \bibnamefont {DeGrave}}, \bibinfo {author} {\bibfnamefont {K.~T.}\ \bibnamefont {Debiec}}, \bibinfo {author} {\bibfnamefont {D.~S.}\ \bibnamefont {Cerutti}}, \bibinfo {author} {\bibfnamefont {D.~A.}\ \bibnamefont {Case}}, \bibinfo {author} {\bibfnamefont {W.~S.}\ \bibnamefont {Horne}},\ and\ \bibinfo {author} {\bibfnamefont {L.~T.}\ \bibnamefont {Chong}},\ }\bibfield  {title} {\bibinfo {title} {A twist in the road less traveled: {The} {AMBER} ff15ipq-m force field for protein mimetics},\ }\href {https://doi.org/10.1063/5.0019054} {\bibfield  {journal} {\bibinfo  {journal} {The Journal of Chemical Physics}\ }\textbf
  {\bibinfo {volume} {153}},\ \bibinfo {pages} {064101} (\bibinfo {year} {2020})}\BibitemShut {NoStop}%
\bibitem [{\citenamefont {Eastman}\ \emph {et~al.}(2017)\citenamefont {Eastman}, \citenamefont {Swails}, \citenamefont {Chodera}, \citenamefont {McGibbon}, \citenamefont {Zhao}, \citenamefont {Beauchamp}, \citenamefont {Wang}, \citenamefont {Simmonett}, \citenamefont {Harrigan}, \citenamefont {Stern}, \citenamefont {Wiewiora}, \citenamefont {Brooks},\ and\ \citenamefont {Pande}}]{eastman_openmm_2017}%
  \BibitemOpen
  \bibfield  {author} {\bibinfo {author} {\bibfnamefont {P.}~\bibnamefont {Eastman}}, \bibinfo {author} {\bibfnamefont {J.}~\bibnamefont {Swails}}, \bibinfo {author} {\bibfnamefont {J.~D.}\ \bibnamefont {Chodera}}, \bibinfo {author} {\bibfnamefont {R.~T.}\ \bibnamefont {McGibbon}}, \bibinfo {author} {\bibfnamefont {Y.}~\bibnamefont {Zhao}}, \bibinfo {author} {\bibfnamefont {K.~A.}\ \bibnamefont {Beauchamp}}, \bibinfo {author} {\bibfnamefont {L.-P.}\ \bibnamefont {Wang}}, \bibinfo {author} {\bibfnamefont {A.~C.}\ \bibnamefont {Simmonett}}, \bibinfo {author} {\bibfnamefont {M.~P.}\ \bibnamefont {Harrigan}}, \bibinfo {author} {\bibfnamefont {C.~D.}\ \bibnamefont {Stern}}, \bibinfo {author} {\bibfnamefont {R.~P.}\ \bibnamefont {Wiewiora}}, \bibinfo {author} {\bibfnamefont {B.~R.}\ \bibnamefont {Brooks}},\ and\ \bibinfo {author} {\bibfnamefont {V.~S.}\ \bibnamefont {Pande}},\ }\bibfield  {title} {\bibinfo {title} {{OpenMM} 7: {Rapid} development of high performance algorithms for molecular dynamics},\ }\href
  {https://doi.org/10.1371/journal.pcbi.1005659} {\bibfield  {journal} {\bibinfo  {journal} {PLOS Computational Biology}\ }\textbf {\bibinfo {volume} {13}},\ \bibinfo {pages} {e1005659} (\bibinfo {year} {2017})}\BibitemShut {NoStop}%
\bibitem [{\citenamefont {Piana}\ \emph {et~al.}(2011)\citenamefont {Piana}, \citenamefont {Lindorff-Larsen},\ and\ \citenamefont {Shaw}}]{piana_robust_2011}%
  \BibitemOpen
  \bibfield  {author} {\bibinfo {author} {\bibfnamefont {S.}~\bibnamefont {Piana}}, \bibinfo {author} {\bibfnamefont {K.}~\bibnamefont {Lindorff-Larsen}},\ and\ \bibinfo {author} {\bibfnamefont {D.~E.}\ \bibnamefont {Shaw}},\ }\bibfield  {title} {\bibinfo {title} {How robust are protein folding simulations with respect to force field parameterization?},\ }\href {https://doi.org/10.1016/j.bpj.2011.03.051} {\bibfield  {journal} {\bibinfo  {journal} {Biophysical Journal}\ }\textbf {\bibinfo {volume} {100}},\ \bibinfo {pages} {L47} (\bibinfo {year} {2011})}\BibitemShut {NoStop}%
\bibitem [{\citenamefont {Abraham}\ \emph {et~al.}(2015)\citenamefont {Abraham}, \citenamefont {Murtola}, \citenamefont {Schulz}, \citenamefont {Páll}, \citenamefont {Smith}, \citenamefont {Hess},\ and\ \citenamefont {Lindahl}}]{abraham_gromacs_2015}%
  \BibitemOpen
  \bibfield  {author} {\bibinfo {author} {\bibfnamefont {M.~J.}\ \bibnamefont {Abraham}}, \bibinfo {author} {\bibfnamefont {T.}~\bibnamefont {Murtola}}, \bibinfo {author} {\bibfnamefont {R.}~\bibnamefont {Schulz}}, \bibinfo {author} {\bibfnamefont {S.}~\bibnamefont {Páll}}, \bibinfo {author} {\bibfnamefont {J.~C.}\ \bibnamefont {Smith}}, \bibinfo {author} {\bibfnamefont {B.}~\bibnamefont {Hess}},\ and\ \bibinfo {author} {\bibfnamefont {E.}~\bibnamefont {Lindahl}},\ }\bibfield  {title} {\bibinfo {title} {{GROMACS}: {High} performance molecular simulations through multi-level parallelism from laptops to supercomputers},\ }\href {https://doi.org/10.1016/j.softx.2015.06.001} {\bibfield  {journal} {\bibinfo  {journal} {SoftwareX}\ }\textbf {\bibinfo {volume} {1-2}},\ \bibinfo {pages} {19} (\bibinfo {year} {2015})}\BibitemShut {NoStop}%
\end{thebibliography}%


\section*{End Matter}

\phantomsection\label{em:ffi}
\paragraph{Flux--fidelity identity.}
Let $\Omega'=\Omega\setminus(\bar A\cup\bar B)$, let $\mathbf{n}_A$ and
$\mathbf{n}_B$ be the basin normals pointing into $\Omega'$, and extend $q$ by its
boundary values, $q\equiv0$ on $A$ and $q\equiv1$ on $B$. Then
$\mathbf{J}=\rho\mathbf{D}\nabla q$ vanishes in the basins, so
$\langle u,q\rangle_{\mathbf{D}}=\int_{\Omega'}\nabla u\!\cdot\!\mathbf{J}\,
d\mathbf{x}$ for any $u$ of finite Dirichlet energy, and Green's first identity
gives
\begin{equation}
\langle u,q\rangle_{\mathbf{D}}
=\underbrace{-\!\int_{\Omega'}\!u\,\nabla\!\cdot\!\mathbf{J}\,d\mathbf{x}}_{=\,0}
-\int_{\partial A}\!u\,\mathbf{J}\!\cdot\!\mathbf{n}_A\,dS
-\int_{\partial B}\!u\,\mathbf{J}\!\cdot\!\mathbf{n}_B\,dS ,
\label{eq:em_green}
\end{equation}
the volume term vanishing because $\nabla\!\cdot\!\mathbf{J}=0$. On $\partial A$
and $\partial B$ the normal current is outgoing and incoming respectively, with
total $\nu_{AB}$ through either [Eq.~(\ref{eq:nuab})], so dividing each surface
integral by $\nu_{AB}$ gives averages against unit-mass measures,
$\langle f\rangle^{\mathrm{flux}}_{\partial A}=\nu_{AB}^{-1}\int_{\partial A}
f\,\mathbf{J}\!\cdot\!\mathbf{n}_A\,dS$ and
$\langle f\rangle^{\mathrm{flux}}_{\partial B}=-\nu_{AB}^{-1}\int_{\partial B}
f\,\mathbf{J}\!\cdot\!\mathbf{n}_B\,dS$, and Eq.~(\ref{eq:em_green}) becomes
Eq.~(\ref{eq:ffi}). Since both measures have unit mass, $\mathcal{F}$ annihilates
constants, and $\mathcal{F}[u]\in[-1,1]$ for $u\in[0,1]$ because
$\mathbf{J}\!\cdot\!\mathbf{n}_A\ge0$ on $\partial A$ and
$\mathbf{J}\!\cdot\!\mathbf{n}_B\le0$ on $\partial B$ for cleanly separated
basins. For such $u$, $\mathcal{F}[u]=1$ requires $u=0$ and $u=1$ almost everywhere
with respect to the two flux measures.

\phantomsection\label{em:principle}
\paragraph{The principle without boundary conditions.}
Assume $\nu_{AB}>0$ and let
$\mathcal{U}=\{u:\mathcal{E}[u]<\infty,\ \mathcal{F}[u]\neq0\}$. For
$u\in\mathcal{U}$, Cauchy--Schwarz gives
$\langle u,q\rangle_{\mathbf{D}}^2\le\mathcal{E}[u]\,\mathcal{E}[q]$;
substituting Eq.~(\ref{eq:ffi}) and $\mathcal{E}[q]=\nu_{AB}$ and dividing by
$\nu_{AB}$,
\begin{equation}
\nu_{AB}\le\frac{\mathcal{E}[u]}{\mathcal{F}[u]^2},\qquad u\in\mathcal{U}.
\label{eq:em_bound}
\end{equation}
The quotient is well defined on $\mathcal{U}$: $\mathcal{E}[u]=0$ implies $u$
constant, and constants have $\mathcal{F}=0$. The bound is attained at $u=q$, where
$\mathcal{F}[q]=1$ and $\mathcal{E}[q]=\nu_{AB}$; the infimum therefore equals
$\nu_{AB}$, which is Eq.~(\ref{eq:principle}). Equality in Cauchy--Schwarz
requires $\nabla u$ and $\nabla q$ parallel in the
$\langle\cdot,\cdot\rangle_{\mathbf{D}}$ inner product, i.e.
$u=\lambda q+\mathrm{const}$ with $\lambda\neq0$; such $u$ has
$\mathcal{F}[u]=\lambda$ and $\mathcal{E}[u]=\lambda^2\nu_{AB}$, whose quotient is
$\nu_{AB}$ for every $\lambda$. The minimiser is a family, which is why the ansatz
of Eq.~(\ref{eq:ansatz}) contains a constant. No boundary condition was used: only
the flux-weighted average of the boundary violation enters, through $\mathcal{F}$.
The classical context of Eq.~(\ref{eq:principle}) and its prior
art~\cite{lorpaiboon_exact_2026} are discussed in the main text. For a finite
$V_M$ the maximisation in Eq.~(\ref{eq:ratio}) is over a scale-invariant quotient,
hence $\mathcal{R}_M<\infty$, and monotonicity under $V_M\subset V_{M+1}$ is
immediate.

\phantomsection\label{em:solve}
\paragraph{The optimum, its calibration, and its price.}
With $\mathcal{E}[\bar q]=\mathbf{w}^\top G\mathbf{w}$ and
$\mathcal{F}[\bar q]=\mathbf f^\top\mathbf{w}$,
$\max_{\mathbf{w}}(\mathbf f^\top\mathbf{w})^2/(\mathbf{w}^\top G\mathbf{w})
=\mathbf f^\top G^{-1}\mathbf f=\mathcal{R}_M$, attained on
$\mathbf{w}\propto G^{-1}\mathbf f$. Normalising to
$\mathbf f^\top\mathbf{w}=1$ and replacing $\mathbf f$ by $\hat{\mathbf f}$
gives Eq.~(\ref{eq:wstar}) and
$\mathcal{E}[\bar q^\star]=1/\hat{\mathcal{R}}_M$. The bias, fixed by
$\mu_A[\bar q^\star]=0$, does not affect $\mathcal{R}_M$: both $\mathcal{E}$ and
$\mathcal{F}$ annihilate constants, so enlarging the trial space by them leaves the
quotient of Eq.~(\ref{eq:ratio}) unchanged.

Minimising the Dirichlet error instead gives the Galerkin optimum
$\mathbf{w}_{\mathrm{Gal}}=\nu_{AB}G^{-1}\mathbf f$, the same ray rescaled by the
unknown flux. Its fidelity is $\nu_{AB}\mathcal{R}_M\le1$, so
$\bar q_{\mathrm{Gal}}$ under-spans $[0,1]$ by the factor by which the bound is
loose. The unit-fidelity choice costs
\begin{equation}
\mathcal{E}[\bar q^\star-q]-\mathcal{E}[\bar q_{\mathrm{Gal}}-q]
=\mathcal{R}_M\,\Delta^2,\qquad \Delta\equiv\frac{1}{\mathcal{R}_M}-\nu_{AB}.
\label{eq:em_cost}
\end{equation}
Since $\mathcal{E}[\bar q^\star-q]=\Delta$ by Eq.~(\ref{eq:apost}), the excess is
a fraction $1-\nu_{AB}\mathcal{R}_M\in[0,1]$ of the error, so the calibration
never dominates it. Both results are exact in $\mathbf f$ and inherit the
bias of Eq.~(\ref{eq:phihat}) in $\hat{\mathbf f}$.

\phantomsection\label{em:basis}
\paragraph{The slice profile.}
With hard boundaries at the edges $s_A,s_B$ of the two projected states, the
$\kappa=0$ minimiser of Eq.~(\ref{eq:Ekappa}) is
$q_j(s)=\int_{s_A}^{s}e^{\beta F_j}dy\big/\!\int_{s_A}^{s_B}e^{\beta F_j}dy$ in
the projected free energy
$F_j=-\beta^{-1}\log\rho_j$~\cite{berezhkovskii_one-dimensional_2005,
krivov_one-dimensional_2006,krivov_diffusive_2008}, which the
tail pinning of the main text renders unusable. The penalty of
Eq.~(\ref{eq:Ekappa}) replaces it, and the Euler--Lagrange equation of that
functional~\cite{supplemental} is the committor of a one-dimensional diffusion
killed into $A$ and $B$ wherever their densities carry weight, at cost
$\mathcal{O}(n_{\mathrm{bins}})$ per direction. Substitute the
$\kappa\to\infty$ classifier of Eq.~(\ref{eq:classifier}) into the basin moments.
The region where only $A$ has support contributes nothing, since $q_j=0$ there;
writing $\rho^B_j/(\rho^A_j+\rho^B_j)=1-\rho^A_j/(\rho^A_j+\rho^B_j)$ in the
second moment leaves the same harmonic-mean overlap integral in both,
$\mathcal{I}_j=\int\rho^A_j\rho^B_j/(\rho^A_j+\rho^B_j)\,ds$ over
$\mathrm{supp}\rho^A_j\cap\mathrm{supp}\rho^B_j$. Hence
$a_j=\mathcal{I}_j/m^A_j$, $b_j=1-\mathcal{I}_j/m^B_j$ and
Eq.~(\ref{eq:overlap}). The overlap vanishes iff the projected supports are
disjoint, so a slice has unit empirical fidelity exactly when its projection
separates the states. Under sufficiently heavy overlap $\hat{\mathcal{F}}_j$ falls
below zero, which the sign-indefinite solve of Eq.~(\ref{eq:wstar}) accommodates.

\ifarxiv
  \clearpage
  \onecolumngrid
  \begin{center}
    \large\bfseries Supplemental Material
  \end{center}
  \vspace{1em}
  \twocolumngrid
  
\setcounter{equation}{0}
\renewcommand{\theequation}{S\arabic{equation}}
\setcounter{figure}{0}
\renewcommand{\thefigure}{S\arabic{figure}}
\setcounter{table}{0}
\renewcommand{\thetable}{S\arabic{table}}
\renewcommand{\theHequation}{S\arabic{equation}}
\renewcommand{\theHfigure}{S\arabic{figure}}
\renewcommand{\theHtable}{S\arabic{table}}

\section*{Reaction-diffusion slice committors}\label{em:rd}

Each slice profile minimises the penalised one-dimensional energy of
Eq.~(\ref{eq:Ekappa}) of the main text, reproduced here,
\begin{equation}
\mathcal{E}_\kappa[q_j]
= \!\int\!\rho_j\,(q_j')^2\,ds
+ \kappa\!\!\int\!\rho_j^A\,q_j^2\,ds
+ \kappa\!\!\int\!\rho_j^B\,(1-q_j)^2\,ds,
\label{eq:E_kappa}
\end{equation}
on the projected line with natural Neumann conditions at the domain edges.
The first term penalises $|q_j'|^2$ weighted by the equilibrium marginal
$\rho_j$; the second and third drive $q_j$ towards $0$ and $1$ with strength
$\kappa$ distributed by the projected state densities. Varying
Eq.~(\ref{eq:E_kappa}) and integrating the gradient term by parts gives its
Euler--Lagrange equation, the reaction-diffusion problem solved once per
direction,
\begin{equation}
\begin{gathered}
\frac{d}{ds}\!\left[\rho_j(s)\,\frac{dq_j}{ds}\right]
=\kappa\bigl[\rho_j^A(s)\,q_j-\rho_j^B(s)\,(1-q_j)\bigr],\\
\rho_j\,q_j'\big|_{\partial}=0 ,
\end{gathered}
\label{eq:rd_ode}
\end{equation}
in which the two penalty terms act as absorbers that pull $q_j$ towards $0$
where $A$ projects and towards $1$ where $B$ projects, while the left-hand side
is untouched: where neither state carries weight the profile obeys the bare
one-dimensional committor equation $(\rho_j q_j')'=0$. As $\kappa\to\infty$ the penalty dominates wherever either state carries weight,
and the profile approaches the Bayesian classifier
$q_j\to\rho_j^B/(\rho_j^A+\rho_j^B)$ of Eq.~(\ref{eq:classifier}). This limit is
the absorption probability of a diffusion in the projected potential
$F_j(s)=-\beta^{-1}\log\rho_j(s)$, killed at the state-conditional rates
$\kappa\rho_j^A/\rho_j$ into $A$ and $\kappa\rho_j^B/\rho_j$ into $B$. Its basin
moments collapse to one minus the harmonic-mean overlap of the projected state
densities, Eq.~(\ref{eq:overlap}), derived in the End Matter (\emph{The slice
profile}); the overlap vanishes iff the projected supports are disjoint, where
$\hat{\mathcal{F}}_j=1$, and $\hat{\mathcal{F}}_j$ falls below zero under sufficiently heavy overlap.
This overlap underlies the empirical-estimator argument below
(\hyperref[em:emp]{Empirical fidelity on a molecular system}).

\paragraph{Slice dissipation.}
The bare dissipation of the RD profile $q_j^{\mathrm{RD}}$ solving
Eq.~(\ref{eq:rd_ode}), $I_j = \int\rho_j\,[(q_j^{\mathrm{RD}})']^2\,ds$, is computed
numerically; it enters the Gram matrix through the diagonal
$G_{jj} = (\bm\theta_j^\top\mathbf{D}\bm\theta_j)\,I_j$
(\hyperref[em:diag]{Diagonal limit}).

\paragraph{Tridiagonal discretisation.}
On a uniform grid $\{s_i\}_{i=0}^{n-1}$ with spacing $\Delta s$,
Eq.~(\ref{eq:rd_ode}) yields the symmetric positive-definite tridiagonal
system
\begin{equation}
\begin{split}
&-\rho_{i-\frac{1}{2}}\,q_{i-1}
+ \bigl[\rho_{i-\frac{1}{2}} + \rho_{i+\frac{1}{2}}
       + \kappa\,\Delta s^2\,(\rho_i^A + \rho_i^B)\bigr]\,q_i \\
&\qquad - \rho_{i+\frac{1}{2}}\,q_{i+1} = \kappa\,\Delta s^2\,\rho_i^B,
\end{split}
\label{eq:rd_tridiag}
\end{equation}
with $\rho_{i\pm\frac{1}{2}} = (\rho_i + \rho_{i\pm 1})/2$, solved in
$\mathcal{O}(n_\text{bins})$ by the Thomas algorithm.

\section*{Constrained solve and weights}

\paragraph{Error decomposition and constrained solve.}
The sliced committor gradient is
$\nabla\bar q = \sum_j w_j\,q_j'(\bm\theta_j\!\cdot\!\mathbf{x})\,\bm\theta_j$,
the bias contributing nothing since $\nabla c = 0$, so
$\mathcal{E}[\bar q] = \mathbf{w}^\top G\,\mathbf{w}$ with $G$ from
Eq.~(\ref{eq:gram}) of the main text. Reducing each cross-term through the
flux--fidelity identity Eq.~(\ref{eq:ffi}) and substituting the basin-moment
fidelity of Eq.~(\ref{eq:phihat}) (justified in \hyperref[em:emp]{Empirical
fidelity on a molecular system}) turns the Dirichlet error into the explicit
quadratic
\begin{equation}
\mathcal{E}[\bar q - q]
= \mathbf{w}^\top G\,\mathbf{w}
- 2\,\mathcal{E}[q]\,(\mathbf{b}-\mathbf{a})^\top\mathbf{w}
+ \mathcal{E}[q],
\label{eq:error_quad_supp}
\end{equation}
in which the unknown $\mathcal{E}[q]=\nu_{AB}$ multiplies only the linear term.
The two weak boundary conditions
$\mu_A[\bar q] = c + \mathbf{a}^\top\mathbf{w} = 0$ and
$\mu_B[\bar q] = c + \mathbf{b}^\top\mathbf{w} = 1$ eliminate the bias through
$c = -\mathbf{a}^\top\mathbf{w}$ and collapse to the single constraint
$(\mathbf{b}-\mathbf{a})^\top\mathbf{w} = 1$; on that surface the linear term is
the constant $-2\,\nu_{AB}$, so
$\mathcal{E}[\bar q - q] = \mathbf{w}^\top G\,\mathbf{w} - \nu_{AB}$ and minimising
the computable energy is identical to minimising the true error. The minimiser is
the closed form of Eq.~(\ref{eq:wstar}) with
$\mathcal{E}[\bar q^\star] = 1/\hat{\mathcal{R}}_M$ (derived in the End Matter of
the main text, \emph{The optimum, its calibration, and its price}).

\paragraph{Range and sign of the fidelity.}
The basin moments are means of $q_j\in[0,1]$ over state samples, so
$a_j,b_j\in[0,1]$ and the fidelity $\hat{\mathcal{F}}_j = b_j - a_j\in[-1,1]$ without
further conditioning. A negative $\hat{\mathcal{F}}_j$ marks a slice whose projected
committor is anti-correlated with the true committor on the equilibrium
samples, with higher mean in $A$ than in $B$; such slices are admitted and
corrected by the sign-indefinite solve, not discarded. The constrained
solve of Eq.~(\ref{eq:wstar}) likewise permits sign-indefinite
$w_j^\star$: a negative entry subtracts a slice that its Gram partners
overcount, the standard outcome of a least-squares projection with a
non-diagonal $G$. Imposing $\mathbf{w}\ge0$ would bias the optimum, and would not
by itself keep $\bar q$ in $[0,1]$ either: with $\mathbf{w}\ge0$ the ansatz lies in
$[c,\,c+\sum_j w_j]$, and the calibration $c=-\mathbf{a}^\top\mathbf{w}$ makes
$c\le0$. Because the ansatz is non-conforming, $\bar q$ does leave the interval:
on the molecular systems a third to a half of samples fall outside, almost all of
them inside a basin, where the profiles saturate and the projected states overlap.
We therefore clip $\bar q\mapsto\max(0,\min(1,\bar q))$ on evaluation. The clip
does not affect the weight optimisation, which is algebraic in $\mathbf{w}$ and
$G$, and it does not affect the flux read-off, since clipped samples are carried to
$\bar q=0$ or $1$ and so fall outside the band of Eq.~(\ref{eq:k_qstrat}).

\paragraph{Diagonal limit.}\label{em:diag}
For $G = \mathrm{diag}(G_{11},\ldots,G_{MM})$ with
$G_{jj} = (\bm\theta_j^\top\mathbf{D}\bm\theta_j)\,I_j$, the solve of
Eq.~(\ref{eq:wstar}) is componentwise,
\begin{equation}
w_j^\star = \frac{1}{\hat{\mathcal{R}}_M}\,\frac{b_j - a_j}{G_{jj}},
\qquad
\hat{\mathcal{R}}_M = \sum_j\frac{(b_j - a_j)^2}{G_{jj}}.
\label{eq:diag_limit_supp}
\end{equation}
Each slice carries
weight proportional to its basin-moment gap over its dissipation; a flat
slice ($b_j = a_j$) is suppressed regardless of its $G_{jj}$, and the projected
diffusivity $\bm\theta_j^\top\mathbf{D}\bm\theta_j$ enters only the
denominator.

\paragraph{A certified bound on the reactive flux.}\label{em:bound}
Equation~(\ref{eq:principle}) holds for \emph{every} trial function, so any $u$
with $\mathcal{F}[u]\neq0$ caps the flux,
\begin{equation}
\nu_{AB}\;\le\;\frac{\mathcal{E}[u]}{\mathcal{F}[u]^2} ,
\label{eq:cap}
\end{equation}
with no boundary conditions, no restriction to $[0,1]$, no nesting of trial
spaces, and no sign condition on $\mathcal{F}$, which enters squared. Over a
finite $V_M$ the minimum of the right-hand side is $1/\mathcal{R}_M$,
Eq.~(\ref{eq:ratio}). Two things follow that the estimate below does not
provide. First, a ceiling on the flux and hence on the rate, certified from the
samples and the state definitions alone. Second, an admissibility test: any
independent estimate of $\nu_{AB}$ that exceeds the smallest available cap is
inconsistent with the same data.

The cap needs $\mathcal{F}[u]$, so it is certified where reactive segments can
be harvested (\hyperref[em:emp]{Empirical fidelity on a molecular system}). On
AIB9 the deployed $M=512$ committor has $\mathcal{F}[\bar q]=0.9935$ and
$\mathcal{E}[\bar q]=0.01795$, giving $\nu_{AB}\le0.01819$; the plateau-free
$1/\mathcal{R}_M$ with the measured $\mathbf f$ is $0.01820$, looser by $0.05\%$
and so effectively the same bound. The two coincide because the ridge selected by
Eq.~(\ref{eq:ridge_cv}) is light enough here that it barely depresses
$\mathcal{E}[\bar q]$ below the unregularised optimum, while $\mathcal{F}$ stays
near one. At $M=1024$,
$\mathcal{F}[\bar q]=0.9923$ and $\mathcal{E}[\bar q]=0.01522$ tighten the cap to
$\nu_{AB}\le0.01545$.
One caveat on the units. Both $\mathcal{E}$ and $\hat\nu_{AB}$ carry the
arbitrary constant $\mathbf{D}$, and so does the cap of Eq.~(\ref{eq:cap}).
(That shared factor is also what lets $\hat e$ below cancel $\mathbf{D}$ and so
be available on the peptides.) The cap is a physical ceiling on the rate only
where $\mathbf{D}$ is independently known, as on the two-dimensional benchmark.
On AIB9 the rate is read by the $\mathbf{D}$-free forward tracing, so there the
cap's content is dimensionless: a statement about $\hat\nu_{AB}/\nu_{AB}$.

\paragraph{Evaluating the estimate without a reference.}\label{em:apost}
Equation~(\ref{eq:apost}) needs two numbers and neither requires $q$. The first
is the ansatz energy $\mathcal{E}[\bar q]=\mathbf{w}^\top G\,\mathbf{w}$, which
equals $1/\hat{\mathcal{R}}_M$ at the unregularised optimum. The second is the
reactive flux, read from the isocommittor-flux profile of Eq.~(\ref{eq:Fhat})
across the saddle band $\bar q\in[0.2,\,0.8]$. Both are averages of
$\rho\,\nabla\bar q^\top\mathbf{D}\,\nabla\bar q$ over the same samples and
differ only in the region of $\bar q$ they cover, so a constant $\mathbf{D}$
multiplies them alike and cancels from the relative error
\begin{equation}
\hat e \;=\; \frac{\mathcal{E}[\bar q]}{\hat\nu_{AB}} - 1
\;\simeq\; \frac{\mathcal{E}[\bar q-q]}{\nu_{AB}} .
\label{eq:ehat}
\end{equation}
Because $\mathbf{D}$ drops out, $\hat e$ is available wherever the committor is,
and in particular on the two peptides, for which no diffusion constant is
estimated anywhere in this work. Table~\ref{tab:apost} reports it for all four
systems at their deployed $M$, together with the flatness of the profile, which
tests whether the plateau the read-off presumes exists at all. Both energies are
expressed in each system's own feature metric and so are comparable only within
a row; $\hat e$ and the flatness are dimensionless. Substituting $\hat\nu_{AB}$
for $\nu_{AB}$ is the one step the samples cannot check on their own, so $\hat e$
is an estimate where Eq.~(\ref{eq:cap}) is a bound. Two structural limits fix its
endpoints. If every slice separates the states, $\hat{\mathcal{F}}_j = 1$ along
every direction, the constraint is met without distortion and the residual
$1/\hat{\mathcal{R}}_M - \nu_{AB}$ is purely representational and vanishes iff $q$
lies in the slice span; if instead the states fail to separate, $b_j\to a_j$ for
every $j$, so $\hat{\mathcal{R}}_M\to0$ and the optimal energy diverges, a
representation gap curable only by enriching the direction set.

\paragraph{A stopping criterion from the next rung.}\label{em:ladder}
The cap makes the estimate's own inputs into a bound. Writing the error
expansion in full, $\mathcal{E}[\bar q_M-q]=\mathcal{E}_M-2\nu_{AB}\mathcal{F}_M
+\nu_{AB}$, and capping $\nu_{AB}$ by Eq.~(\ref{eq:cap}) applied to a
\emph{finer} fit $\bar q_{M'}$,
\begin{equation}
\frac{\mathcal{E}[\bar q_M-q]}{\nu_{AB}}\;\ge\;
\frac{\mathcal{E}_M}{\mathcal{E}_{M'}}\mathcal{F}_{M'}^2-2\mathcal{F}_M+1 .
\label{eq:ladder}
\end{equation}
At unit fidelity this is $L_M=\mathcal{E}_M/\mathcal{E}_{M'}-1$, and in absolute
terms $\mathcal{E}_M-\mathcal{E}_{M'}\le\mathcal{E}[\bar q_M-q]$: the energy one
more refinement recovers is a lower bound on the Dirichlet error of the fit
before it, in the units of $\nu_{AB}$ that the target accuracy is set in. That
is the stopping criterion of the main text, and it needs no reference committor,
no plateau and no fidelity measurement, only a second fit from the sweep already
being run. Nesting is not required either: any trial function with a lower
energy will serve, since Eq.~(\ref{eq:cap}) is unconditional. What will not
serve is the fit itself, $u=\bar q_M$, which returns the vacuous
$\mathcal{E}[\bar q_M-q]/\nu_{AB}\ge(1-\mathcal{F}_M)^2$; a fit cannot bound its
own error, which is also why Eq.~(\ref{eq:cap}) cannot double as an estimate of
$\nu_{AB}$: the cap is tight only for $u=\lambda q+\mathrm{const}$.

Because $L_M$ assumes exactly what Eq.~(\ref{eq:apost}) already assumes, the two
are comparable, and $L_M>\hat e$ says the plateau term is the one at fault.
Table~\ref{tab:apost_2d} tests both forms on the two-dimensional benchmark,
where $\nu_{AB}$, $\nabla q$ and $\mathcal{F}$ are all known. Over the twenty
nested pairs of the two sweeps, at $N=10^5$ and $N=4\times10^5$, the
unit-fidelity bound $L_M\le e$ holds in every one, as does the measured-fidelity
form $L_M^{\mathcal{F}}$ of Eq.~(\ref{eq:ladder}). The unit-fidelity form's only
failure mode, a fidelity drift between rungs, occurs in neither sweep.
Read as a test of $\hat e$, $L_M>\hat e$ flags $\hat e$ as understating
the error with $0$ false positives and $4$ false negatives, and it is the absence of
false positives that licenses reading a flag as a verdict on the read-off. One is the
$M=1$ blind spot discussed below, in the $N=4\times10^5$ sweep, where nothing flags
it; the other three are marginal,
at $M=64$, $128$ and $256$ of the $N=10^5$ sweep, which
Table~\ref{tab:apost_2d} does not tabulate, where $\hat e$ understates $e$ by less
than a factor of two.

\paragraph{What the estimate says on each system.}
Each reading below normalises $\hat\Phi$, the sampled flux profile $\hat F$ of
Eq.~(\ref{eq:Fhat}) below, by its own plateau mean over
$\bar q\in[0.2,0.8]$ and quotes its value in the two extreme bins, at
$\bar q\!\to\!0$ and $\bar q\!\to\!1$ (Table~\ref{tab:apost}). The two ends fix the
sign: $\mathcal{E}[\bar q]=\int_0^1\hat\Phi\,dz$ runs over the whole domain while
$\hat\nu_{AB}$ is read only inside the band, so ends above the plateau make $\hat e$
positive and ends below it make $\hat e$ negative.
\emph{Two-dimensional benchmark.} Both ends sit above one ($1.19$ and
$1.11$), the mild basin-side excess that Eq.~(\ref{eq:k_qstrat}) is designed to
filter, giving $\hat e=+0.0066$. The ladder does not flag it, $L_M$ being
slightly negative once the energy has converged in $M$: nesting would force
$L_M\ge0$ for the exact minimiser over the slice span, but the regularised solve is
not that minimiser, so the difference between rungs fluctuates about zero with the
direction draw. This rung is one of the
marginal false negatives counted above, since the true $e$ is twice $\hat e$ here.
\emph{Villin.} The ends rise above the plateau ($1.20$ and $1.72$), so $\hat e>0$;
it decreases on the next rung, from $0.21$ at the deployed $M=2048$ to $0.17$ at
$M=4096$, which also supplies the rung that bounds the error below ($L_M=0.093$).
\emph{AIB9.} The ends collapse to $0.31$ and $0.41$ of the plateau value. This
pulls the whole-domain energy below the plateau and drives $\hat e$ negative
at $M=512$. The dip is not an artefact of the boundary quantile of
Table~\ref{tab:aib9_params}: refitting with no boundary truncation leaves
it in place ($0.38$ and $0.55$, $\hat e=-0.085$).
\emph{Chignolin (control).} Its two ends straddle the plateau, the only system where
they do, and the folded-side excess dominates the integral; there is no plateau to
read at all (\emph{Chignolin: the plateau premise fails}).

Villin and chignolin both clear their ladder bounds, villin by a factor of $2$, and
AIB9 alone is flagged. None of these orderings need follow the RMSE column
($0.062$ for AIB9 against $0.21$
for villin): the
Dirichlet error and the pointwise error against a forward-tracing reference are
different quantities, and Eq.~(\ref{eq:apost}) estimates only the first.

\paragraph{Diagnosing the negative gap on AIB9.}\label{em:apost_aib9}
The gap is a sum of three terms,
$\mathcal{E}[\bar q]-\hat\nu_{AB}
= \mathcal{E}[\bar q-q] + 2\nu_{AB}(\mathcal{F}[\bar q]-1)
+ (\nu_{AB}-\hat\nu_{AB})$, and which one is responsible can be settled because
the fidelity is \emph{linear}. Since $\mathcal{F}$ kills constants,
$\mathcal{F}[\bar q]=\sum_j w_j\mathcal{F}_j$, and the constraint fixes
$\sum_j w_j\hat{\mathcal{F}}_j=1$, so
\begin{equation}
1-\mathcal{F}[\bar q] \;=\; \sum_j w_j\bigl(\hat{\mathcal{F}}_j-\mathcal{F}_j\bigr),
\label{eq:fid_deficit}
\end{equation}
which needs no reference committor, only the per-slice reactive-flux fidelities
evaluated on the deployed slices. Measured there
(\hyperref[em:emp]{Empirical fidelity on a molecular system}), the weighted
deficit is only $0.0065$ with a bootstrap interval $[0.0045,0.0090]$ over the
harvested segments, giving $\mathcal{F}[\bar q]=0.9935$.

The three candidate terms then separate. (i)~The energy
$\mathcal{E}[\bar q]$ is a sample average with no read-off choices, and
reproduces the solver's own $\mathbf{w}^\top G\mathbf{w}$ to a relative
$10^{-16}$; it is exact. (ii)~The fidelity term is small: overturning the
cap-versus-plateau inequality below would require $\mathcal{F}\le0.949$, a
deficit about eight times the measured one, and the $73$ observed crossings put
$\bar q$ at a mean of $0.004$ on $\partial A$ and $0.997$ on $\partial B$, hence
$\mathcal{F}=0.993$ by a route independent of the weighted deficit above.
Only a reactive flux concentrated on the least favourable
crossings could reach $0.949$, which is the caveat quantified below.
(iii)~The remaining term is
the read-off, which Eq.~(\ref{eq:cap})
settles without reference to the error at all: the cap is
$\mathcal{E}[\bar q]/\mathcal{F}[\bar q]^2=0.01819$, whereas the plateau reads
$\hat\nu_{AB}=0.01995$. On AIB9 the read-off therefore overestimates $\nu_{AB}$
by at least $9.7\%$. The read-off, not the fidelity substitution, drives
$\hat e$ negative: fidelity contributes at most $2(1-\mathcal{F})=0.013$ of the
measured $\hat e=-0.100$, about an eighth.

The finding's scope is narrow. It concerns
Eq.~(\ref{eq:k_qstrat}) on this system, not $\bar q$, whose
transition-region RMSE of $0.062$ against the forward-tracing reference is
measured independently and stands. Taking the $M=1024$ fit as the trial function
in Eq.~(\ref{eq:cap}) gives $\nu_{AB}\le0.01545$, hence
$r=\hat\nu_{AB}/\nu_{AB}\ge1.29$ at $M=512$ and, through
Eq.~(\ref{eq:e_decomp}) below, $e\ge0.175$. The $M=512$ trial space was
therefore not exhausted, and the stopping criterion says so directly: the
$M=1024$ rung recovers $0.0027$ of energy, which is the margin that produced
that bound.
It does not follow that the finer fit is pointwise better. Its RMSE is $0.0679$,
worse than $0.0615$ at $M=512$, which is the same
Dirichlet-versus-pointwise distinction as above: the ladder bounds the first and
is not a predictor of the second. For rates the direction of the bias is the
useful part, since a read-off that is too high makes a rate too fast. Villin's and
chignolin's read-offs clear their ladder bounds, so the rate reported in the main
text is untouched.

\paragraph{Closing the gap: the two-bias decomposition.}
Written additively in units of $\nu_{AB}$ the identity separates the measured
gap from the two biases,
\begin{equation}
e \;=\; \hat e\,r \;+\; (r-1) \;+\; 2\bigl(1-\mathcal{F}[\bar q]\bigr),
\qquad r=\hat\nu_{AB}/\nu_{AB} .
\label{eq:e_decomp}
\end{equation}
On the two-dimensional benchmark the two biases carry \emph{opposite} signs and
largely cancel: at $M=256$, $N=10^5$ they are $-0.042$ and $+0.048$ against a
gap term of $+0.006$, summing to the true $e=0.0128$. That cancellation
makes $\hat e$ usable there, and it is not guaranteed. On AIB9 both biases are
positive and therefore add, and at $r\ge1.29$ the plateau term $+0.29$ exceeds
the fidelity term $+0.013$ by a factor of $22$.

\paragraph{Where the decomposition can be closed.}
For villin and chignolin only the unit-fidelity form is available, because
closing Eq.~(\ref{eq:e_decomp}) needs $\mathcal{F}$, and $\mathcal{F}$ needs
reactive segments harvested on the deployed slices. Villin's trajectory supplies
about fifteen committed folding events against AIB9's $73$, and chignolin's
umbrella sampling supplies none at all, its trajectories being biased. The
decomposition is therefore complete only where unbiased reactive paths are
plentiful, which is the one place the method itself does not need them; the
ladder bound of Eq.~(\ref{eq:ladder}) remains available everywhere.

\paragraph{Three limits on the AIB9 overestimate.}
The overestimate is specific to the $[0.2,\,0.8]$ window: it falls to roughly
$9\%$ on $[0.3,\,0.7]$ and $5.6\%$ on $[0.4,\,0.6]$, though the sign never turns. It is specific to AIB9: on the two-dimensional benchmark
the same read-off \emph{under}-estimates at converged $M$
($\hat\nu_{AB}/\nu_{AB}=0.88$ to $0.98$), so this is not a general bias of
Eq.~(\ref{eq:k_qstrat}). And it is not free of the flux measure on the state
boundary, since a reactive flux concentrated on the least favourable $1\%$ of
the rim would give a deficit of $0.070$, above the $0.051$ that overturning the
inequality requires, and so reverse it; the $73$
observed crossings argue against that, but that is
a $73$-event empirical statement rather than a proof.

\paragraph{Chignolin: the plateau premise fails.}
Chignolin's profile does not plateau but decays monotonically, from $9.9$ times the
band mean at the folded end to $0.134$ at the unfolded one, so the band carries $27\%$ of the
Dirichlet mass where flux conservation would put $60\%$, the flatness is $0.71$,
nearly four times the largest of the other three systems, and the read-off moves
by $22\%$ across the nested windows. The narrower flatness-selected band of
Table~\ref{tab:chignolin_params} does not repair it and reads higher
($\hat e = 1.76$ over $\bar q\in[0.34,0.58]$ against $1.54$ over $[0.2,0.8]$).
The reported
$\hat e\approx1.5$ should therefore be read as the
diagnostic it is, that the umbrella-sampled chignolin committor carries large
basin-side Dirichlet error in the $86$-dimensional torsion metric, and not as a
calibrated error bar. Part of that error is attributable to the boundary
treatment: chignolin is the one system evaluated without the basin clamp, so its
$\mathcal{E}[\bar q]$ retains residual basin gradients that the clamp removes
elsewhere, which inflates $\hat e$ but leaves the band untouched. This does not
bear on the chignolin rate of the main text, which is read from the
flatness-selected median precisely to be insensitive to the basin-side
contamination that can dominate the volume integral.

\paragraph{Auditing the plateau read-off.}
The plateau substitution of Eq.~(\ref{eq:ehat}) is the one step the samples
cannot self-check (\hyperref[em:apost]{Evaluating the estimate without a
reference}); the two-dimensional benchmark can, because the finite-difference
reference supplies both $\nu_{AB}$ and $\nabla q$ (Table~\ref{tab:apost_2d}). Across the $14$ of
$22$ configurations in which $\hat\nu_{AB}$ falls within $10\%$ of $\nu_{AB}$,
the ratio $\hat e/e$ has median $1.11$ and lies in $[0.16,\,2.1]$, and $\hat e$ is
positive in every one. The two worst cases are the $M=16$ rung of each sweep, where
the trial space is still coarse and the true error is an order of magnitude above
its converged value; from $M=32$ upward the ratio stays within a factor of $2.1$.
The exact-flux form
$\mathcal{E}[\bar q]-\nu_{AB}$ instead turns negative in $12$ of the $22$, once
the true error drops below the bias incurred by the fidelity substitution of
Eq.~(\ref{eq:phihat}); the plateau form does not, because that bias enters
$\mathcal{E}[\bar q]$ and $\hat\nu_{AB}$ alike and partly cancels. Moving the
read-off window between $[0.2,\,0.8]$, $[0.3,\,0.7]$ and $[0.4,\,0.6]$ shifts
$\hat\nu_{AB}$ by $2.5\%$ at the median and $15\%$ at worst.

\paragraph{Where the estimate fails.}
At $M\le4$ the estimate goes blind, as Table~\ref{tab:apost_2d}
shows: the trial space is one or two ridges, the profile is flat for a reason
unrelated to accuracy, and $\hat e$ collapses towards zero while the true error is
at its largest ($e\approx29$). The stopping criterion shares the blind spot only at
$M=1$, where $L_M$ collapses to $8\times10^{-5}$, a false
all-clear rather than a false alarm; at $M=2$ and $M=4$ it still flags the estimate. The flatness is at its \emph{lowest} at $M\le2$, so
it is a necessary and not a sufficient guard: it catches a non-conserving profile,
as on chignolin, but not one that conserves flux along the wrong coordinate. $\hat\nu_{AB}$ itself catches both: it falls by a factor of about $30$
between $M=1$ and $M=16$ and only then settles, since a flux estimate still
drifting with $M$ shows the trial space, not the read-off, to be the limitation.
Reporting $\hat\nu_{AB}(M)$ beside $\hat e$ and $L_M$ therefore closes the gap at
no cost, all three read from the same sweep in $M$ that the stopping criterion of
the main text already calls for.

\begin{table*}[t]
\caption{\label{tab:apost}A posteriori Dirichlet-error estimate, Eq.~(\ref{eq:apost}), evaluated on the samples alone. $\mathcal{E}[\bar q]$ is the ansatz energy $\mathbf{w}^\top G\mathbf{w}$ and $\hat\nu_{AB}$ the isocommittor-flux plateau over $\bar q\in[0.2,0.8]$; both carry the same arbitrary constant $\mathbf{D}$, which cancels in $\hat e=\mathcal{E}[\bar q]/\hat\nu_{AB}-1$, the estimated Dirichlet error relative to $\nu_{AB}$. Energies are therefore comparable only within a row. The spread is the relative variation of $\hat\nu_{AB}$ over the nested read-off windows $[0.2,0.8]$, $[0.3,0.7]$ and $[0.4,0.6]$. RMSE is against the forward-tracing reference on transition frames, where one exists. The flatness $\sigma/\mu$ of $\hat\Phi$ across the band tests the premise that a plateau exists at all; chignolin's profile does not plateau, so its $\hat e$ is a diagnostic and not an error bar (see text). The negative AIB9 entry is traced to the read-off in the text. $L_M=\mathcal{E}[\bar q_M]/\mathcal{E}[\bar q_{M'}]-1$ is the ladder lower bound on the same relative error, read from the next rung $M'$ of the sweep at unit fidelity.}
\begin{ruledtabular}
\begin{tabular}{lccccccccccccc}
System & $d$ & $N$ & $M$ & $\hat{\mathcal{R}}_M$ & $1/\hat{\mathcal{R}}_M$ & $\mathcal{E}[\bar q]$ & $\hat\nu_{AB}$ & $\hat e$ & $M'$ & $L_M$ & spread & flatness & RMSE \\
\hline
Wolfe--Quapp (2D) & 2 & $100{,}000$ & 256 & 155.0 & 0.00645 & 0.00638 & 0.0063 & 0.0066 & 512 & -0.001 & 0.010 & 0.19 & --- \\
AIB9 & 52 & $270{,}000$ & 512 & 55.61 & 0.0180 & 0.0180 & 0.0199 & -0.100 & 1024 & 0.18 & 0.046 & 0.09 & 0.0615 \\
villin HP-35 & 350 & $198{,}750$ & 2048 & 79.27 & 0.0126 & 0.0124 & 0.0102 & 0.21 & 4096 & 0.093 & 0.068 & 0.08 & 0.208 \\
chignolin & 86 & $630{,}063$ & 256 & 18.16 & 0.0551 & 0.0529 & 0.0208 & 1.5 & 512 & 0.30 & 0.22 & 0.71 & --- \\
\end{tabular}
\end{ruledtabular}
\end{table*}

\begin{table*}[t]
\caption{\label{tab:apost_2d}A posteriori estimate and ladder bound on the two-dimensional benchmark, where $\nu_{AB}$, $\nabla q$ and $\mathcal{F}$ are all known from the finite-difference reference ($N=400{,}000$). $e=\mathcal{E}[\bar q-q]/\nu_{AB}$ is the true relative Dirichlet error and $\hat e$ its plateau estimate. Each rung is also the coarse member of a nested pair $(M,2M)$: $L_M=\mathcal{E}_M/\mathcal{E}_{2M}-1$ is the deployable unit-fidelity ladder bound and $L_M^{\mathcal{F}}=(\mathcal{E}_M/\mathcal{E}_{2M})\mathcal{F}_{2M}^2-2\mathcal{F}_M+1$ the same bound with the measured fidelities $\mathcal{F}_M,\mathcal{F}_{2M}$; both should fall below $e$. For the $M\le4$ rungs see \emph{Where the estimate fails}.}
\begin{ruledtabular}
\begin{tabular}{cccccccccc}
$M$ & $\hat\nu_{AB}/\nu_{AB}$ & $\mathcal{F}_M$ & $\mathcal{F}_{2M}$ & $L_M$ & $L_M^{\mathcal{F}}$ & $e$ & $\hat e$ & $\hat e/e$ & flatness \\
\hline
1 & 29.8 & 1.005 & 1.005 & 7.78e-05 & -0.000258 & 28.8 & 0.00181 & 6.3e-05 & 0.037 \\
2 & 29.7 & 1.005 & 1.003 & 2.09 & 2.1 & 28.8 & 0.00217 & 7.5e-05 & 0.035 \\
4 & 7.19 & 1.003 & 0.9939 & 5.1 & 5.02 & 8.63 & 0.34 & 0.039 & 0.14 \\
8 & 1.48 & 0.9939 & 0.9918 & 0.467 & 0.456 & 0.591 & 0.0632 & 0.11 & 0.12 \\
16 & 1.06 & 0.9918 & 0.9946 & 0.0779 & 0.0828 & 0.092 & 0.0148 & 0.16 & 0.13 \\
32 & 0.987 & 0.9946 & 0.9946 & 0.00126 & 0.00109 & 0.00862 & 0.0115 & 1.3 & 0.1 \\
64 & 0.983 & 0.9946 & 0.994 & 0.00115 & 0.000134 & 0.00755 & 0.0134 & 1.8 & 0.11 \\
128 & 0.984 & 0.994 & 0.9946 & -0.000768 & 0.000344 & 0.00745 & 0.0121 & 1.6 & 0.12 \\
256 & 0.982 & 0.9946 & 0.9943 & 0.000389 & -5.28e-05 & 0.00714 & 0.0146 & 2.0 & 0.12 \\
512 & 0.983 & 0.9943 & 0.9942 & 0.000332 & -1.38e-05 & 0.00722 & 0.0134 & 1.9 & 0.12 \\
1024 & 0.982 & 0.9942 & --- & --- & --- & 0.00727 & 0.0136 & 1.9 & 0.12 \\
\end{tabular}
\end{ruledtabular}
\end{table*}

\section*{Direction sampling}

The variational framework is agnostic to how the direction set
$\{\bm\theta_j\}_{j=1}^M$ is drawn: any distribution on $S^{d-1}$ gives a
valid estimator, with the Gram matrix $G$ adapting to the sampling
statistics. Sample efficiency is not: the fraction of the sphere within angle
$\theta$ of any fixed $k$-dimensional subspace shrinks as $\theta^{d-k}$ for
$\theta\ll1$, so isotropic draws concentrate, exponentially in $d$, on
directions orthogonal to any low-dimensional structure of $q$.

\paragraph{Fisher-discriminant axis.}
The committor changes most steeply along directions that separate reactant
from product, so a natural label-only proxy for a good slice direction is the
Fisher linear-discriminant (LDA) axis between the state-labelled samples,
\begin{equation}
\begin{aligned}
\mathbf{v}_{\mathrm{LDA}}
 &\;\propto\; (\Sigma_w + \eta_{\mathrm{LDA}} I)^{-1}(\bm\mu_B - \bm\mu_A),\\
\eta_{\mathrm{LDA}} &= \varepsilon_{\mathrm{LDA}}\,\frac{\mathrm{tr}\,\Sigma_w}{d},
\end{aligned}
\label{eq:lda_axis}
\end{equation}
normalised to unit length, with $\bm\mu_A,\bm\mu_B$ the per-state feature
means and $\Sigma_w$ the pooled within-state covariance. The shrinkage ridge
$\eta_{\mathrm{LDA}}$ (set by $\varepsilon_{\mathrm{LDA}} = 10^{-2}$ for the two
peptides and $10^{-1}$ for chignolin) regularises
$\Sigma_w^{-1}$ in the high-dimensional, finite-sample regime. The axis is
computed once in the full torsion feature space and uses only the state
assignment the method already requires: no lagged or dynamical input. Locating a
dimension-reduction direction from the response labels rather than from the
predictor is the device of sliced inverse regression~\cite{li_sliced_1991}; the
sense of ``sliced'' there is distinct from the random-direction projections that
name the present method.

\paragraph{Power-spherical broadening.}
A single axis would yield $M$ identical slices, so the axis of
Eq.~(\ref{eq:lda_axis}) is broadened into a direction set by a power-spherical mixture~\cite{decao_power_2020}: a fraction
$\alpha$ of the directions are isotropic on $S^{d-1}$, a coverage floor, and
the remaining $1-\alpha$ are drawn from a power-spherical law concentrated
about the axis,
\begin{equation}
\begin{aligned}
p(\bm\theta) &= \alpha\,\mathcal{U}_{S^{d-1}}(\bm\theta)
 + (1-\alpha)\,Z_\xi^{-1}\,(1 + \mathbf{v}_{\mathrm{LDA}}\!\cdot\!\bm\theta)^{\xi},\\
\xi &= \frac{\mu\,(d-1)}{1-\mu}.
\end{aligned}
\label{eq:lda_sample}
\end{equation}
The concentration is set by a dimension-invariant mean cosine
$\mu = \mathbb{E}[\mathbf{v}_{\mathrm{LDA}}\!\cdot\!\bm\theta]\in[0,1)$:
$\mu\to0$ recovers the isotropic law and $\mu\to1$ collapses onto the axis,
while the reparameterisation $\xi(\mu,d)$ holds the angular spread fixed as
$d$ grows. A draw takes a cosine $t = 2u-1$ with
$u\sim\mathrm{Beta}\bigl(\xi+\tfrac{d-1}{2},\,\tfrac{d-1}{2}\bigr)$ and a
uniform perpendicular unit vector $\mathbf{e}_\perp\perp\mathbf{v}_{\mathrm{LDA}}$,
giving $\bm\theta = t\,\mathbf{v}_{\mathrm{LDA}} + \sqrt{1-t^2}\,\mathbf{e}_\perp$.
The pair $(\mu,\alpha)$ trades concentration against coverage, and we select it
by the same held-out criterion as the ridge, Eq.~(\ref{eq:ridge_cv}), over a coarse
grid of twelve to thirteen pairs per system spanning
$\mu,\alpha\in[0.2,0.8]$. The justification is the
same as for $\varepsilon$: the cap is a bound on $\nu_{AB}$ for whatever trial
space produced it, so it ranks direction sets as legitimately as it ranks
regularisers, and no reference committor enters the choice. The selected
pairs are $(\mu,\alpha)=(0.4,0.2)$ for AIB9, $(0.8,0.2)$ for villin and
$(0.6,0.4)$ for chignolin.

\paragraph{Isotropic baseline.}
Drawing $\bm\theta_j$ uniformly on $S^{d-1}$ remains a valid choice. In this
case the normalised off-diagonals of $G$ decay as
$|G_{jk}/\sqrt{G_{jj}G_{kk}}| = \mathcal{O}(1/\sqrt d)$ and the constrained
Gram solve reduces to the closed-form diagonal limit of
Eq.~(\ref{eq:diag_limit_supp}) (\hyperref[em:diag]{Diagonal limit}), each slice
weighted by its basin-moment gap $b_j - a_j$ over its dissipation $G_{jj}$. The
construction pays the $\theta^{d-k}$ waste described above; it is appropriate
when the committor varies over a non-negligible fraction of $S^{d-1}$, as
for AIB9, where in practice it is well approximated by the broad-cone
discriminant sampler of Eq.~(\ref{eq:lda_sample}) running on the same data.

\paragraph{Aggregation of basin moments.}
AIB9 and villin compute the basin moments $a_j,b_j$ by the per-state
sample mean. The per-sample integrand can be bimodal: samples in
$\mathrm{supp}\,\rho_j^A\cap\mathrm{supp}\,\rho_j^B$ contribute the local
Bayesian split of Eq.~(\ref{eq:classifier}), while state-bulk
samples contribute near $0$ or $1$. The framework also admits robust alternatives, such as a per-state
high quantile to suppress overlap-region outliers, though the systems studied
here did not require them.

\paragraph{Empirical fidelity on a molecular system.}\label{em:emp}
The basin-moment substitution $\hat{\mathcal{F}}_j=b_j-a_j$ replaces the flux-weighted
boundary average $\mathcal{F}_j$ by an equilibrium one. Because the equilibrium
measure is weighted toward the basin interior while the reactive current
concentrates on the saddle-facing boundary, $\hat{\mathcal{F}}_j$ is expected to
slightly overestimate $\mathcal{F}_j$. We quantify this on two systems. On the 2D
rotated Wolfe--Quapp benchmark the exact $\mathcal{F}_j$ is available numerically:
we integrate the reactive current $\mathbf{J}=\rho\,\mathbf{D}\,\nabla q$ of the
finite-difference committor over the state boundaries
$\partial A,\partial B$, giving the flux-weighted average of each lifted slice
$\tilde q_j$ directly. Across the $256$ directions the equilibrium moment
overestimates by a mean of $0.015$, on $97\%$ of directions against this
finite-difference oracle. The
reactive-trajectory estimator, applied to the boundary-crossing frames of $244$
overdamped-Langevin $A\!\to\!B$ reactive segments, recovers the exact $\mathcal{F}_j$
to within $0.001$. Each
boundary-crossing frame is interpolated to the exact state boundary along its
crossing step; on the 2D benchmark this removes an interior bias and brings the
empirical estimate onto the exact $\mathcal{F}_j$. On AIB9 the exact $\mathcal{F}_j$ is not
available, but the equilibrium trajectory supplies committed $A\!\to\!B$ folding
events (and as many reverse); the same estimator gives $\mathcal{F}_j$ from the
boundary-crossing frames. Applied to the slices that
Table~\ref{tab:aib9_params} deploys, and to the $73$ reactive segments that the
state definitions of that table select, the equilibrium moment overestimates the
reactive-flux fidelity by a mean of $0.0151$, and does so on $83\%$ of
directions. The magnitude matches the two-dimensional one rather than exceeding
it, though the two measurements use different direction samplers and so do not on
their own settle how the bias scales with dimension.

Beyond the per-slice number, two facts matter. First, the constrained
solve places its weight away from the biased directions (mechanism and
ground-truth check in \hyperref[em:substitution]{Why the substitution does not
propagate}): the weighted deficit of Eq.~(\ref{eq:fid_deficit}) is only $0.0065$
[95\% CI $0.0045$, $0.0090$], a factor of $2.3$ smaller than the per-slice
mean, so the combination has fidelity $\mathcal{F}[\bar q]=0.9935$ even though its
slices are visibly biased. Second,
the bias does not propagate to the committor: re-solving
Eq.~(\ref{eq:wstar}) with $\mathbf f$ in place of $\hat{\mathbf f}$ moves the
estimate by a transition-region RMSE of $0.0044$ and moves the error against the
forward-tracing reference only from $0.0615$ to $0.0613$, so the approximation the
method actually uses is not degraded. The same holds at $M=1024$ ($0.0051$ and
$0.0679\!\to\!0.0671$), so the larger flux-substitution entry in the AIB9
control of Table~\ref{tab:villin_ablation} reflects that table's unconstrained
full-Gram solver rather than a growth of the bias with $M$.

\paragraph{Why the substitution does not propagate.}\label{em:substitution}
The picture above is confirmed by an independent flux-form finite-volume solve of
the committor with geometric-mean face conductances, itself a genuine ground truth
rather than an approximation: two discrete identities hold to a relative $10^{-11}$,
with the flux out of $A$, the flux into $B$, and $\sum_f C_f(\Delta q_f)^2$ all
agreeing and equal to $\nu_{AB}$. Applied to both Wolfe--Quapp and a second exact
system, the 2D double well, it gives one picture throughout
(Fig.~\ref{fig:substitution}).
Three facts, all read off this ground truth, explain why the per-slice bias does
not reach the committor.

First, the error is an oscillation of the test function, not a property of the
current. The two fidelities average the identical lifted slice $\tilde q_j$
against two normalised measures, so
\begin{equation}
|\hat{\mathcal{F}}_j-\mathcal{F}_j|\;\le\;
\mathrm{osc}_A\,\tilde q_j+\mathrm{osc}_B\,\tilde q_j ,
\label{eq:osc_bound}
\end{equation}
with $\mathrm{osc}_S\,\tilde q_j$ the range of $\tilde q_j$ over state $S$; the
bound makes no reference to where the current sits and held in every one of the
$256$ directions of both systems. The absorbers that define the slice profile
(\hyperref[em:rd]{Reaction-diffusion slice committors}) make that oscillation
small by pinning $\tilde q_j$ flat inside each basin through a boundary layer of
width $\sim\kappa^{-1/2}$: the separated-direction error falls with stiffness
(fitted $\kappa^{-0.42}$ over $\kappa\in[10^2,10^5]$), while the two measures do
not resemble each other at all [Fig.~\ref{fig:substitution}(a): the flux measure
piles onto the exit face, the equilibrium measure is centred].

Second, the error separates by geometry. Along the directions whose projected
basins are disjoint --- $229/256$ on Wolfe--Quapp, $210/256$ on the double
well --- the bias is $\sim10^{-3}$ ($1.1\times10^{-3}$ and $6.6\times10^{-3}$
respectively); along the residual overlapping directions no profile can be flat
on both basins at once, and the bias saturates near $0.15$, independent of
$\kappa$ [Fig.~\ref{fig:substitution}(b)]. The overestimate is signed,
$\hat{\mathcal{F}}_j\ge\mathcal{F}_j$ on $95$--$96\%$ of directions against this
finite-volume oracle.

Third, the constrained solve discards exactly the overlapping directions, so the
per-slice bias does no harm. A profile that varies across a
basin pays Dirichlet energy for it, so the per-slice error correlates with the
slice dissipation $G_{jj}$ (Pearson $+0.79$ on Wolfe--Quapp, $+0.65$ on the
double well), and the solve weights each slice by $1/G_{jj}$ in the diagonal
limit of Eq.~(\ref{eq:diag_limit_supp}) (\hyperref[em:diag]{Diagonal limit}). The
large errors therefore land on the smallest weights
[Fig.~\ref{fig:substitution}(c)], and the combination fidelity is accurate even
where the slices are not: $\mathcal{F}[\bar q]=\mathbf{w}^\top\mathbf f=1.0000$ on
Wolfe--Quapp and $0.9963$ on the double well, against a target of one. That is the
same cancellation the AIB9 weighted deficit ($0.0065$) shows directly above. The
molecular systems live in this regime. Two dimensions is the easy case: most
random directions separate the basins. In the $d=52$ to $350$ feature spaces
almost every direction overlaps, so the guarantee rests on the weight
suppression, not on per-slice accuracy.

\begin{figure*}[t]
\centering
\includegraphics[width=\textwidth]{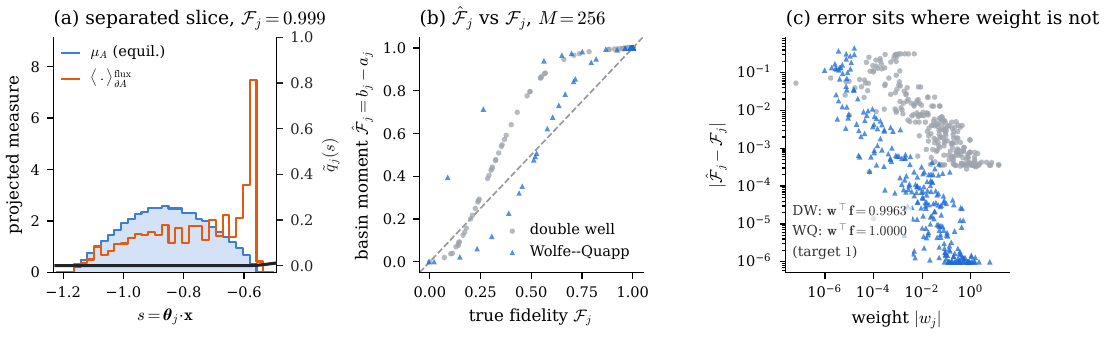}
\caption{\label{fig:substitution}The basin-moment substitution on exact
two-dimensional ground truth (rotated Wolfe--Quapp and 2D double well; committor
and flux-weighted boundary measures from a flux-form finite-volume solve, $M=256$
random directions). (a)~Along a separated direction the equilibrium basin measure
$\mu_A$ and the flux-weighted boundary measure
$\langle\cdot\rangle^{\mathrm{flux}}_{\partial A}$ differ substantially (left
axis), yet the slice profile $\tilde q_j$ (black, right axis) is flat across the
basin, so the two averages of $\tilde q_j$ nearly agree. (b)~The basin moment
$\hat{\mathcal{F}}_j=b_j-a_j$ against the exact flux fidelity $\mathcal{F}_j$: a
signed overestimate that is $\sim10^{-3}$ where the projected basins separate and
grows only where they overlap. (c)~Per-slice error against the assigned weight
$|w_j|$: the large errors carry the smallest weights, so the constrained solve
suppresses them and the combination fidelity $\mathbf{w}^\top\mathbf f$ reaches
$1.0000$ (WQ) and $0.9963$ (DW). The separated-direction error falls with
absorber stiffness as $\kappa^{-1/2}$ (fitted $\kappa^{-0.42}$ over
$\kappa\in[10^2,10^5]$); the overlapping error is $\kappa$-independent.}
\end{figure*}

\section*{Implementation and parameters}

\paragraph{Software.}
All computations use JAX~\cite{jax2018github} with JIT compilation and
\texttt{vmap} vectorisation across directions; random directions are drawn with
the JAX pseudo-random generator using a fixed per-system seed (see the
parameter tables: Table~\ref{tab:2d_params} for the two-dimensional
validation, Tables~\ref{tab:aib9_params} and~\ref{tab:villin_params} for the
peptides, Table~\ref{tab:chignolin_params} for chignolin). Projected densities $\rho_j$,
$\rho_j^A$, $\rho_j^B$ are estimated by quantile-binned histograms over the
full projected range with an adaptive density floor
$\max(n_\text{min}/(N\Delta s_i),\,\rho_{\min})$ (per-bin width $\Delta s_i$,
$\rho_{\min}=10^{-6}$ by default, raised to $10^{-3}$ for the two-dimensional
benchmark, Table~\ref{tab:2d_params}) that prevents tail bins with few samples from
producing artificially high free energies. The two peptides use equal-count
(quantile) bins, so $\Delta s_i$ varies along the projected coordinate and the first
term of the floor is evaluated per bin; the two-dimensional benchmark and chignolin
use equal-width bins. The RD
system~(\ref{eq:rd_tridiag}) is assembled once per direction and solved by
the Thomas algorithm. Where a \emph{boundary quantile} $<1$ is listed, the
projected state densities $\rho_j^A,\rho_j^B$ are truncated at that quantile of
their own projected distribution before entering Eq.~(\ref{eq:rd_ode}), so that the
absorbers act on the bulk of each state rather than on the tail its projection
inherits from the nuisance directions. The basin moments $a_j,b_j$ are accumulated over the
state samples. The constrained solve of Eq.~(\ref{eq:wstar}), including
the bias $c^\star = -\mathbf{a}^\top\mathbf{w}^\star$, is obtained by
Cholesky factorisation of $G$ with a Tikhonov ridge $\varepsilon$, which
suppresses conditioning artefacts at large $M$ without distorting
well-conditioned blocks.

The size of the ridge is selected from the data rather than
fixed: we split the samples into $K=5$ folds, fit $\mathbf{w}$ on $K-1$ of them
at each $\varepsilon$, and evaluate the cap
Eq.~(\ref{eq:cap}) on the held-out fold,
\begin{equation}
\varepsilon^\star = \arg\min_\varepsilon\;
\frac{\mathbf{w}^\top G_{\mathrm{test}}\,\mathbf{w}}
     {\bigl((\mathbf{b}-\mathbf{a})_{\mathrm{test}}^\top\mathbf{w}\bigr)^2}.
\label{eq:ridge_cv}
\end{equation}
Both factors are out of sample, so the score is comparable across $\varepsilon$,
and each is the quantity by which Eq.~(\ref{eq:cap}) bounds $\nu_{AB}$, evaluated
out of sample; minimising it is therefore the same variational principle the
weights already obey, applied to the one remaining free scalar. Folds are
contiguous blocks taken within each of $A$, $B$ and the transition region:
contiguous because molecular-dynamics frames are serially correlated, so a random
split puts near-duplicate frames on both sides and every held-out score comes out
optimistic; stratified so that every fold contains samples of both basins, without
which the held-out moment gap is undefined. The search runs over a geometric grid
of $\varepsilon$ spanning $[10^{-10},10^{2}]$ times the Gram scale
$\mathcal{A}=M\,\overline{\mathrm{diag}\,G}$, which is extensive in $M$ because $G$
sums over the $M$ directions, so that $\varepsilon/\mathcal{A}$ is comparable
across $M$ and across systems. The selection is the plain minimiser. The rule costs
one further pass over the samples plus one eigendecomposition of $G$ per fold,
which then gives every grid point at once.

The solve is direct, hence deterministic given the direction draw. After
weighted assembly, $\bar q$ is clipped to $[0,1]$ and, on the two-dimensional
benchmark and the two peptides, clamped to $0$ on $A$ and $1$ on $B$; chignolin is
the exception (Table~\ref{tab:chignolin_params}). Code and data availability are
detailed in the Data Availability statement of the main text.

\paragraph{Hardware and cost.}
Every number reported here was produced on a single workstation, a 24-thread
Intel Core i9-12900K with the BLAS thread pool capped at eight, running JAX on
the CPU in float64; no GPU was used. Table~\ref{tab:timings} gives the cost of
the committor itself, timed around the slice build and the constrained solve at
the deployed settings, with trajectory reading, featurisation, TICA, MBAR
reweighting and the reference committor all excluded. JAX compiles on first
call, so each system was fit twice at identical shapes; the compile column is
the difference between the two, and is paid once per shape rather than per
direction.

\begin{table}[t]
\caption{\label{tab:timings}Wall-clock cost of the sliced committor alone, at the
deployed settings of Tables~\ref{tab:2d_params}--\ref{tab:chignolin_params}, on
one CPU workstation. \emph{Compile} is the one-off JAX compilation; \emph{fit} is
one complete committor, the slice build plus the constrained solve of
Eq.~(\ref{eq:wstar}). Repeat fits agreed to within $0.3$~s.}
\begin{ruledtabular}
\begin{tabular}{lrrrrr}
System & $N$ & $d$ & $M$ & Compile & Fit \\
\hline
Wolfe--Quapp (2D) & $100{,}000$ & 2 & 256 & $2.1$~s & $3.3$~s \\
AIB9 & $270{,}000$ & 52 & 512 & $2.9$~s & $27.7$~s \\
villin HP-35 & $198{,}750$ & 350 & 2048 & $3.0$~s & $97.9$~s \\
chignolin & $630{,}063$ & 86 & 256 & $2.2$~s & $20.7$~s \\
\end{tabular}
\end{ruledtabular}
\end{table}

\begin{table}[h]
\caption{\label{tab:2d_params}Parameters for the 2D validation
experiment (Fig.~\ref{fig:schematic}). The rotated Wolfe--Quapp potential is
$V = x_r^4 + y_r^4 - 2x_r^2 - 4y_r^2 + x_r y_r + 0.3\,x_r + 0.1\,y_r$ on
coordinates rotated by $\theta = -0.15\pi$
($x_r = x\cos\theta - y\sin\theta$, $y_r = x\sin\theta + y\cos\theta$),
with states centred at $(-1.717, 0.783)$ and $(1.676, -0.813)$ on the domain
$[-2.5, 2.5]^2$; samples are drawn by grid-based Boltzmann inverse-CDF
sampling.}
\begin{ruledtabular}
\begin{tabular}{lc}
Parameter & Value \\
\hline
Number of samples $N$ & 100{,}000 \\
Number of directions $M$ & 256 \\
Direction sampling & isotropic on $S^{d-1}$ \\
Histogram bins $n_\text{bins}$ & 200 \\
Binning method & equal-width \\
Density floor & $10^{-3}$ \\
Min.\ samples per bin $n_\text{min}$ & 1 \\
Inverse temperature $\beta$ & 1.0 \\
State radius & 0.3 \\
Potential & rotated Wolfe--Quapp \\
Weighting & constrained solve, Eq.~(\ref{eq:wstar}) \\
Tikhonov ridge $\varepsilon$ (on $G$) & held-out cap, Eq.~(\ref{eq:ridge_cv}) \\
1D solver & RD ($\kappa = 10^{24}$) \\
Boundary clamp & applied \\
Affine rescale of $\bar q$ to $[0,1]$ & applied \\
Finite-difference grid & $300 \times 300$ \\
PDE convergence tolerance & $10^{-9}$ \\
Random seed & 42 \\
\end{tabular}
\end{ruledtabular}
\end{table}

\begin{table*}[t]
\caption{\label{tab:aib9_params}Parameters for AIB9 peptide
analysis [Fig.~\ref{fig:molecular}(a--c)].}
\begin{ruledtabular}
\begin{tabular}{lc}
Parameter & Value \\
\hline
\multicolumn{2}{c}{\textit{Sliced committor}} \\
Number of frames $N$ & 270{,}000 (subsample 10) \\
Feature representation & sine and cosine torsions (52-D) \\
Number of directions $M$ & 512 \\
Direction sampling & Fisher LDA, Eq.~(\ref{eq:lda_sample}) \\
LDA mean cosine $\mu$ & 0.4 \\
LDA mixture weight $\alpha$ & 0.2 \\
LDA shrinkage $\varepsilon_{\mathrm{LDA}}$ & $10^{-2}$ \\
$(\mu,\alpha)$ selected by & held-out cap, Eq.~(\ref{eq:ridge_cv}) \\
Histogram bins $n_\text{bins}$ & 2000 \\
Binning method & quantile \\
Min.\ samples per bin $n_\text{min}$ & 10 \\
Boundary quantile & 0.99 \\
Inverse temperature $\beta$ & 1.0 \\
1D solver & RD ($\kappa = 10^{12}$) \\
Weighting & constrained solve, Eq.~(\ref{eq:wstar}) \\
Tikhonov ridge $\varepsilon$ (on $G$) & held-out cap, Eq.~(\ref{eq:ridge_cv}) \\
Basin-moment aggregator & sample mean \\
Boundary clamp & applied \\
Affine rescale of $\bar q$ to $[0,1]$ & applied \\
Random seed & 0 \\
\hline
\multicolumn{2}{c}{\textit{Empirical committor}} \\
Method & forward-tracing \\
Visualisation grid & $50 \times 50$ \\
\hline
\multicolumn{2}{c}{\textit{Simulation}} \\
Temperature & 500 K \\
Total simulation time & 2.7 ms \\
Force field & AMBER ff15ipq-m \\
Langevin friction $\gamma$ & 1.0 ps$^{-1}$ \\
\hline
\multicolumn{2}{c}{\textit{State definitions (TICA space)}} \\
State $A$ centre $(\mathrm{TICA}_1, \mathrm{TICA}_2)$ & $(-1.327, -0.062)$ \\
State $A$ semi-axes & $(0.098, 0.191)$ \\
State $B$ centre $(\mathrm{TICA}_1, \mathrm{TICA}_2)$ & $(1.144, 0.085)$ \\
State $B$ semi-axes & $(0.095, 0.188)$ \\
\end{tabular}
\end{ruledtabular}
\end{table*}

\begin{table*}[t]
\caption{\label{tab:villin_params}Parameters for villin HP-35
analysis [Fig.~\ref{fig:molecular}(d--f)].}
\begin{ruledtabular}
\begin{tabular}{lc}
Parameter & Value \\
\hline
\multicolumn{2}{c}{\textit{Sliced committor}} \\
Number of frames $N$ & 198{,}750 \\
Feature representation & sine and cosine torsions (all backbone + sidechain, 350-D) \\
Number of directions $M$ & 2048 \\
Direction sampling & Fisher LDA, Eq.~(\ref{eq:lda_sample}) \\
LDA mean cosine $\mu$ & 0.8 \\
LDA mixture weight $\alpha$ & 0.2 \\
LDA shrinkage $\varepsilon_{\mathrm{LDA}}$ & $10^{-2}$ \\
$(\mu,\alpha)$ selected by & held-out cap, Eq.~(\ref{eq:ridge_cv}) \\
Histogram bins $n_\text{bins}$ & 500 \\
Binning method & quantile \\
Min.\ samples per bin $n_\text{min}$ & 10 \\
Boundary quantile & 0.98 \\
Inverse temperature $\beta$ & 1.0 \\
1D solver & RD ($\kappa = 10^{12}$) \\
Weighting & constrained solve, Eq.~(\ref{eq:wstar}) \\
Tikhonov ridge $\varepsilon$ (on $G$) & held-out cap, Eq.~(\ref{eq:ridge_cv}) \\
Basin-moment aggregator & sample mean \\
Boundary clamp & applied \\
Affine rescale of $\bar q$ to $[0,1]$ & applied \\
Random seed & 42 \\
\hline
\multicolumn{2}{c}{\textit{Empirical committor}} \\
Method & forward-tracing \\
Visualisation space & leading two TICA components \\
TICA lag for visualisation (frames) & 500 \\
\hline
\multicolumn{2}{c}{\textit{Simulation}} \\
System & WT villin HP-35 (PDB 1YRF) \\
Temperature & 345 K \\
Total simulation time & 398 $\mu$s \\
Trajectory source & D.E.\ Shaw Research~\cite{piana_protein_2012} \\
Force field & Amber ff99SB*-ILDN \\
\hline
\multicolumn{2}{c}{\textit{State definitions (TICA1, TICA2)}} \\
State $A$ (folded) centre & $(-1.36, -0.07)$ \\
State $A$ semi-axes & $(0.25, 0.46)$ \\
State $B$ (unfolded) centre & $(0.48, -0.07)$ \\
State $B$ semi-axes & $(0.35, 0.65)$ \\
\end{tabular}
\end{ruledtabular}
\end{table*}

\paragraph{Villin sampling statistics and reference reliability.}
The villin HP-35 forward-tracing reference rests on limited folding statistics.
Of the $N = 198{,}750$ equilibrium frames, $16.3\%$ fall in the folded native
basin $A$ and $38.4\%$ in the unfolded basin $B$; over the $398~\mu\mathrm{s}$
trajectory the reactive channel between them is crossed by only $\approx 15$
committed folding events ($15$ $B\!\to\!A$ and $16$ $A\!\to\!B$, consecutive
same-state runs collapsed), a mean time between folding events of
$\approx 27~\mu\mathrm{s}$. The $q\!\to\!0$ (folded) side of the reference
therefore carries a statistical uncertainty of order $1/\sqrt{15}\approx 26\%$.
The reactive neck into the folded basin is also poorly resolved: on the
$50\times50$ TICA grid, $38\%$ of non-empty transition-region bins fall below
the $n_\text{min} = 10$ reliability threshold and are masked, and of the
well-sampled transition bins $81\%$ drain to the unfolded state versus only
$5\%$ to the folded state. This incomplete equilibrium sampling of the folding
channel, rather than the sliced-committor construction alone, is a leading
contributor to the larger nominal villin transition-region RMSE ($0.21$ versus $0.062$
for AIB9).

\paragraph{Villin error attribution and regime diagnostic.}
To attribute the larger villin RMSE rather than assert its cause, we isolate the three
candidate mechanisms on villin and on AIB9 (the well-resolved control), both
built with the same discriminant sampler and $M=1024$ (Table~\ref{tab:villin_ablation}).
The transition-region RMSE is throughout a per-frame average over transition
samples, but the reference differs by system: for AIB9 a forward-tracing reference
bin-cleaned on the leading TICA plane ($50\times50$ bins, at least five labelled
frames per bin, sparser bins masked), for villin the raw per-frame labels. The
entries below are therefore not directly comparable with the values quoted above:
this table fixes a common $M=1024$ and the unconstrained full-Gram solve so the
three mechanisms can be isolated on one footing, whereas the deployed committors
use the constrained solve of Eq.~(\ref{eq:wstar}) at the settings of
Tables~\ref{tab:aib9_params} and~\ref{tab:villin_params}.
\emph{(i) Representation gap.} The enriched moment-gap conditioning
$\mathrm{cond}_{\mathcal{E}}=\hat{\mathcal{R}}_M/\max(S_A,S_B)$ is healthy for both
systems ($0.84$ villin, $0.98$ AIB9), so the slice basis can represent the
basin moments; the gap has not collapsed. \emph{(ii) Gram conditioning.} With
the deployed sampler $G$ is in fact \emph{better} conditioned for villin
($\mathrm{cond}(G)\approx1.4\times10^2$) than for the accurate AIB9
($\approx1.4\times10^5$), and removing the Tikhonov ridge shifts the villin RMSE
by under $0.001$: conditioning is not the accuracy bottleneck. \emph{(iii)
Fidelity substitution.} Replacing the equilibrium fidelity by the flux estimator
changes the villin RMSE by $\approx0.001$. None of the three explains the gap to
AIB9. The state separability along the reactive axis separates the regimes: the Mahalanobis distance between the projected basins is $12.9$ for villin
versus $65.9$ for AIB9. This quantity, together with $\mathrm{cond}_{\mathcal{E}}$
and the attained energy $1/\hat{\mathcal{R}}_M$, is computable from the labelled
samples alone. It is therefore an a-priori indicator of the good regime that needs no
ground truth: villin's low value flags the weak-signal, undersampled regime
before any reference committor is consulted.

\begin{table}[h]
\caption{\label{tab:villin_ablation}Villin error attribution versus the AIB9
control (discriminant sampler, $M=1024$). RMSE is the transition-region error
against the per-frame forward-tracing reference; RMSE$_{\varepsilon=0}$ removes the
Tikhonov ridge; RMSE$_{\mathrm{flux}}$ uses the reactive-flux fidelity.
$\mathrm{cond}_{\mathcal{E}}=\hat{\mathcal{R}}_M/\max(S_A,S_B)$ with
$S_A=\mathbf{a}^\top G^{-1}\mathbf{a}$ and $S_B=\mathbf{b}^\top G^{-1}\mathbf{b}$
the inverse-Gram self-overlaps of the basin-moment vectors;
$1/\hat{\mathcal{R}}_M=\mathcal{E}[\bar q^\star]$ is the attained ansatz energy;
$d_M$ is the state-separability (Mahalanobis) distance between the projected
basins. This table is a separate ablation run with the unconstrained full-Gram
solver, so its $1/\hat{\mathcal{R}}_M$ is not comparable with the deployed-solver
values of Table~\ref{tab:apost} at the same $M$.}
\begin{ruledtabular}
\begin{tabular}{lcc}
 & villin & AIB9 \\
\colrule
RMSE & $0.26$ & $0.066$ \\
$d_M$ (Mahalanobis) & $12.9$ & $65.9$ \\
$\mathrm{cond}_{\mathcal{E}}$ & $0.84$ & $0.98$ \\
$1/\hat{\mathcal{R}}_M$ & $0.038$ & $0.020$ \\
$\mathrm{cond}(G)$ & $1.4\times10^{2}$ & $1.4\times10^{5}$ \\
RMSE$_{\varepsilon=0}$ & $0.26$ & $0.065$ \\
RMSE$_{\mathrm{flux}}$ & $0.26$ & $0.080$ \\
\end{tabular}
\end{ruledtabular}
\end{table}

\section*{Kinetic-rate computation from umbrella sampling}

\paragraph{Current-conservation estimator.}
For reversible Langevin dynamics on configuration $\mathbf{x}$ with equilibrium
density $\rho(\mathbf{x})\propto e^{-\beta V(\mathbf{x})}$ and position-dependent
scalar diffusivity $D(\mathbf{x})$, the committor $q(\mathbf{x})$ between basins
$A$ and $B$ satisfies $\nabla\!\cdot\!(\rho D\,\nabla q) = 0$ with $q|_A = 0$,
$q|_B = 1$. Here $D(\mathbf{x})$ generalises, in the
umbrella-sampling setting, the constant diffusion tensor $\mathbf{D}$ of the
main text. The flux--fidelity identity and the variational principle carry over
unchanged, since neither uses constancy of $\mathbf{D}$; the closed form of
Eq.~(\ref{eq:gram}) does not, since it separates into
$\bm\theta_j^\top\mathbf{D}\bm\theta_k$ times a scalar average only for constant
$\mathbf{D}$. The reactive flux is
$\nu_{AB} = \langle D\,|\nabla q|^2\rangle_\rho$ and the rate is
$k_{A\to B} = \nu_{AB}/p_A$~\cite{vanden-eijnden_transition_2006,hummer_transition_2004};
the same decomposition extends to sequences of events beyond a single
$A\!\to\!B$ reaction~\cite{lorpaiboon_augmented_2022}.
Because the reactive current $\mathbf{J} = \rho D\,\nabla q$ is
divergence-free, the net flux through any iso-committor surface
$\Gamma(q') = \{\mathbf{x} : q(\mathbf{x}) = q'\}$ equals the same constant $\nu_{AB}$,
independent of $q'$. Stratifying the volume integrand on the committor value
gives the $q$-resolved flux density
\begin{equation}
\hat F(q') \;=\; \bigl\langle D\,|\nabla q|^2\bigr\rangle_{q=q'}\,\rho(q'),
\label{eq:Fhat}
\end{equation}
which is the level-set flux $\Phi(z)$ of the main text evaluated at $z=q'$, and
equals $\nu_{AB}$ at every $q'$ for the exact committor. For an
approximate committor $\bar q$, Eq.~(\ref{eq:Fhat}) varies with $q'$; the
$q$-stratified rate estimator reads the rate off the saddle-band median,
\begin{equation}
\hat k_{A\to B} \;=\; \frac{1}{p_A}\;\operatorname*{med}_{q'\in[0.2,\,0.8]} \hat F(q'),
\label{eq:k_qstrat}
\end{equation}
following the same iso-surface logic used by Berezhkovskii \& Szabo for
splitting-probability diffusion~\cite{berezhkovskii_diffusion_2013}. The band
$q'\in[0.2,\,0.8]$ is the default saddle window; where the plateau is narrower,
a flatness criterion (smallest $\sigma/\mu$ of $\hat F$) selects a sub-band
within it, as for chignolin ($q\approx[0.34,\,0.58]$;
Table~\ref{tab:chignolin_params}).

Unlike milestoning~\cite{faradjian_computing_2004,bello-rivas_exact_2015},
weighted-ensemble and trajectory-stratification
schemes~\cite{zuckerman_weighted_2017,strahan_bad-neus_2024}, or infrequent
metadynamics~\cite{tiwary_metadynamics_2013}, all of which read the rate off
dynamics, Eq.~(\ref{eq:k_qstrat}) needs no trajectories, only configurations
reweightable to equilibrium and a diffusion estimate.

\paragraph{Why basin-noise breaks the volume integral.}
The volume integrand $\rho\,D\,|\nabla\bar q|^2$ is concentrated at the saddle
for the exact committor, but any finite-sample $\bar q$ has residual
$|\nabla\bar q|^2$ in the basins. There it is multiplied by Boltzmann weight
$\sim\!e^{\beta\Delta F^{\ddagger}}$ relative to the saddle, so even
sub-percent residual basin gradients dominate the volume integral once
$\Delta F^{\ddagger}\gtrsim5\,k_{\mathrm{B}}T$. Eq.~(\ref{eq:k_qstrat})
filters this by construction: basin-side bins ($q'<0.2$, $q'>0.8$) do not
enter the median. This is why every rate quoted here is read from the stratified
flux plateau and never from the volume integral.

\paragraph{Diffusion and reweighting.}
For umbrella-sampling input, projected densities are reweighted to the
unbiased ensemble with MBAR~\cite{shirts_statistically_2008} and the
diffusion along the bias coordinate is estimated per window from
$D_Q = \mathrm{Var}(Q)/\tau_{\mathrm{int}}(Q)$ in the short-time
limit~\cite{hummer_position-dependent_2005}, with the integrated
autocorrelation time $\tau_{\mathrm{int}}$ obtained from the Geyer
initial-positive sequence~\cite{geyer_practical_1992}. The Hummer estimate is
rigorous only along the restrained coordinate $Q$, where the harmonic bias
leaves the potential-independent $D$ untouched. We therefore map it to the
configurational scalar $D(\mathbf{x})$ entering the flux, by the change of
variables $D_0 = D_Q/\langle|\nabla Q|^2\rangle$. Here $\langle\cdot\rangle$ is
the MBAR-reweighted window average, and $\langle|\nabla Q|^2\rangle$ is evaluated
in the same feature space as $\nabla\bar q$, as the squared coefficient of a
ridge-regularised weighted linear fit of $Q$ to the features. Taking both gradients
in one metric makes the rate well defined: under a rescaling of the
features the two mean squared gradients scale together, so
$\langle D_0\,|\nabla\bar q|^2\rangle$ is invariant, whereas pairing a
collective-variable diffusivity with a feature-space gradient is not. For a
Cartesian-like bias coordinate ($|\nabla Q|\equiv1$, as in the 2D tests) the map
is the identity and $D_0 = D_Q$.

The same estimator is applied to chignolin (Sec.~``Chignolin (CLN025) system
setup and analysis details'', three pooled umbrella-sampling replicates of
$50$~ns per window). The resulting
folding and unfolding rates are compared in the main text against two independent
sets of $340$~K equilibrium simulations, the Anton rates of Lindorff-Larsen \emph{et
al.}~\cite{lindorff-larsen_how_2011} and the $120~\mu\mathrm{s}$ reference rates of
Lazzeri \emph{et al.}~\cite{lazzeri_molecular_2023}; the absolute values and ratios
are collected in Table~\ref{tab:chignolin_rates}.

%
\begin{table}[h]
\caption{\label{tab:chignolin_rates}Chignolin (CLN025) folding and unfolding
rates at $340$~K: the sliced-committor estimate versus two independent sets of
equilibrium simulations. Our rates are pooled over three replicates and carry a factor
$\sim\!2$--$3$ uncertainty.
The Lindorff-Larsen \emph{et al.}~\cite{lindorff-larsen_how_2011} rates
correspond to folding and unfolding times of $0.6$ and $2.2~\mu\mathrm{s}$.}
\begin{ruledtabular}
\begin{tabular}{lccc}
 & $k_{\mathrm{fold}}$ & $k_{\mathrm{unfold}}$ & our$/$ref \\
 & ($\mu\mathrm{s}^{-1}$) & ($\mu\mathrm{s}^{-1}$) & (fold, unfold) \\
\colrule
This work & $2.2$ & $0.20$ & -- \\
Equilibrium~\cite{lazzeri_molecular_2023} & $2.5(5)$ & $0.28(5)$ & $0.9\times,\ 0.7\times$ \\
Equilibrium~\cite{lindorff-larsen_how_2011} & $1.7$ & $0.45$ & $1.3\times,\ 0.4\times$ \\
\end{tabular}
\end{ruledtabular}
\end{table}

\section*{AIB9 system setup and analysis details}

Reference equilibrium samples come from a 2.7~ms molecular dynamics
simulation of AIB9 at 500~K using the AMBER ff15ipq-m force
field~\cite{bogetti_twist_2020} in OpenMM~\cite{eastman_openmm_2017}, with a
Langevin integrator at friction $\gamma = 1.0$~ps$^{-1}$. From this trajectory
we retain every tenth saved frame, $N = 270{,}000$ in all
(Table~\ref{tab:aib9_params}). All backbone
torsion angles ($\phi$, $\psi$, $\omega$) are encoded as sine and
cosine pairs, yielding a 52-dimensional feature vector per frame; state membership
is determined in the leading two components of a TICA model fit to these features
(lag time $1$ frame), where states $A$ and $B$ are ellipsoidal basins about the two
dominant free-energy minima (Table~\ref{tab:aib9_params}).

For the empirical reference committor we exploit the time-series structure
of the equilibrium trajectory: for each frame, we trace forward in time
until the system first enters state $A$ or state $B$, assigning $q = 0$ or
$q = 1$ accordingly (frames that never reach either state are discarded; fewer
than $0.1\%$ for AIB9).
The per-frame committor values are binned onto a $50\times50$ grid in
the leading TICA plane. Because the trajectory contains many transitions between $A$
and $B$, this forward-tracing estimate converges to the true committor for
sufficiently long simulation.

\section*{Chignolin (CLN025) system setup and analysis details}

\paragraph{Umbrella-sampling data.}
Reference samples are 21 umbrella windows along the Best--Hummer fraction of
native contacts $Q$~\cite{best_native_2013} (114 native pairs, sharpness $\beta_Q = 50~\mathrm{nm}^{-1}$,
slack $\lambda = 1.8$), simulated for CLN025 with the CHARMM22* force
field~\cite{piana_robust_2011,lindorff-larsen_how_2011} in
GROMACS~\cite{abraham_gromacs_2015} at $340$~K. Window centres are spaced by
$\Delta Q = 0.05$ with a harmonic bias of strength $5000~\mathrm{kJ\,mol^{-1}}$
per unit $Q^2$; each window contributes $50$~ns of production, saved every
$1$~ps and analysed at a $5$~ps stride. Three independent replicates were run
and all $21$ windows carry data, for $N = 630{,}063$ frames pooled across the
replicates, reweighted to the unbiased ensemble with
MBAR~\cite{shirts_statistically_2008} at $340$~K. States are $A$ (folded,
$Q > 0.85$) and $B$ (unfolded, $Q < 0.30$).

\paragraph{Sliced committor.}
The committor is estimated directly in an $86$-dimensional torsion-angle
representation (backbone and side-chain torsions, sine- and cosine-encoded), the
same class of representation used for AIB9 and villin, with
$M = 256$ directions concentrated on the Fisher linear-discriminant (LDA) axis
between the folded and unfolded states (broadened by a power-spherical mixture as
for the peptides, but with its own concentration, Table~\ref{tab:chignolin_params})
and $n_\text{bins} = 100$ equal-width
bins per slice (Table~\ref{tab:chignolin_params}). Weights use the constrained
solve of Eq.~(\ref{eq:wstar}) with the MBAR sample weights. Unlike the peptides,
$\bar q$ is here only clipped to $[0,1]$ and is \emph{not} clamped to the basin
values on evaluation, so $\bar q$ attains $0.023$ on average in $A$ and $0.961$ in
$B$ rather than exactly $0$ and $1$. The clamp is redundant on this path because
the rate is read from the saddle band, which excludes the basins. TICA is used only
for the two-dimensional display in Fig.~\ref{fig:chignolin}.

\begin{table*}[t]
\caption{\label{tab:chignolin_params}Parameters for chignolin (CLN025) analysis
(Fig.~\ref{fig:chignolin}).}
\begin{ruledtabular}
\begin{tabular}{lc}
Parameter & Value \\
\hline
\multicolumn{2}{c}{\textit{Sliced committor}} \\
Number of frames $N$ & 630{,}063 (21 windows $\times$ 3 replicates) \\
Feature representation & torsion angles (86-D, sine/cosine) \\
Number of directions $M$ & 256 \\
Direction sampling & Fisher LDA, Eq.~(\ref{eq:lda_sample}) \\
LDA mean cosine $\mu$ & 0.6 \\
LDA mixture weight $\alpha$ & 0.4 \\
LDA shrinkage $\varepsilon_{\mathrm{LDA}}$ & $10^{-1}$ \\
Histogram bins $n_\text{bins}$ & 100 \\
Binning method & equal-width \\
Min.\ samples per bin $n_\text{min}$ & 10 \\
Inverse temperature $\beta$ & 1.0 \\
1D solver & RD ($\kappa = 10^{12}$) \\
Weighting & constrained solve, Eq.~(\ref{eq:wstar}) + MBAR \\
Tikhonov ridge $\varepsilon$ (on $G$) & held-out cap, Eq.~(\ref{eq:ridge_cv}) \\
$(\mu,\alpha)$ selected by & held-out cap, Eq.~(\ref{eq:ridge_cv}) \\
Boundary treatment & $\bar q$ clipped to $[0,1]$, not clamped; rate read from the saddle band \\
Random seed & 0 \\
\hline
\multicolumn{2}{c}{\textit{Umbrella sampling}} \\
Force field & CHARMM22* \\
Simulation engine & GROMACS \\
Temperature & 340 K \\
Collective variable & Best--Hummer $Q$ (114 native pairs) \\
$Q$ sharpness $\beta_Q$ / slack $\lambda$ & $50~\mathrm{nm}^{-1}$ / $1.8$ \\
Number of windows & 21 (all valid) \\
Replicates & 3 (pooled) \\
Window-centre spacing $\Delta Q$ & 0.05 \\
Harmonic bias $k_\text{umb}$ & $5000~\mathrm{kJ\,mol^{-1}}\,Q^{-2}$ \\
Production per window & 50 ns \\
Frame interval (analysis) & 5 ps \\
Reweighting & MBAR (T = 340 K) \\
\hline
\multicolumn{2}{c}{\textit{State definitions}} \\
State $A$ (folded) & $Q > 0.85$ \\
State $B$ (unfolded) & $Q < 0.30$ \\
\hline
\multicolumn{2}{c}{\textit{Rate estimation}} \\
Diffusion estimator & $D_Q = \mathrm{Var}(Q)/\tau_{\mathrm{int}}(Q)$ per window \\
Change of variables & $D_0 = D_Q/\langle|\nabla Q|^2\rangle$ (feature-space linear-response $\nabla Q$) \\
Aggregation & flux-plateau median (flatness-selected) \\
Plateau window & $q \approx [0.34,\,0.58]$ ($\sigma/\mu \approx 0.25$) \\
Rate identity & $k_{A\to B} = \nu_{AB}/p_A$ \\
$p_A$ (committor-weighted) & $0.92$ \\
Reactive flux $\nu_{AB}$ & $1.8\times10^{-7}~\mathrm{ps^{-1}}$ \\
Hummer diffusion $D_Q$ & $1.3\times10^{-6}~\mathrm{ps^{-1}}$ \\
Folding / unfolding rate & $2.2$ / $0.20~\mu\mathrm{s}^{-1}$ \\
Rate uncertainty (3 replicates) & factor $\sim\!2$--$3$ \\
\end{tabular}
\end{ruledtabular}
\end{table*}

\fi

\end{document}